\documentclass[11pt,graphicx,amsmath]{article}
\usepackage{amsmath}
\usepackage{graphicx}
\usepackage{bm}
\usepackage[dvips]{color}
\usepackage{amssymb}
\usepackage{amsfonts}
\usepackage{comment}
\usepackage{cite}
\usepackage{todonotes}
\usepackage{caption}
\usepackage{subcaption}

\def\be{\begin{equation}}
\def\ee{\end{equation}}
\def\nn{\nonumber}

\def\ba{\begin{eqnarray}}
\def\ea{\end{eqnarray}}
\def\bl#1\el{\begin{align}#1\end{align}}

\def\l{\left}
\def\r{\right}

\title{ Regularized stress tensor of vector fields  in de Sitter  space }
\author{\small
             Yang  Zhang\thanks{yzh@ustc.edu.cn} , \,
           Xuan Ye   \thanks{yyyyy@ustc.edu.cn}   ,
            \\
 \small  Department of  Astronomy,
         CAS Key Laboratory for Researches in Galaxies and Cosmology, \\
 \small  School of Astronomy and Space Sciences, \\
 \small  University of Science and Technology of China, Hefei, Anhui, 230026, China \\
 }

 \date{}

\def\be{\begin{equation}}
\def\ee{\end{equation}}
\def\nn{\nonumber}

\allowdisplaybreaks

\large

\begin{document}

\maketitle

\begin{abstract}
\large

We study the Stueckelberg field in de Sitter space,
which is a massive vector field
with the gauge fixing (GF) term $\frac{1}{2\zeta} (A^\mu\, _{;\, \mu})^2$.
We obtain the vacuum stress tensor,
which  consists of the transverse,
longitudinal,  temporal, and  GF  parts,
and each contains various UV divergences.
By the minimal subtraction rule,
we regularize  each part of the stress tensor
to its pertinent adiabatic order.
The transverse  stress tensor is regularized
to the 0th adiabatic order,
the longitudinal, temporal, and GF stress tensors
are regularized to the 2nd adiabatic order.
The resulting total regularized vacuum stress tensor is convergent
and maximally-symmetric,
has a positive energy density,
and respects the covariant conservation,
and thus can be identified as the cosmological constant
that drives the de Sitter inflation.
Under the Lorenz condition $A^\mu\, _{;\, \mu}=0$,
the regularized Stueckelberg  stress tensor reduces to
the regularized Proca  stress tensor
that contains only the transverse and longitudinal modes.
In the massless limit,
the regularized Stueckelberg stress tensor becomes zero,
and is the same as that of the Maxwell field with the GF term,
and no trace anomaly exists.
If the order of adiabatic regularization were lower than our prescription,
some divergences would remain.
If the order were higher, say,
under the conventional 4th-order regularization,
more terms than necessary would be subtracted off,
leading to an unphysical negative energy density
and the trace anomaly simultaneously.

\end{abstract}

\


    Inflationary universe, 98.80.Cq ;

    Quantum fields in curved spacetimes 04.62.+v ;

    Mathematical and relativistic  aspects of cosmology, 98.80.Jk;

    Relativity and gravitation,  95.30.Sf

\large

\section{Introduction}

The stress tensor of a quantum field in the vacuum state
generally has ultraviolet (UV) divergences.
For instance in the Minkowski spacetime,
the zero point vacuum energy is UV divergent,
and is removed through the conventional normal-ordering of field operators.
In curved spacetimes,
the vacuum stress tensor may not be simply dropped,
as its finite part can be the source of the Einstein equation
and generate gravitational field
 \cite{UtiyamaDeWitt1962,DeWitt1975,DeWittBrehme1960,FeynmanHibbs1965}.
An important example is the de Sitter inflation
which can be  driven by the vacuum stress tensor
of some quantum fields in the early universe.
The UV divergent part  must be properly removed from the stress tensor
before considering its physical  consequences.
The adiabatic regularization is a useful method
 to remove UV divergences in  $k$-space
\cite{ParkerFulling1974,FullingParkerHu1974,
HuParker1978,BLHu1978,Birrell1978,
Bunch1978,Bunch1980,AndersonParker1987,BunchParker1979,
BirrellDavies1982,ParkerToms,Parker2007,Markkanen2018,
WangZhangChen2016,ZhangWangJCAP2018},
respecting  the covariant conservation of energy
to each adiabatic order.
The issue of infrared divergences
of  a massless  field \cite{Youssef2014}
is beyond the scope of this paper.

In our previous work \cite{ZhangYeWang2020},
assuming the minimal subtraction rule \cite{ParkerFulling1974},
we have performed the adiabatic regularization on
the coupling  scalar fields in de Sitter space,
and found that
for the conformally-coupling scalar field
the 0th-order regularization
is sufficient to remove all UV divergences,
and for the minimally-coupling the 2nd-order regularization
is sufficient.
In both cases, the resulting regularized spectral stress tensor
is UV and infrared (IR) convergent and positive,
and respects the covariant conservation.
The regularized stress tensor is finite and maximally-symmetric,
has a positive energy density,
and can be regarded as the cosmological constant \cite{Weinberg1989}
that drives the de Sitter inflation.
In the massless limit, the regularized stress tensor is zero
for both the conformally-  and minimally-coupling scalar fields,
and this  is consistent with the results
of the massless scalar fields \cite{ZhangWangYe2020}.
To scrutiny  these results,
we have taken an alternative approach,
performed the point-splitting regularization in $x$-space
 \cite{ZhangYe2022PointSpl} on the scalar fields,
and obtained the same regularized stress tensor as those from
the adiabatic regularization \cite{ZhangYeWang2020,ZhangWangYe2020}.
These experiences on the scalar fields with various couplings
help us to do regularization on other types of fields.
For the Maxwell field with the GF term
in the de Sitter space \cite{ZhangYe2022} and respectively
in the radiation-dominant (RD)  stage \cite{YeZhang2024},
we derived the exact solutions, implemented the covariant canonical quantization,
and obtained the Maxwell vacuum stress tensor.
It is found that the longitudinal and temporal stress tensors cancel
to zero in the Gupta-Bleuler (GB) states
\cite{ZhangYe2022,YeZhang2024,Gupta1977,ItzyksonZuber},
only the transverse and the GF parts remain.
By the minimal subtraction rule,
we regularized the transverse vacuum stress tensor
by the 0th-order regularization,
the GF vacuum stress tensor by the 2nd-order,
and the total regularized  Maxwell stress tensor is zero.
This tells that different components of a vector field generally require
different order of regularization.
For the Maxwell field with the GF term
in the matter-dominant (MD) stage  \cite{YeZhang2024},
the GF stress tensor was regularized
by the 4th-order, instead of the 2nd-order.
This tells that, for a given field in different curved spacetimes,
in general, one needs different orders of regularization.

Refs.\cite{AllenJacobson1986,Youssef2011,FrobHiguchi2014} studied
the two-point functions of the Proca field or of the Stueckelberg field
in de Sitter space,  but did not calculate the stress tensor.
By the DeWitt-Schwinger formulation \cite{DeWitt1975},
Refs.\cite{DowkerCritchley1977,BrownCassidy1977,Endo1984}
performed the dimensional regularization
on the Stueckelberg stress tensor,
and claimed the trace anomaly in the massless limit.
However, the DeWitt-Schwinger integration is  singular and thus
undefined in the massless case \cite{DeWitt1975,ZhangYeWang2020}.
Ref.\cite{ChimentoCossarini1990}
applied the 4th-order adiabatic regularization
on the Proca and Stueckelberg stress tensors
in a general flat Robertson-Walker (fRW)  spacetime,
and pointed out new divergences in the 4th-order subtraction terms
for the longitudinal Proca stress tensor.
Ref.\cite{Salas2023} also applied the 4th-order adiabatic regularization
to  Proca stress tensor and analyzed these new divergences
in the longitudinal part
in connection with a minimally-coupling scalar field.
Ref.\cite{ChuKoyama2017} performed the 4th-order regularization
on the trace of the Stueckelberg stress tensor in de Sitter space.
In regard to the subtraction terms involved,
the 4th-order adiabatic regularization is
equivalent to the dimensional
regularization \cite{DowkerCritchley1977,BrownCassidy1977,Endo1984}.
The trace anomaly in the massless limit
claimed in Refs.\cite{DowkerCritchley1977,BrownCassidy1977,Endo1984,
ChimentoCossarini1990,ChuKoyama2017,Salas2023}
was based upon the 4th-order regularization
of a massive vector field,
and is in conflict with the zero trace of
the regularized stress tensor of
the corresponding massless vector field  \cite{ZhangYe2022,YeZhang2024}.
There must be something inconsistent.

In this paper we  study  the Stueckelberg field
which is a massive vector field with the GF term
in de Sitter space.
The vacuum stress tensor consists of
the transverse, longitudinal, temporal, and GF parts.
Instead of the 4th-order regularization,
we shall regularize each part of the stress tensor
to its pertinent adiabatic order
according to  the minimal subtraction rule  \cite{ParkerFulling1974}
(specifically, the 0th-order for the transverse part,
the 2nd-order for the longitudinal, temporal, and GF parts),
and obtain the convergent spectral  stress tensor,
which, after $k$-integration, yields
the maximally-symmetric regularized stress tensor
with a positive energy density.
In particular, in the massless limit,
the regularized Stueckelberg  stress tensor is zero,
and agrees with that of the Maxwell field with the GF term \cite{ZhangYe2022},
and the conflict disappears.
We shall also show that under relevant conditions
the Stueckelberg vacuum stress tensor reduces to the Proca vacuum stress tensor,
and respectively to the Maxwell vacuum stress tensor with the GF term.
Moreover,  we shall examine
the conventional 4th-order regularization in great details,
and calculate the 4th-order subtraction terms
for each part of the Stueckelberg stress tensor,
and demonstrate that the 4th-order regularization leads to
an unphysical,  negative total energy density
and the so-called trace anomaly simultaneously.
In addition,
we also show that, in generic  fRW spacetimes,
 the new divergences in the 4th-order terms
occur not only in the longitudinal part
\cite{ChimentoCossarini1990,Salas2023},
but also in the temporal and GF parts as well.

The paper is organized as follows.
Sect. 2  gives the solutions of the Stueckelberg field in de Sitter space.
Sect. 3  presents the covariant canonical quantization
         and the vacuum stress tensor.
In Sect. 4, we perform pertinent regularization
         on each part of the  vacuum stress tensor
         and give the regularized Stueckelberg and Proca stress tensors.
Sect. 5  gives the reduction of the Stueckelberg stress tensor
         to the Proca stress tensor,
         and respectively, in the massless limit,
         to the Maxwell stress tensor with the GF term.
Sect. 6  demonstrates that the negative total energy density and the trace anomaly
         are caused by the improper 4th-order regularization.
Sect. 7 gives conclusions and  discussions.
In Appendix A,  based on the WKB solutions,
we present the adiabatic subtraction terms up to the 4th-order
 and the new divergences,
and the trace anomaly carried by the 4th-order terms above the 2nd-order ones.
Appendix B  gives the massless limit of
the solutions of the longitudinal and temporal components.
We shall work in units $\hbar=c=1$.

\section{ Solution of Stueckelberg field in de Sitter space }

The Proca field is a massive vector field
with the Lagrangian density
\bl
{\cal L}_{Pr} &
 = \sqrt{-g}\Big( -\frac14 g^{\mu\rho}g^{\nu\sigma}
    F_{\mu\nu} F_{\rho\sigma}
     -\frac12 m^2 g^{\mu\nu}A_\mu A_\nu
        \Big) \, ,
\el
with  $F_{\mu\nu} = A_{\mu, \nu}- A_{\nu, \mu}$,
and its field equation is given by
\bl\label{Procaeq}
F^{\mu\nu}\, _{;\nu} +  m^2 A^\mu =0 ,
\el
where `` $;$ "  denotes the covariant differentiation
with respect to the curved spacetime background.
Due to the mass term,  the Proca field has no gauge invariance,
unlike  the Maxwell field \cite{ZhangYe2022}.
The covariant four-divergence of \eqref{Procaeq} leads to
\bl\label{Lsunb}
A^\mu\, _{; \, \mu} =0 ,
\el
which is the Lorenz condition,
so only three components of $A^\mu$
are independent dynamical degrees of freedom.
In the massless limit $m=0$,
the propagators of the Proca field are singular \cite{ItzyksonZuber}.
To allow for the massless limit,
one adds a gauge fixing (GF) term by hand,
and works with the Stueckelberg field with the Lagrangian density
\cite{ItzyksonZuber,ChuKoyama2017,ChimentoCossarini1990}
\bl\label{lagr}
{\cal L}_{St} &
 = \sqrt{-g}\Big( -\frac14 g^{\mu\rho}g^{\nu\sigma}
    F_{\mu\nu} F_{\rho\sigma}
     -\frac12 m^2 g^{\mu\nu}A_\mu A_\nu
    -\frac{1}{2\zeta} (A^\mu_{~ ; \mu} )^2
     \Big) \, ,
\el
and the Stueckelberg field equation
\bl
 F^{\mu\nu }\, _{;\, \nu}  + m^2  A^\mu
    - \frac{1}{\zeta}  g^{\mu\nu} (A^\sigma_{~ ; \sigma} )_{;\, \nu}=0 ,
  \label{Fmmeq}
\el
where $\zeta $ is the GF parameter.
$\zeta=1$ is referred to as the Feynman gauge, $\zeta=0$ as the Landau gauge,
and $\zeta =\infty$  as the unitary gauge.
Due to the mass term, the Stueckelberg field is not a gauge field either.
The original Stueckelberg theory contains an additional scalar field
$B$ to maintain the gauge invariance \cite{Ruiz2003}.
For the purpose of studying the regularization scheme
of a generic vector field which can reduce to
Maxwell field and Proca field,
we consider only the simplest Stueckelberg field \eqref{lagr}
without a scalar $B$ field.
All the four components $A^\mu$ of
the Stueckelberg field are dynamical degrees of freedom,
since the condition \eqref{Lsunb} is not imposed,
unlike the Proca field.
The covariant four-divergence of \eqref{Fmmeq} leads to
$(A^\mu\, _{;\, \mu })^{;\nu}_{~~  ;\nu }
   - \zeta m^2  A^\mu\, _{;\, \mu } =0  $,
which tells that
$A^\mu\, _{;\, \mu}$ satisfies the equation of
a minimally-coupling scalar  field with a mass $\zeta m^2$.
The Stueckelberg field is interesting in that,
under pertinent conditions, it reduces to the Proca field,
and,  respectively,
to the Maxwell field with GF term in the limit $m=0$  \cite{ZhangYe2022},
as we shall show in Sect 5.
The Stueckelberg field  studied in this paper is different from
the massive gauge field studied in Ref.\cite{BelokogneFolacci2016}.

In a fRW  spacetime
\be \label{metricmm}
ds^2=a^2(\tau)[- d\tau^2 + \delta_{ij}   dx^idx^j] ,
\ee
the Stueckelberg equation \eqref{Fmmeq} is written  as
\bl\label{equaAmzeta}
   \eta^{\sigma \rho} \partial_\sigma  \partial_\rho A_\nu &
      +\Big(  \frac{1}{\zeta} -1 \Big)
      \partial_\nu (\eta^{\rho\sigma} \partial_\sigma  A_\rho)
      -  m^2 a^2 A_\nu
        \nn \\
& + \frac{1}{\zeta}
  \Big[ \delta_{\nu 0}
  \Big( - D \eta^{\rho\sigma } \partial_\sigma  A_\rho  + D^2 A_0 - D' A_0  \Big)
         -  D  \partial_\nu A_0  \Big]    =0 ,
\el
where $\eta_{\mu\nu} = \text{diag}(-1,1,1,1)$
is the  Minkowski  metric, and $D \equiv 2 a'/a$
with $'$ denoting the derivative with respect to the conformal time.
(For easy comparison we use notations close to
Refs.\cite{ChuKoyama2017,FrobHiguchi2014,ZhangYe2022}.)
The $i$-component is written as
\bl
A_i  & = B_i +\partial_i A,
\el
where $B_i$ are the transverse components satisfying  $\partial_i B_i =0$,
and $A$ is the longitudinal component.
For convenience, we shall work with the Fourier $k$-modes.
Eq.\eqref{equaAmzeta} in $k$-modes is decomposed into
 the following three equations
\bl
\partial_0^2 B_i + (k^2 +m^2 a^2)   B_i   =0  ,\label{Bieq1}
    \\
- \partial_0^2 A -\frac{1}{\zeta} k^2 A - m^2 a^2 A
   + (1- \frac{1}{\zeta} )\partial_0  A_0 - \frac{1}{\zeta} D A_0
     =0 , \label{0aA}
     \\
- \frac{1}{\zeta}\partial_0^2  A_0  -k^2 A_0 - m^2 a^2 A_0
  +\frac{1}{\zeta} ( D^2  -D' ) A_0
  + k^2 \big( (1- \frac{1}{\zeta} )\partial_0  A + \frac{1}{\zeta}  D A \big)
            =0  .   \label{Aa0m}
\el
where  we also use $B_i$, $A$ and $A_0$ to represent
their $k$-modes when no confusion arises.
The components $A$ and $A_0$ are mixed up in \eqref{0aA} \eqref{Aa0m}.
We need the canonical momenta for canonical quantization,
which are defined by
\bl\label{pi0m}
\pi_A^\mu & = \frac{\partial \cal L}{\partial (\partial_0 A_\mu)}
 =     \eta^{\mu \sigma}
     (\partial_0  A_\sigma - \partial_\sigma A_0 )
     - \frac{1}{\zeta}  \eta^{0 \mu}
     \Big( \eta^{\rho\sigma}  \partial_\sigma  A_\rho - D A_0 \Big).
\el
The  $0$- and $i$-components are
\bl   \label{pi0defm}
\pi_A^0   &  = \frac{1}{\zeta} \,  a^2  A^\mu\, _{; \, \mu}
    = \frac{1}{\zeta} \Big( -(\partial_0 + D) A_0 +\nabla^2  A \Big) ,
\\
\pi_A^i  &  =  \delta^{i j} \Big(\partial_0  A_j - \partial_j A_0 \Big)
   = w^i +\partial^i\pi_A    ,
               \label{piImdef}
\el
where
\bl
w^ i &  = \partial_0  B_i ,
    \label{wi}
 \\
\pi_A & = \partial_0   A  -   A_0 \, .
      \label{piacm}
\el
With $\nabla^2 = -k^2$,
the $k$-mode of \eqref{pi0defm} is
\be
\pi_A^0    =  - \frac{1}{\zeta}
  \big( (\partial_0 + D) A_0 + k^2  A \big) ,
  \label{pi0def}
\ee
which is contributed by the GF term.
We also use $\pi_A$, $\pi_A^0$
to represent their $k$-modes.

In this paper we consider the de Sitter space with
the scale factor
\be
a(\tau) = - \frac{1}{H \tau}, ~~~~ -\infty < \tau <\tau_1 ,
\ee
where $H$ is a constant, and $\tau_1$ is the ending time of inflation.
 $D =-2/\tau$ in de Sitter space.
The solutions of eqs.\eqref{Bieq1} \eqref{0aA} \eqref{Aa0m} in de Sitter space
are easier to derive than those of
the Maxwell equations  \cite{ZhangYeWang2020}.
The equation \eqref{Bieq1} of the transverse components $B_i$
is the same as the rescaled conformally-coupling
massive scalar field  \cite{ZhangYeWang2020},
and has  the  positive frequency solution
\be \label{f1f2}
B_i \propto f^{(\sigma)}_k (\tau)
 \equiv  \sqrt{\frac{\pi}{2}} \sqrt{\frac{x}{2k}}
     e^{i \frac{\pi}{2}(\nu+\frac12)}
     H^{(1)}_{\nu}(x) ,
  ~~~~~  \nu =\sqrt{\frac14 -\frac{m^2}{ H^2}} ,
\ee
with $x \equiv -k\tau$
and  $\sigma =1,2$ for two transverse polarizations.
In the massless limit $m=0$, $\nu=\frac 12$,
\eqref{f1f2} reduces to
$f^{(\sigma)} =\frac{1}{\sqrt{2k}} e^{-i k\tau}$.

The basic equations \eqref{0aA} \eqref{Aa0m} are mixed up for $A$ and $A_0$.
By differentiation and combination,
eqs.\eqref{0aA} \eqref{Aa0m} are written as
the following  third order  equations
\bl
\big(\partial_0^2  - D \partial_0  + k^2 + m^2 a^2 \big)
\Big(\partial_0 A  - A_0 \Big)  & =0 ,
          \label{piAeqmass2}
  \\
\big( \partial_0^2  - D \partial_0  - D'
       +k^2 + \zeta m^2 a^2 \big) \Big((\partial_0 + D) A_0 +  k^2 A \Big) &  =0  .
  \label{pi0Aeqmass2}
\el
With the help of \eqref{piAeqmass2} \eqref{pi0Aeqmass2},
eqs.\eqref{0aA} \eqref{Aa0m}  can be converted into
the separate, fourth order differential equations
of $A$ and $A_0$ as the following
\bl
&  - \partial_0^2 A -\frac{1}{\zeta} k^2 A - m^2 a^2 A
\nn
\\
&  + \Big[ (1- \frac{1}{\zeta} )\partial_0 - \frac{1}{\zeta} D  \Big]
 \Bigg[ \Big[ -\zeta(k^2 + m^2 a^2)
   + ( D^2 -D') -\frac{1}{\zeta -1} D^2 + (k^2 + m^2 a^2) \Big]^{-1}
   \nn \\
&  \times \Big[ \big(\partial_0^2  - D \partial_0  + k^2 + m^2 a^2 \big)
  \partial_0 A
 -(\zeta- 1 ) k^2 \partial_0  A - k^2 D A
 \nn
 \\
 & +  \frac{\zeta}{\zeta -1} D  \partial_0^2 A
   + \frac{1}{\zeta -1} k^2 D A
   + \frac{\zeta}{\zeta -1}  m^2 a^2 D A \Big]  \Bigg]
     =0  \, , \label{0aA3}
\el
\bl
& - \frac{1}{\zeta}\partial_0^2  A_0  -k^2 A_0 - m^2 a^2 A_0
  +\frac{1}{\zeta} ( D^2  -D' ) A_0
\nn
\\
& + \Big[(1- \frac{1}{\zeta} ) k^2 \partial_0 + \frac{1}{\zeta} k^2 D \Big]
  \Bigg[ \Big( (\frac{1}{\zeta}-1)  k^4 +  k^2  D' - \frac{1}{\zeta -1} k^2 D^2
  + (1 - \zeta) k^2 m^2 a^2  \Big)^{-1}
 \nn \\
&   \times
  \Big[ \big( \partial_0^2 - D \partial_0 - D' +k^2 + \zeta m^2 a^2 \big)
   \Big( (\partial_0 + D) A_0  \Big)
 +  k^2 \Big( (1- \frac{1}{\zeta} )\partial_0  A_0 - \frac{1}{\zeta} D A_0 \Big)
\nn \\
&  - \frac{1}{(\zeta - 1) } D  \partial^2_0 A_0
 +  \frac{1}{( \zeta - 1 )} D  (D^2 - D') A_0
- \frac{\zeta}{\zeta - 1} D  k^2 A_0
-  \frac{\zeta}{\zeta - 1} D  m^2 a^2 A_0 \Big] \Bigg]  =0 \, .
\label{st4eqA0}
\el
Eqs. \eqref{0aA3} \eqref{st4eqA0} have the positive frequency solutions
\bl
A&=\frac{1}{a^2 m^2}\Big( - c_1 \partial_0  f^{(3)}_k + c_2 f^{(0)}_k \Big)
\nn \\
& = \Big[ c_1 e^{\frac{i\pi}{2} (\nu-\nu_0)}
\big( 2x H^{(1)}_{\nu-1}(x)-(1+2\nu)
          H^{(1)}_{\nu}(x) \big)
+ c_2 2xH^{(1)}_{\nu_0}(x)  \Big]\frac{\sqrt{\pi x}}{4k^{3/2}}
\frac{H}{m}e^{i\frac{\pi}{2}(\nu_0+\frac12)},
 \label{Asol}
\\
A_0 & =\frac{1}{a^2  m^2}\Big( c_2 (\partial_0  f^{(0)}_k -  D f^{(0)}_k )
      +  k^2  c_1 f^{(3)}  \Big)
      \nn \\
& =   \Big[c_1 2x H^{(1)}_{\nu}(x)
- c_2  e^{i\frac\pi2(\nu_0-\nu)}
\big( 2xH^{(1)}_{\nu_0-1}(x)-(-3+2\nu_0)H^{(1)}_{\nu_0}(x) \big) \Big]
\frac{\sqrt{\pi x}}{4k^{1/2}}\frac{H}{m}e^{i\frac\pi2(\nu+\frac12)}  ,
  \label{A0sol}
\el
where  $c_1$ and $c_2$ are constants, and
\bl \label{pisoldestt}
f^{(3)}_k \equiv  \frac{m}{k} a(\tau) Y^{(L)}_k(\tau) ,
\\
\label{pia0}
 f^{(0)}_k \equiv   m a(\tau) Y^{(0)}_k(\tau) ,
\el
with
\bl \label{Y3sol}
Y^{(L)}_k(\tau) & = \sqrt{\frac{\pi}{2}} \sqrt{\frac{x}{2k}}
     e^{i \frac{\pi}{2}(\nu+\frac12)}
     H^{(1)}_{\nu}(x) ,
~~~ ~~  \nu = \sqrt{\frac14 -\frac{m^2}{ H^2}}  ,
\\
Y^{(0)}_k(\tau) & =  \sqrt{\frac{\pi}{2}} \sqrt{\frac{x}{2k}}
     e^{i \frac{\pi}{2}(\nu_0+\frac12)}
     H^{(1)}_{\nu_0}(x),
~~~~ \nu_0 =\sqrt{\frac94 - \zeta \frac{m^2}{ H^2}} .
\label{Y0solmo}
\el
In the massless limit $m=0$,
\bl \label{YLsolm0}
 Y^{(L)}_k(\tau)   & =\frac{1}{\sqrt{2k}} e^{-i k\tau} ,
 \\
Y^{(0)}_k(\tau)  & =  \sqrt{\frac{\pi}{2}} \sqrt{\frac{x}{2k}}
           e^{i \pi}  H^{(1)}_{\frac32}(x)
  = \frac{1}{\sqrt{2k}} (1- \frac{i}{k \tau})e^{- i k\tau}  .
  \label{Y0solm0}
\el
Notice that the solution \eqref{Y0solm0} is the same as that of
a minimally-coupling massless scalar field in de Sitter space.
In particular,
the parameters $\zeta$ and $m^2$ occur in a combination $\zeta m^2$
in the mode $Y^{(0)}_k$,
for which the limit $m=0$ is equivalent to the Landau gauge $\zeta=0$.

The $c_1$ and $c_2$ parts of \eqref{Asol} \eqref{A0sol}
are  respectively two independent positive-freq solutions,
and their complex conjugates are
two independent negative-freq solutions,
together,
they form the complete set of solutions of   \eqref{0aA} \eqref{Aa0m}.
The  Stueckelberg field solutions \eqref{Asol} \eqref{A0sol}
reduce to the solutions of the Proca field,
under the Lorenz condition \eqref{Lsunb}, ie, $\pi_A^0  =0$.
In the massless limit $m=0$,  eqs.\eqref{0aA3}  \eqref{st4eqA0}
and the solutions \eqref{Asol} \eqref{A0sol}
reduce to those of the Maxwell field
with the GF term \cite{ZhangYe2022} (see appendix B).

The solutions \eqref{Asol} \eqref{A0sol} can be derived
alternatively as the following.
Eqs.\eqref{piAeqmass2} \eqref{pi0Aeqmass2} can be written as
\bl
\big(\partial_0^2  - D \partial_0  + k^2 + m^2 a^2 \big) \pi_A  & =0 ,
          \label{piAeqmass}
  \\
\big( \partial_0^2  - D \partial_0  - D'
       +k^2 + \zeta m^2 a^2 \big) \pi^0_A   &  =0  ,
  \label{pi0Aeqmass}
\el
which have  the positive frequency solution
\bl \label{pisoldestt}
\pi_A= c_1 f^{(3)}_k ,
\\
\label{pia0}
\pi_A^0 = c_2  f^{(0)}_k .
\el
From the basic  eqs.\eqref{0aA} \eqref{Aa0m},
we express the fields $A$ and $A_0$
in terms of the canonical momenta
\bl \label{ApiA}
A  & = \frac{1}{a^2 m^2}\Big( -\partial_0 \pi_A   +\pi^0_A \Big),
\\
A_0 & = \frac{1}{a^2  m^2}\Big(   \partial_0 \pi^0_A - D \pi^0_A
      +  k^2  \pi_A \Big) .
      \label{A_0}
\el
Plugging the solutions  \eqref{pisoldestt}  \eqref{pia0}
into \eqref{ApiA} \eqref{A_0} yields
 the solutions \eqref{Asol} \eqref{A0sol} consistently.

\section{ Covariant canonical quantization
       and stress tensor of   Stueckelberg field }

We shall implement the  canonical quantization
and require the field operators
satisfy the equal-time covariant canonical  commutation relations
\bl
[A_\mu(\tau,{\bf x}), \pi_A^\nu (\tau,{\bf x}') ]
   &  = i  \delta^{\nu}_{\mu} \delta^{(3)}({\bf x}- {\bf x}')   ,
   \label{commAmupi}
   \\
[A_\mu(\tau,{\bf x}), A_\nu (\tau,{\bf x}') ]    &  =0,
 \label{AAcommmunu}   \\
[\pi_{A}^{\mu} (\tau,{\bf x}), \pi_A^\nu (\tau,{\bf x}') ]
   &  =0   ,    \label{commpipi}
\el
the $ij$ components of \eqref{commAmupi}
is decomposed into
\bl \label{AiPiAmassive}
[A_i,\pi_A^j] =[B_i,  w^j] + [\partial_i A, \partial^j \pi_A ].
\el
The transverse fields and canonical momenta are written as
\bl  \label{expBi2}
B_i   ({\bf x},\tau) & =  \int\frac{d^3k}{(2\pi)^{3/2}}
 \sum_{\sigma=1}^2 \epsilon^\sigma_{i}(k)
  \left[ a^{(\sigma)} _{\bf k}  f_k^{(\sigma)}(\tau) e^{i\bf{k} \cdot\bf{x}}
  +a^{(\sigma) \dagger}_{\bf k} f_k^{(\sigma)*}(\tau) e^{-i\bf{k}\cdot\bf{x}}
    \right]  ,
\\
w^i (\tau, {\bf x})
&  =  \int \frac{d^3k}{(2\pi)^{3/2}}
 \sum_{p=1}^2 \epsilon^\sigma_i(k)
 \Big[ a_{\bf k}^{(\sigma)} f_k^{(\sigma)'}(\tau) e^{i \bf k\cdot x}
   + a_{\bf k}^{(\sigma)\dagger} f_k^{(\sigma)*'}(\tau)
      e^{- i \bf k\cdot x} \Big] , \label{wjexp1}
\el
where $f_k ^{(\sigma)}$ with $\sigma=1,2$ are the transverse modes   (\ref{f1f2}),
the polarization vector satisfies
\bl   \label{polarexpm0}
\sum_{i=1,2,3}  k^i \epsilon^\sigma _i(k) =0,
\\
\sum_{i=1,2,3} \epsilon^\sigma _{i}(k ) \epsilon^{\sigma'} _{i}(k)
    = \delta^{\sigma \sigma'},   \label{polarexpmsum}
    \\
\sum_{\sigma=1,2}   \epsilon^\sigma_i(k) \epsilon^\sigma_j(k)
   = \delta_{ij}-\frac{k_i k_j}{k^2} \, .
\el
The longitudinal and temporal
canonical momenta $\pi_A$ and $\pi_A^0$ are expanded as
\bl
\pi_A
&=\int\frac{d^3k}{(2\pi)^{\frac32}}\Big((a_{\bf k} ^{(0)}
\pi_{A1k}+a_{\bf k} ^{(3)}\pi_{A2k})
   e^{i\bf{k} \cdot\bf{x} }+h.c.\Big),
      \label{113}
   \\
\pi_A^0
&=\int\frac{d^3k}{(2\pi)^{\frac32}}\Big((a_{\bf k} ^{(0)}
\pi_{A1k}^{0}
+a_{\bf k} ^{(3)}\pi_{A2k}^{0})e^{i\bf{k} \cdot\bf{x}}+h.c.\Big),
      \label{pi0}
\el
where
\bl
\pi_{A1k}&=c_1 f^{(3)}_k,\label{moderesolutionpiA}
\\
\pi_{A2k}&=m_1 f^{(3)}_k ,\label{moderesolutionpiA}
\\
\pi_{A1k}^{0}&=c_2 f^{(0)}_k ,\label{moderesolutionpiA0}
\\
\pi_{A2k}^{0}&=m_2 f^{(0)}_k  ,\label{moderesolutionpiA0}
\el
with the modes  $f^{(3)}_k$ and $f^{(0)}_k$
given by (\ref{pisoldestt}) (\ref{pia0}),
and $c_1$, $c_2$, $m_1$, $m_2$ being coefficients to be determined.
The commutator of creation and annihilation operators are
\bl \label{aa0commu}
[a_{\bf k}^{(\mu)}, a_{\bf k}^{(\nu)\dagger}]
= \eta^{\mu\nu}  \delta^{(3)}({\bf k-k'}) .
\el
The expansions \eqref{113} \eqref{pi0} contain
both $a^{(3)}_{\bf k}$ and $a^{(0)}_{\bf k}$,
and are more general than that in Ref.\cite{FrobHiguchi2014,ChuKoyama2017}.
From the relations (\ref{ApiA}) (\ref{A_0})  follow
\bl
A&=\int\frac{d^3k}{(2\pi)^{\frac32}}
\Big((a^{(0)}_{\bf k } A_{1k}+a^{(3)}_{\bf k} A_{2k})
e^{i\bf{k} \cdot\bf{x}}+h.c.\Big),\label{Amexp}
\\
A_0&=\int\frac{d^3k}{(2\pi)^{\frac32}}
\Big((a^{(0)}_{\bf k }A_{01k}+a^{(3)}_{\bf k } A_{02k})
e^{i\bf{k} \cdot\bf{x}}+h.c.\Big) ,  \label{A0mexp}
\el
where
\bl
A_{1k}  &= \frac{1}{a^2 m^2}\Big( - c_1 \partial_0  f^{(3)}_k
   + c_2 f^{(0)}_k  \Big)
     ,\label{A1so}
\\
A_{2k}
&=\frac{1}{a^2 m^2}\Big( - m_1 \partial_0  f^{(3)}_k + m_2 f^{(0)}_k \Big),
      \label{A2sol}
\\
A_{01k}
&= \frac{1}{a^2  m^2}\Big( c_2 (\partial_0  f^{(0)}_k -  D f^{(0)}_k)
      +  k^2  c_1 f^{(3)}_k  \Big)  ,\label{A01s}
\\
A_{02k}
&=\frac{1}{a^2  m^2}\Big( m_2 (\partial_0  f^{(0)}_k -  D f^{(0)}_k)
      +  k^2  m_1 f^{(3)}_k  \Big) . \label{A02sole}
\el
The operators  $A$ and $A_0$
given by  \eqref{Amexp} \eqref{A0mexp}
satisfy the basic equations \eqref{0aA} \eqref{Aa0m}.
From  the  commutation relations
\eqref{commAmupi}  \eqref{AAcommmunu} \eqref{commpipi} \eqref{aa0commu},
we obtain the following constraints on the coefficients
\bl
&|m_1|^2-|c_1|^2=1 , \label{antherconij}
\\
&|c_2|^2-|m_2|^2=1 , \label{constraint001}
\\
&c_1^*c_2-m_1^*m_2=0 . \label{constraint002}
\el
There  are infinitely many choices of coefficients
satisfying \eqref{antherconij} \eqref{constraint001} \eqref{constraint002},
for instance,  a simple choice,
\be
(c_1,~c_2,~m_1,~m_2)=(0,1,1,0) ,  \label{chucoff}
\ee
corresponds to the one taken in Ref.\cite{FrobHiguchi2014,ChuKoyama2017},
and another simple choice is
$c_1 =1$, $m_1=\sqrt 2$, $c_2= \sqrt 2$, $ m_2=1$.
We have checked that the vacuum stress tensor is the same
for all the different choices of the coefficients
satisfying  \eqref{antherconij} \eqref{constraint001} \eqref{constraint002}.

From  the  action $S = \int {\cal L}_{St}  d^4 x$
the Stueckelberg stress tensor,
$T _{\mu\nu}  = -\frac{2}{\sqrt{-g}}\frac{ \delta S}{\delta g^{\mu\nu}}$,
is given by
\bl \label{Tmunuelctro}
T  _{\mu\nu}
& = F_{\mu\lambda} F_{\nu}\, ^{\lambda}
      -\frac14 g_{\mu\nu} F_{\sigma \lambda} F^{\sigma \lambda}
      -\frac12 g_{\mu\nu} m^2 A^\sigma A_\sigma + m^2 A_\mu A_\nu
          \nn \\
& +  \frac{1}{\zeta} \Big[\frac{1}{2  }  g_{\mu\nu} (A^\sigma\, _{;\sigma})^2
  +  g_{\mu\nu} A^\lambda  A^\sigma\, _{; \sigma , \lambda}
  - A^\sigma\, _{; \sigma , \mu}A_\nu
  - A^\sigma\, _{; \sigma , \nu}A_\mu  \Big] .
\el
The energy density is $\rho = - T^0\,_0$,
 the pressure is $p = \frac13 T^j\, _j$,
and the  trace  is
\bl \label{trace}
T^\mu\, _\mu =-\rho + 3p
&  = \frac{2}{\zeta} \big( A^\sigma\, _{;\sigma} A^\rho \big)_{; \, \rho}
    -  m^2 A^\sigma A_\sigma ,
\el
which is contributed by the GF and mass terms.
For convenience,
we group the stress tensor into the following three parts.
The transverse stress tensor is
\bl \label{Bstresstensor}
\rho^{TR} & =  \frac{1}{2} a^{-4}
\Big( B_j ' B_j' + B_{i,j} B_{i,j} + m^2 a^2 B_i  B_i \Big) ,
\\
p^{TR} & =  \frac{1}{6} a^{-4}
\Big( B_j ' B_j' + B_{i,j} B_{i,j} - m^2 a^2 B_i  B_i \Big) .
\label{Bpr}
\el
The longitudinal and temporal (LT) stress tensor  is
\bl
\rho^{LT} & =  \frac{1}{2} a^{-4}
          \Big( A'_{,j} A'_{,j} + m^2 a^{2} A_{,i} A_{,i}
           +A_{0,j}A_{0,j}  +m^2 a^{2} A_0 A_0
          -  2 A_{0,j}  A_{,j 0}   \Big), \label{longittemporalrho2}
           \\
p^{LT}   &  =  \frac12   a^{-4} \frac13
          \Big( A'_{,j} A'_{,j} - m^2  a^2  A_{,i}  A_{,i}
           +A_{0,j}A_{0,j} +  3 m^2 a^2 A_0 A_0
           -  2 A_{0,j}  A_{,j 0} \Big) , \label{longittemporal2}
\el
which has cross terms $  A_{0,j}  A_{,j0}$.
The GF   stress tensor is
\bl
\rho^{GF}
 = &  - \frac{1}{a^4} \Big[  \frac{1}{2} \zeta  ( \pi_A^0  )^2
      + A_0 \Big(  \partial_0 \pi_A^{0}  - D \pi_A^0  \Big)
      + A_{,j} \pi^0_{A\, ,j}   \Big] , \label{rhoGFvev}
\\
 p^{GF}  = &     \frac{1}{a^4}  \Big[\frac{1}{2} \zeta (\pi_A^0)^2
  - A_0 \Big(\partial_0 \pi_A^{0} - D \pi_A^0 \Big)
  + \frac13 A_{,j} \pi^0_{A\, ,j}  \Big] . \label{GFpre}
\el

\begin{figure}[htb]
\centering
\subcaptionbox{}
    {%
        \includegraphics[width = .41\linewidth]{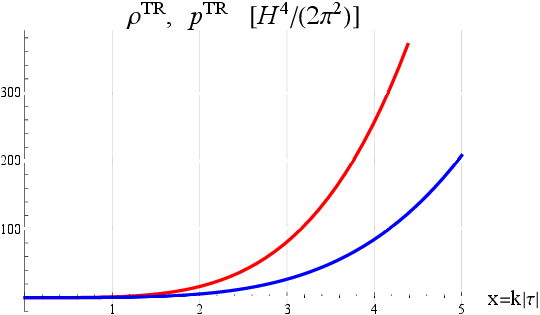}
}
\subcaptionbox{}
    {%
        \includegraphics[width = .41\linewidth]{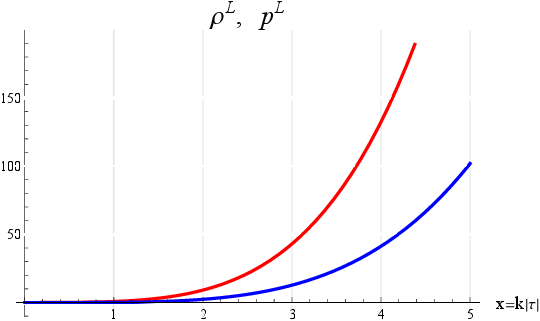}
    }
\subcaptionbox{}
    {%
        \includegraphics[width = .41\linewidth]{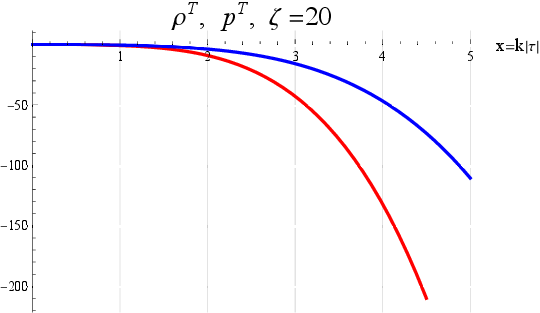}
    }
\subcaptionbox{}
    {%
        \includegraphics[width = .41\linewidth]{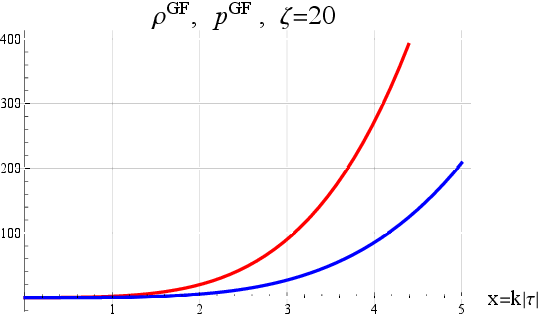}
    }
\caption{ The unregularized spectral stress tensor.
The unit of $\rho_k$ and $p_k$ are $H^4/2\pi^2$.
(a) the transverse;
(b) the longitudinal;
(c) the temporal;
(d) the GF.
Red: the spectral energy density, Blue: the spectral pressure.
For illustration,
   $\frac{m^2}{H^2}=0.1$,  $\zeta=20$,
   and $|\tau|=1$ are taken in Figure 1 through Figure 6,
   except otherwise  specified.
}
\label{unregular}
\end{figure}

We need  the expectation value of $T_{\mu\nu}$
in the quantum state $|\psi \rangle$,
which is built upon the Bunch-Davies vacuum state
defined by $a_{\bf k}^{(\mu)} |0\rangle = 0$.
By use of the expansion $B_i$ in (\ref{expBi2}),
the transverse  stress tensor in the state $|\psi \rangle$
is given  by the following
\bl \label{rhomasslessdeSitter}
\langle \psi|  \rho^{TR} |\psi \rangle
=&   \int^{\infty}_0   \rho^{TR}_k   \frac{d k}{k}
  + \int\frac{d k}{k}   \rho^{TR}_k
   \sum_{\sigma=1,2}
  \langle \psi|a_{\bf k}^{(\sigma)\dag} a_{\bf k}^{(\sigma)} |\psi \rangle,
  \\
\langle \psi|  p^{TR} |\psi \rangle
=&   \int^{\infty}_0   p^{TR}_k   \frac{d k}{k}
  + \int\frac{d k}{k}   p^{TR}_k
   \sum_{\sigma=1,2}
  \langle \psi|a_{\bf k}^{(\sigma)\dag} a_{\bf k}^{(\sigma)} |\psi \rangle,
\label{pressdeSitter}
\el
where
\bl
\rho^{TR}_k  & =  \frac{k^3}{2\pi^2 a^4}
  \Big[ |f_k^{(1)'}(\tau)|^2 + k^2 |f_k^{(1)}(\tau)|^2
  + m^2 a^2 |f_k^{(1)}(\tau)|^2 \Big] ,
   \label{spTRPRho}
 \\
p^{TR}_k & = \frac{k^3}{2\pi^2 a^4}
   \frac13 \bigg[  |f_k^{(1)'}(\tau)|^2
      +  k^2 |f_k^{(1)}(\tau)|^2
      -  m^2 a^2  |f_k^{(1)}(\tau)|^2  \bigg]   ,
      \label{spTRPRess}
\el
are the transverse  spectral  energy density and spectral pressure,
 independent of $\zeta$.
With the two transverse mode functions  $f^{(1)}_k=f^{(2)}_k$,
the transverse spectral stress tensor \eqref{spTRPRho} \eqref{spTRPRess}
is twice that of
the conformally-coupling massive scalar field \cite{ZhangYeWang2020}.
In \eqref{rhomasslessdeSitter} \eqref{pressdeSitter}
the first terms
\bl
\int^{\infty}_0 \rho^{TR}_k  \frac{dk}{k}
  & \equiv   \langle 0| \rho^{TR}  |0 \rangle  ,
~~~~~~
 \int^{\infty}_0 p^{TR}_k  \frac{dk}{k}
    \equiv \langle 0| p^{TR}  |0 \rangle    ,
    \label{transvrhopress}
\el
are the vacuum part,
and the remaining terms are the particle  part.
In this paper, we consider only  the stress tensor
in the Bunch-Davies vacuum state.
The properties of vacuum states in de Sitter space were discussed
 in Refs. \cite{Allen1985,Borchers1999}.
Hereafter we focus the vacuum stress tensor,
which  contains UV divergences.
By their  structure,
$\rho^{TR}_k$ and $p^{TR}_k  $ are twice
that of the conformally-coupling massive scalar field \cite{ZhangYeWang2020},
$p^{TR}_k \ne \frac13 \rho^{TR}_k$ due to  the mass term.
It is checked that
the transverse spectral stress tensor respects the covariant conservation
\bl \label{csvt}
\rho^{TR}_{k}\, '  +3 \frac{a'}{a} (\rho^{TR}_{k}+p^{TR}_{k})=0 .
\el
Fig.\ref{unregular} (a) shows
the unregularized transverse  $\rho^{TR}_k$  and $p^{TR} _k$,
which are UV divergent $\propto k^4$ at high $k$.
These UV divergences in the stress tensor will be removed
by appropriate regularization in the next section.

By  use of \eqref{Amexp} \eqref{A0mexp}
and the constraints \eqref{antherconij} \eqref{constraint001} \eqref{constraint002},
 the  LT stress tensor in the vacuum state is
\bl
\langle0| \rho^{LT} |0 \rangle
= \int \rho^{LT} _k \frac{dk }{k} ,
~~~~
\langle0| p^{LT} |0 \rangle
= \int p^{LT} _k \frac{dk }{k}  ,
             \label{longip}
\el
where the LT spectral stress tensor is further written as
\bl
\rho^{LT}_k  = \rho^{L}_k + \rho^{T}_k    ,
~~~~~
p^{LT}_k    = p^{L}_k + p^{T}_k     ,
\el
with  the longitudinal part
\bl
\rho^{L}_k  & = \frac{k^3}{4 \pi^2a^4}
\Big[  |( \partial_0 +\frac{D}{2} ) Y^{(L)}_k|^2
               + k^2  |Y^{(L)}_k|^2
               + m^2 a^2 |Y^{(L)}_k|^2   \Big],
  \label{rholongitudinal}
 \\
p^{L}_k  & = \frac{k^3}{4 \pi^2a^4} \frac13
 \Big[    - |( \partial_0 +\frac{D}{2} ) Y^{(L)}_k|^2
     +  3 k^2 |Y^{(L)}_k|^2  +  m^2 a^2 | Y^{(L)}_k |^2 \Big] ,
     \label{Lonpr}
\el
 independent of $\zeta$,
and the temporal part
\bl
\rho^{T}_k  & = \frac{k^3}{4 \pi^2a^4}
\Big[-| ( \partial_0 - \frac{D}{2} ) Y^{(0)}_k |^2
 - k^2 |Y^{(0)}_k|^2 \Big] ,
   \label{rhotemporal}
     \\
p^{T}_k  & = \frac{k^3}{4 \pi^2a^4} \frac13
 \Big[ - 3| ( \partial_0 -\frac{D}{2} ) Y^{(0)}_k|^2
          + k^2 |Y^{(0)}_k |^2  \Big] ,
    \label{Temppr}
\el
depending on $\zeta$ through  $\zeta m^2 $
in the mode $Y^{(0)}_k$.
The longitudinal stress tensor \eqref{rholongitudinal} \eqref{Lonpr}
and the temporal stress tensor \eqref{rhotemporal} \eqref{Temppr} do not
resemble that of  a minimally-
nor a conformally-coupling scalar field \cite{ZhangYeWang2020}.
We find that the longitudinal stress tensor respects
the covariant conservation
\bl
\rho^{L}_{k}\, ' +3 \frac{a'}{a} (\rho^{L}_{k}+p^{L}_{k})=0 ,
\el
but the temporal stress tensor generally does not
\bl
&\rho^{T}_{k}\, ' +3\frac{a'}{a}(\rho^{T}_{k}+p^{T}_k)
\nn
\\
&=\frac{k^3}{4 \pi^2a^4}   \zeta m^{2} a^{2}
\Big( Y^{(0)*}_k (\partial_0-\frac{D}{2} )Y^{(0)}_k
 + Y^{(0)}_k (\partial_0-\frac{D}{2} )Y^{(0)*}_k  \Big)
 \ne 0 ,
    \label{3822}
\el
for $\zeta m^2 \ne 0$.
Fig.\ref{unregular} (b) shows  that
$\rho^{L}_k$ and $p^{L} _k $
are UV divergent, $\propto k^4$ at high $k$.
Fig.\ref{unregular} (c) shows that
 $\rho^{T}_k$ and $p^{T} _k $
are UV divergent and negative, $\propto -k^4$.

\begin{figure}
\centering
\includegraphics[width=0.6 \textwidth]{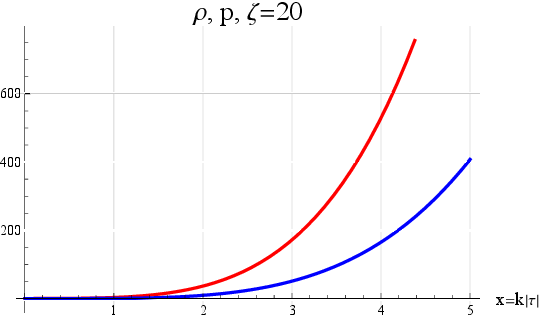}
\caption{ The  unregularized total $\rho_k$ (red)
and $p_k$ (blue) are UV divergent,  $\propto k^4$ at high $k$.}
\label{unregtotalzeta1}
\end{figure}

Similarly,  the GF vacuum stress tensor is given by
\bl  \label{rhoGFVEV}
\langle 0| \rho^{GF}  |0 \rangle
= \int \rho^{GF} _k \frac{dk }{k} ,
~~~~
\langle 0| p^{GF}  |0 \rangle
= \int p^{GF} _k \frac{dk }{k} ,
\el
with the GF  spectral stress tensor
\bl
\rho_k^{GF} &  = \frac{k^3}{2  \pi^2a^4}\Big(
  |(\partial_0-\frac{D}{2})Y^{(0)}_k|^2 + k^2|Y^{(0)}_k|^2
        +\frac12\zeta m^2 a^2 |Y^{(0)}_k|^2   \Big)    , \label{gfrh}
       \\
p_k^{GF}&=\frac{k^3}{2  \pi^2a^4}\Big(
   \big|(\partial_0-\frac{D}{2})Y^{(0)}_k  \big|^2 -\frac13  k^2|Y^{(0)}_k|^2
        - \frac12\zeta m^2 a^2 |Y^{(0)}_k|^2\Big)    ,  \label{sptrpressGFVEV}
\el
also depending on $\zeta$ through $\zeta m^2$ in the mode $Y^{(0)}_k$.
Since both the temporal and GF stress tensors are formed from  $Y^{(0)}_k$,
they  become effectively the massless in the Landau gauge $\zeta=0$,
as mentioned earlier below \eqref{Y0solm0}.
Fig.\ref{unregular} (d) shows that
the unregularized $\rho^{GF}_k$  and $p^{GF} _k$
are UV divergent, $\propto k^4$ at high $k$.
The expressions \eqref{gfrh} \eqref{sptrpressGFVEV}
are not the same as that of a minimally-coupling
massive scalar field \cite{ZhangYeWang2020}.
The GF stress tensor does not respect the covariant conservation
\bl
& \rho^{GF \, '}_{k}+3\frac{a'}{a}(\rho^{GF}_{k}+p^{GF}_k)
\nn \\
 &  =  - \frac{k^3}{4 \pi^2a^4}   \zeta m^{2} a^{2}
\Big( Y^{(0)*}_k (\partial_0-\frac{D}{2} )Y^{(0)}_k
 + Y^{(0)}_k (\partial_0-\frac{D}{2} )Y^{(0)*}_k  \Big)
  \ne 0 .
\el
Nevertheless, the sum of the temporal and GF stress tensors
respects the covariant conservation
\bl
   \rho^{T \, '}_{k}+3\frac{a'}{a}(\rho^{T}_{k}+p^{T}_k)
+ \rho^{GF \, '}_{k}+3\frac{a'}{a}(\rho^{GF}_{k}+p^{GF}_k)
  =0 ,
  \label{temGF}
\el
ie, the $\zeta$-dependent
part of the stress tensor is conserved for arbitrary $\zeta$.
Thus, the total unregularized Stueckelberg vacuum spectral stress tensor
\bl \label{rho4parts}
\rho_k = \rho^{TR}_k + \rho^{L}_k + \rho^{T}_k + \rho^{GF}_k,
 \\
p_k = p^{TR}_k + p^{L}_k + p^{T}_k + p^{GF}_k,
\label{p4parts}
\el
respects the covariant conservation,
and the vacuum spectral trace is
\bl
\langle 0| T^\mu\, _\mu |0\rangle_k
 & = -\rho_k + 3 p_k
\nn
\\
& = \frac{k^3}{2\pi^2 a^4}
  \Big[- 2  m^2 a^2 |f_k^{(1)}(\tau)|^2 \Big]
\nn \\
& + \frac{k^3}{4 \pi^2a^4}
 \Big[ -2  | ( \partial_0 +\frac{D}{2} ) Y^{(L)}_k|^2
     + 2 k^2 |Y^{(L)}_k|^2   \Big]
\nn \\
&  + \frac{k^3}{4 \pi^2a^4}
 \Big[ -2  | ( \partial_0 - \frac{D}{2}  ) Y^{(0)}_k|^2
     + 2 k^2 |Y^{(0)}_k|^2   \Big]
     \nn \\
& + \frac{k^3}{2 \pi^2 a^4}
 \Big[ 2  | ( \partial_0 - \frac{D}{2} ) Y^{(0)}_k|^2
     - 2 k^2 |Y^{(0)}_k|^2
     - 2 \zeta m^2 a^2 |Y^{(0)}_k|^2   \Big]  .
\label{unregstrSt}
\el
Fig.\ref{unregtotalzeta1} shows that
the unregularized total $\rho_k$ and $p_k $
are positive and UV divergent $\propto k^4$ at high $k$.

\section{\bf Adiabatic regularization  of
the spectral stress tensor of the Stueckelberg  field}

The Stueckelberg vacuum stress tensor given in Sect 3
 is UV divergent at high $k$.
In literature the 4th-order regularization was adopted by default
\cite{DowkerCritchley1977,BrownCassidy1977,Endo1984,ChimentoCossarini1990,ChuKoyama2017}
on the stress tensor of a massive vector field.
We shall assume the minimal subtraction rule
that only the minimum number of terms should be subtracted
to yield the convergent stress tensor \cite{ParkerFulling1974}.
Since the four parts of \eqref{rho4parts} \eqref{p4parts}
actually have  different UV divergences,
we shall regularize each part respectively to its pertinent order.
With this scheme of regularization,
we shall be able to get a regularized stress tensor
with the following desired properties:
1) UV and IR convergent,
2) the covariant conservation is respected,
3) a nonnegative total regularized energy density.
We emphasize that
a positive energy density is required by
the de Sitter inflation,
as it is the source of the Friedmann equation.
In the following we present the details of regularization on each part.

{\bf 1) the 0th-order regularization of the transverse stress tensor : }

The transverse mode equation \eqref{Bieq1}
and the transverse stress tensor \eqref{spTRPRho} \eqref{spTRPRess}
have the same form as those
of the conformally-coupling massive scalar field
\cite{ZhangYeWang2020,ZhangYe2022PointSpl}.
The 0th-order regularization is sufficient to remove
the UV divergences of the transverse stress tensor.

\begin{figure}[htb]
\centering
\subcaptionbox{}
    {%
        \includegraphics[width = .41\linewidth]{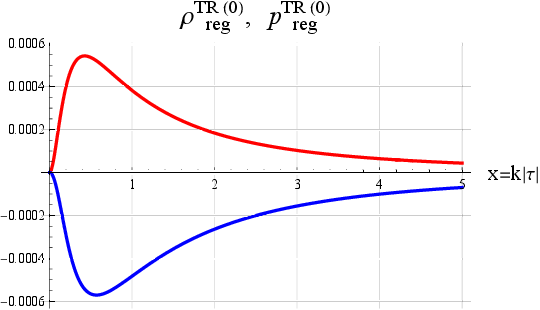}
}
\subcaptionbox{}
    {%
        \includegraphics[width = .41\linewidth]{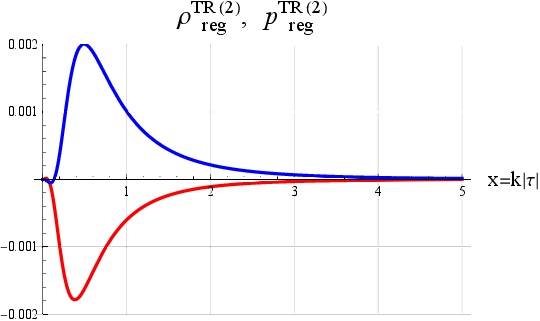}
    }
\subcaptionbox{}
    {%
        \includegraphics[width = .41\linewidth]{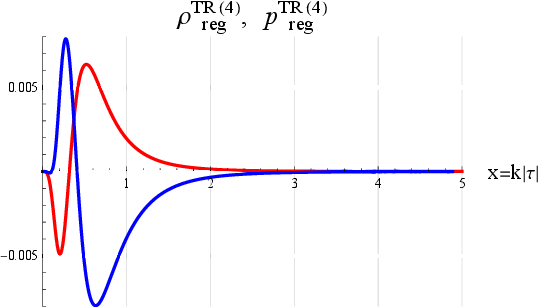}
    }
\caption{ The regularized transverse spectral energy density (red)
          and spectral pressure (blue).
(a) the 0th-order $\rho_{k~reg}^{TR(0)}$
    is  UV convergent and positive.
(b) the 2nd-order $\rho_{k~reg}^{TR(2)} <0$.
(c) the 4th-order  $\rho_{k~reg}^{TR(4)}$ shows a negative segment.
}
\label{Trhop}
\end{figure}

The unregularized transverse spectral
stress tensor \eqref{spTRPRho} \eqref{spTRPRess}
are expanded at high $k$ as the following
\be
\rho^{TR}_{k}= \frac{k^4}{2\pi^2a^4}\Big(1+\frac{m^2}{2 H^2 k^2}
-\frac{m^4}{8 H^4 k^4}+\frac{2 H^2 m^4+m^6}{16 H^6 k^6}
-\frac{5 (12 H^4 m^4+8 H^2 m^6+m^8 )}{128 H^8 k^8}
   +... \Big) ,
  \label{highErhoT}
\ee
\be
p^{TR}_{k}= \frac{k^4}{2\pi^2a^4}\frac13\Big(1-\frac{m^2}{2 H^2 k^2}
+\frac{3 m^4}{8 H^4 k^4}-\frac{5 (2 H^2 m^4+m^6 )}{16 H^6 k^6}
+\frac{35 (12 H^4 m^4+8 H^2 m^6+m^8 )}{128 H^8 k^8}
  +... \Big) .
     \label{highEpT}
\ee
In the above parenthesis
a fixed time $|\tau|=1$ is taken for a simple notation,
and will be adopted in this section.
\eqref{highErhoT} \eqref{highEpT}
 contain the quartic  $k^4$, quadratic $k^2$, and $k^0$ terms
that will cause  UV divergences
in the $k$-integration \eqref{transvrhopress} for $\rho^{TR}$ and $p^{TR}$.
To remove these  divergences,
the 0th-order of adiabatic regularization is sufficient
and given by the following
\be \label{trsadrh}
\rho^{TR(0)}_{k\, reg} =\rho^{TR}_{k} -\rho^{TR}_{k\, A 0},
~~~~
p^{TR(0)}_{k\, reg} =p^{TR}_{k} -p^{TR}_{k\, A 0},
\ee
where
\bl \label{transrho0sb}
\rho^{TR}_{k\, A 0} & = \frac{k^3}{2\pi^2a^4} \, \omega  ,
\\
p^{TR}_{k\, A 0}   &  = \frac{k^3}{2\pi^2a^4}\frac{1}{3}
               \Big( \omega-\frac{m^2a^2}{\omega}\Big) , \label{transp0sb}
\el
are the 0th-order adiabatic subtraction terms
for the transverse stress tensor
(see \eqref{0thtrrho} \eqref{pTMV0th} in Appendix A).
Their  high-$k$ expansions are
\be
\rho^{TR}_{k\, A 0} = \frac{k^4}{2\pi^2a^4}\Big(1+\frac{m^2}{2 H^2 k^2}
-\frac{m^4}{8 H^4 k^4}+\frac{m^6}{16 H^6 k^6}
-\frac{5 m^8}{128 H^8 k^8} )
  +... \Big)  ,  \label{high0thrhoT}
\ee
\be
p^{TR}_{k\, A 0} = \frac{k^4}{2\pi^2a^4}\frac13\Big(1-\frac{m^2}{2 H^2 k^2}
+\frac{3 m^4}{8 H^4 k^4}-\frac{5 m^6}{16 H^6 k^6}
+\frac{35 m^8}{128 H^8 k^8}
  +...  \Big) . \label{high0thpT}
\ee
The resulting regularized transverse spectral stress tensor at high $k$
\bl
&\rho_{k~reg}^{TR(0)}
=\frac{k^3}{2\pi^2a^4}  \Big(\frac{m^4}{8 H^4 k^5}
-\frac{5 \left(3 H^2 m^4+2 m^6\right)}{32 H^6 k^7} +... \Big) ,
        \label{adiatrrho}
\\
& p_{k~reg}^{TR(0)}=\frac{k^3}{2\pi^2a^4} \frac13
 \Big( -\frac{5 m^4}{8 H^4 k^5}+\frac{35
  \left(3 H^2 m^4+2 m^6\right)}{32 H^6 k^7}
   +...  \Big )  , \label{adiatrpr}
\el
 independent of $\zeta$.
Fig. \ref{Trhop} (a) shows that the regularized
$\rho_{k~reg}^{TR(0)}$ is  UV convergent and positive,
and $p_{k~reg}^{TR(0)}$ is UV convergent and  negative.
Thus, the 0th-order regularization removes all the UV divergences
in the transverse stress tensor,
obeys the minimal subtraction rule,
keeps the sign of the transverse spectral energy density unchanged,
and is proper for the transverse stress tensor.
It is checked that
the  transverse 0th-order subtraction terms  \eqref{transrho0sb}
\eqref{transp0sb}  satisfy
\bl
\rho^{TR}_{k\, A0}\, '
   +3 \frac{a'}{a} (\rho^{TR}_{k\, A0} +p^{TR}_{k\, A0})
       =0 ,
\el
so that the 0th-order regularized  transverse stress tensor
respects the covariant conservation
\bl
\rho^{TR(0)}_{k\, reg}\, '
  +3 \frac{a'}{a} (\rho^{TR(0)}_{k\, reg}  +p^{TR(0)}_{k\, reg})
  =0  .
\el

If  the 2nd-order regularization were used,
\bl
\rho_{k~reg}^{TR(2)} = \rho_{k}^{TR}-\rho_{k\, A2}^{TR },
~~~~
p_{k~reg}^{TR(2)} = p_{k}^{TR}-p_{k\, A2}^{TR} ,
\nn
\el
where the 2nd-order adiabatic subtraction terms at high $k$
(see \eqref{2thtrrho} \eqref{pTMV2nd})
\bl
\rho_{k\, A2}^{TR } & = \frac{k^4}{2\pi^2a^4}
\Big(1+\frac{m^2}{2 H^2 k^2}-\frac{m^4}{8 H^4 k^4}
+\frac{2 H^2 m^4+m^6}{16 H^6 k^6}
-\frac{5\left(8 H^2 m^6+m^8\right)}{128 H^8 k^8}
\nn \\
& +\frac{7 m^8\left(20 H^2+m^2\right)}{256 H^{10} k^{10}}
  +... \Big) ,
\\
p_{k\, A2}^{TR } &=  \frac{k^4}{2\pi^2a^4}\frac13
\Big(1-\frac{m^2}{2 H^2 k^2}+\frac{3 m^4}{8 H^4 k^4}
-\frac{5 \left(2 H^2 m^4+m^6\right)}{16 H^6 k^6}
+\frac{35 \left(8 H^2 m^6+m^8\right)}{128 H^8 k^8}
\nn \\
& -\frac{63 m^8 \left(20 H^2+m^2\right)}{256 H^{10} k^{10}}
 +...  \Big) ,
\el
the resulting 2th-order regularized transverse stress tensor
at high $k$ would be
\bl
\rho_{k\, reg}^{TR(2)} &= \frac{k^4}{2\pi^2a^4}
   \Big(-\frac{15 m^4}{32 H^4 k^8}
        +\frac{63 \left(4 H^2 m^4+3 m^6\right)}{64 H^6 k^{10}}
     +... \Big) ,
\\
p_{k \, reg}^{TR(2)}& =  \frac{k^4}{2\pi^2a^4}\frac13
\Big(\frac{105 m^4}{32 H^4 k^8}-\frac{567
\left(4 H^2 m^4+3 m^6\right)}{64 H^6 k^{10}} +...  \Big) .
\el
The difficulty is that $\rho_{k\, reg}^{TR(2)}$ is negative,
as shown in Fig.\ref{Trhop} (b).
Obviously, the 2nd-order regularization subtracts more than necessary,
does not obey the minimum subtraction rule,
and is improper for the transverse stress tensor.

Or, if the conventional 4th-order regularization were adopted,
\bl
\rho_{k~reg}^{TR(4)} = \rho_{k}^{TR} - \rho_{k\, A4}^{TR},
~~~~
p_{k~reg}^{TR(4)} = p_{k}^{TR} - p_{k\, A4}^{TR} ,
\nn
\el
where  the 4th-order subtraction terms   at high $k$
(see  \eqref{tr4thadrho}  \eqref{pTMV4th})
\bl
\rho_{k\, A4}^{TR} &= \frac{k^4}{2\pi^2a^4}
\Big(1+\frac{m^2}{2 H^2 k^2}-\frac{m^4}{8 H^4 k^4}
+\frac{2 H^2 m^4+m^6}{16 H^6 k^6}
-\frac{5 \left(12 H^4 m^4+8 H^2 m^6+m^8\right)}{128 H^8 k^8}
\nn
\\
& ~~ +\frac{7 m^6 \left(108 H^4
+20 H^2 m^2+m^4\right)}{256 H^{10} k^{10}}-\frac{21 m^8
\left(508 H^4+40 H^2 m^2+m^4\right)}{1024 H^{12} k^{12}}
  +...  \Big) ,
\\
 p_{k\, A4}^{TR} & = \frac{k^4}{2\pi^2a^4}\frac13
 \Big(1-\frac{m^2}{2 H^2 k^2}+\frac{3 m^4}{8 H^4 k^4}
 -\frac{5 \left(2 H^2 m^4+m^6\right)}{16 H^6 k^6}+\frac{35
 \left(12 H^4 m^4+8 H^2 m^6+m^8\right)}{128 H^8 k^8}\nn
\\
& ~~ -\frac{63 m^6 \left(108 H^4+20 H^2 m^2+m^4\right)}
{256 H^{10} k^{10}}+\frac{231 m^8 \left(508 H^4+40 H^2 m^2+m^4\right)}
  {1024 H^{12} k^{12}} +... \Big) ,
\el
the 4th-order regularized transverse stress tensor at high $k$
would be
\bl
\rho_{k\, reg}^{TR(4)} &= \frac{k^4}{2\pi^2a^4}
\Big(\frac{63 m^4}{16 H^4 k^{10}}-\frac{189 \left(5 H^2 m^4+4 m^6\right)}
{16 H^6 k^{12}} +...  \Big) ,
\\
p_{k\, reg}^{TR(4)} &= \frac{k^4}{2\pi^2a^4}\frac13
\Big(-\frac{567 m^4}{16 H^4 k^{10}}+\frac{2079
\left(5 H^2 m^4+4 m^6\right)}{16 H^6 k^{12}} +...  \Big) .
\el
Again, $\rho_{k\, reg}^{TR(4)}$ becomes negative at small $k$
as shown in Fig  \ref{Trhop} (c).
So, the 4th-order regularization
is improper  for the transverse stress tensor.

Ref.\cite{ChimentoCossarini1990} noticed
the fact that, for the transverse stress tensor,
only the 0th-order terms are divergent,
and the other higher order terms are actually convergent,
but still adopted the conventional 4th-order regularization.
Ref.\cite{ChuKoyama2017} adopted the 4th-order regularization without
analyzing the divergence behavior of the stress tensor.
These references did not demonstrate explicitly
the resulting regularized spectral stress tensor.

{\bf 2) the 2nd-order regularization of the longitudinal  stress tensor :}

\begin{figure}[htb]
\centering
\subcaptionbox{}
    {%
        \includegraphics[width = .41\linewidth]{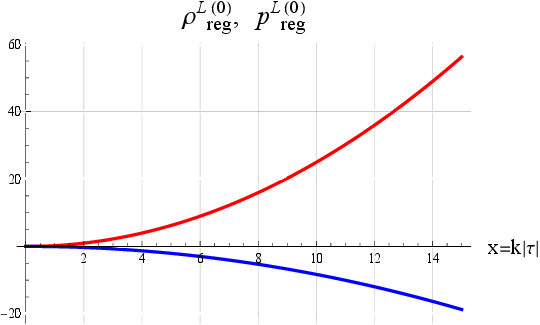}
}
\subcaptionbox{}
    {%
        \includegraphics[width = .41\linewidth]{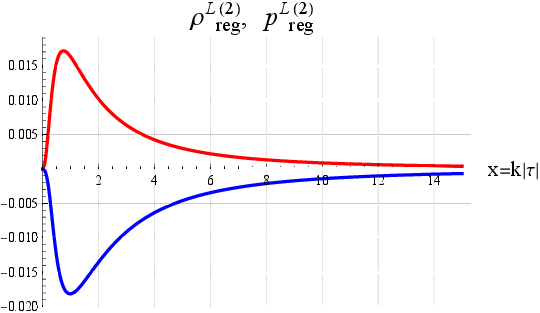}
    }
\subcaptionbox{}
    {%
        \includegraphics[width = .41\linewidth]{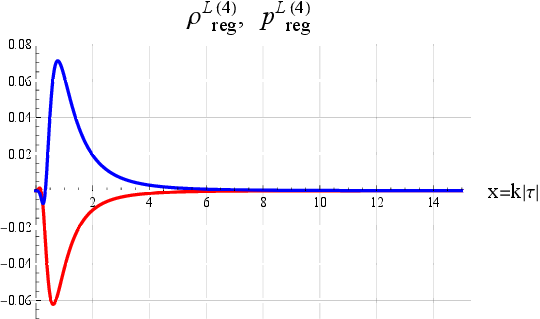}
    }
\caption{ The  regularized longitudinal
       spectral energy density (red) and spectral pressure (blue).
(a) the 0th-order, still divergent.
(b) the 2nd-order $\rho_{k~reg}^{L(2)}$ is convergent and positive.
(c) the 4th-order $\rho_{k~reg}^{L(4)}$ is negative.
}
\label{regrhopL}
\end{figure}

The  unregularized longitudinal stress tensor
\eqref{rholongitudinal} \eqref{Lonpr}
resemble neither the form of
a minimally- nor a conformally-coupling scalar field \cite{ZhangYeWang2020}.
By the minimal subtraction rule
we regularize the longitudinal stress tensor  by trial and error.
The  high-$k$ expansions are
\bl \label{hkexprhoL}
\rho^{L}_{k} = & \frac{k^3}{2\pi^2a^4} \Big( \frac{k}{2}+\frac{1}{4}
(1+\frac{m^2}{H^2})\frac1k-\frac{6 H^2 m^2+m^4}{16 H^4}\frac{1}{k^3}
+\frac{30 H^4 m^2+17 H^2 m^4+m^6}{32 H^6 k^5}
 +... \Big) ,
\\
p^{L}_{k} = &  \frac{k^3}{2\pi^2a^4}\frac13  \Big( \frac{k}{2}
-\frac{ \left(H^2+m^2\right)}{4 H^2}\frac1k
+\frac{18 H^2 m^2+3m^4}{16 H^4}\frac{1}{k^3}
-\frac{5 \left(30 H^4 m^2+17 H^2 m^4+m^6\right)}{32 H^6 k^5}
 +...   \Big) ,
   \label{hkexpprL}
\el
which contain  $k^4$,  $k^2$, and $\ln k$ divergences.
To remove the UV divergences,
the 2nd-order adiabatic regularization is sufficient,
\be \label{Longreg2}
\rho^{L(2)}_{k\, reg} =\rho^{L}_{k} -\rho^{L}_{k\, A 2},
~~~~
p^{L(2)}_{k\, reg}  = p^{L}_{k} -p^{L}_{k\, A 2},
\ee
where  the 2nd-order  longitudinal subtraction terms are
\bl
\rho^{L}_{k\, A 2}
&=\frac{k^3}{2\pi^2a^4}\Big(\frac{k}{2}
  + \frac14 (1+\frac{m^2}{H^2} ) \frac1k -(\frac{m^4}{16 H^4}
    +\frac{3 m^2}{8 H^2}) \frac{1}{k^3}
    + \frac{17 H^2 m^4+m^6}{32 H^6 k^5} +...
   \Big),  \label{highkrhoL2nd}
   \\
p^{L}_{k\, A 2}
&=\frac{k^3}{2\pi^2a^4}\frac13 \Big(\frac{k}{2}
 - \frac14 (1+\frac{m^2}{ H^2}) \frac{1}{k}
  +\frac{18 H^2 m^2+3m^4}{16 H^4}\frac{1}{k^3}
    -\frac{5 \left(17 H^2 m^4+m^6\right)}{32 H^6  k^5 }
  +... \Big) ,
              \label{highkpL2nd}
\el
at high $k$  (see \eqref{2ndrhoL} \eqref{2ndadpL}).
The resulting 2nd order  regularized longitudinal
spectral stress tensor at high $k$ is
\bl
\rho_{k~reg}^{L(2)}&=\frac{k^3}{2\pi^2a^4}\Big(\frac{15 m^2}
{16 H^2 k^5}-\frac{5 \left(84 H^2 m^2+59 m^4\right)}
{64 H^4 k^7} +... \Big) , \label{206}
\\
p_{k~reg}^{L(2)}&=\frac{k^3}{2\pi^2a^4}\frac13
\Big(-\frac{75 m^2}{16 H^2 k^5}+\frac{35 \left(84 H^2 m^2+59 m^4\right)}
{64 H^4 k^7} +... \Big) , \label{2077}
\el
 independent of $\zeta$.
Fig.\ref{regrhopL} (b) shows that
$\rho_{k~reg}^{L(2)}$ is  UV convergent and positive,
and $p_{k~reg}^{L(2)}$ is UV convergent and negative.
So the 2nd-order regularization removes all the UV divergences
for the longitudinal stress tensor,
keeps the sign of the  longitudinal spectral energy density unchanged.
It is checked that
the 2nd-order longitudinal subtraction terms \eqref{2ndrhoL} \eqref{2ndadpL}
satisfies
\bl
\rho^{L}_{k\, A2}\, ' +3 \frac{a'}{a} (\rho^{L}_{k\, A2}+p^{L}_{k\, A2})
    =0 ,
\el
so that the 2nd-order regularized vacuum longitudinal stress tensor
respects the covariant conservation
\bl
\rho^{L(2)}_{k\, reg}\, '
    +3 \frac{a'}{a} (\rho^{L(2)}_{k\, reg} +p^{L(2)}_{k\, reg}) =0  .
\el

If the 0th-order regularization were adopted
for the longitudinal stress tensor,
using the 0th-order subtraction terms \eqref{0thrhot} \eqref{0thprt},
 $\rho_{k~reg}^{L(0)}$  would still be UV divergent
as shown in Fig.\ref{regrhopL} (a).
So, the 0th-order regularization is insufficient
for the longitudinal stress tensor.
Or,  if  the 4th-order regularization were adopted,
using the 4th-order subtraction terms \eqref{4thadrhoL} \eqref{1088},
 $\rho_{k~reg}^{L(4)}$ would become negative as shown in Fig.\ref{regrhopL} (c).
So, the 4th-order regularization
 is improper for the longitudinal stress tensor.

{\bf 3) the 2nd-order regularization of the temporal stress tensor : }

\begin{figure}[htb]
\centering
\subcaptionbox{}
    {%
        \includegraphics[width = .41\linewidth]{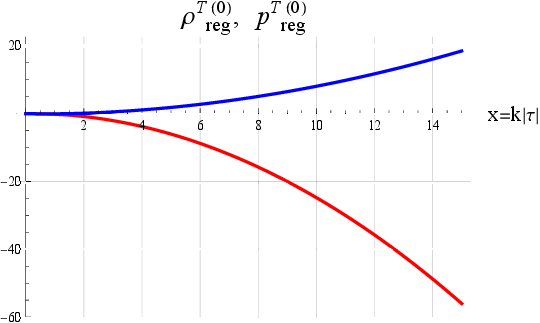}
}
\subcaptionbox{}
    {%
        \includegraphics[width = .41\linewidth]{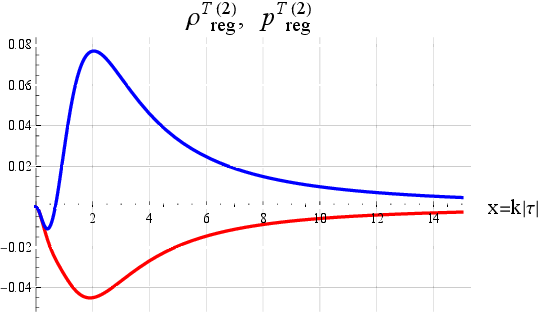}
    }
\subcaptionbox{}
    {%
        \includegraphics[width = .41\linewidth]{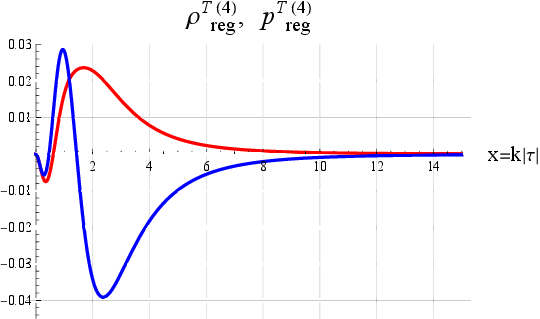}
    }
\caption{ The regularized temporal spectral energy density (red)
and spectral pressure (blue).
(a) the 0th-order.
(b) the 2nd-order.
(c) the 4th-order.
}
\label{regrhoptee}
\end{figure}

The unregularized temporal stress tensor
\eqref{rhotemporal} \eqref{Temppr} resemble neither the form of
a minimally- nor a conformally-coupling scalar field \cite{ZhangYeWang2020}.
The high-$k$ expansions are
\bl
\rho_{k}^{T}
&=\frac{k^3}{2\pi^2a^4} \Big( -\frac{k}{2}-\frac{1}{4 k}+\frac{m^2 \zeta
 \left(2 H^2-m^2 \zeta \right)}{16 H^4 k^3}+\frac{m^2 \zeta
 \left(-10 H^4+H^2 m^2 \zeta +2 m^4 \zeta ^2\right)}{32 H^6 k^5}
 \nn
\\
&~~~-\frac{5 m^2 \zeta
  \left(-112 H^6+4 H^4 m^2 \zeta +20 H^2 m^4 \zeta ^2+3 m^6 \zeta ^3\right)}
   {256 H^8 k^7}     +... \Big) ,
\label{highkErhote}
\\
p_{k}^{T}
& =   \frac{k^3}{2\pi^2a^4}\frac13
\Big( -\frac{k}{2}+\frac1k (\frac{1}{4}-\frac{m^2 \zeta }{2 H^2} )
+\frac{3m^2 \zeta  \left(-2 H^2+m^2 \zeta \right)}{16 H^4k^3}
\nn \\
& ~~~ +\frac{m^2 \zeta  \left(2 H^2-m^2 \zeta \right)
\left(25 H^2+4 m^2 \zeta \right)}{32 H^6 k^5}
\nn
\\
&~~~+\frac{5 m^2 \zeta  \left(-2 H^2+m^2 \zeta \right)
\left(4 H^2+m^2 \zeta \right) \left(98 H^2+5 m^2 \zeta \right)}
{256 H^8 k^7})  +... \Big) ,
\label{highkEpte}
\el
containing $k^4$,  $k^2$, and $\ln k$ divergences,
and being negative.
The 2nd-order adiabatic regularization is sufficient
to remove these UV divergences,
\be\label{tempreg0rho}
\rho^{T(2)}_{k\, reg} =\rho^{T}_{k} -\rho^{T}_{k\, A 2},
~~~~
p^{T(2)}_{k\, reg}  = p^{T}_{k} -p^{T}_{k\, A 2} ,
\ee
where the 2nd-order  temporal subtraction terms are
\bl
\rho^{T}_{k\, A 2} = & \frac{k^3}{2\pi^2a^4}\Big(-\frac{k}{2}-\frac{1}{4k}
-\frac{\zeta ^2 m^4}{16 H^4k^3}+\frac{\zeta  m^2}{8 H^2k^3}
+\frac{\zeta ^2 H^2 m^4+2 \zeta ^3 m^6}{32 H^6 k^5}  +... \Big) ,
\\
p^{T}_{k\, A 2} = & \frac{k^3}{2\pi^2a^4}\frac13 \Big(-\frac{k}{2}
+  (\frac{1}{4}-\frac{m^2 \zeta }{2 H^2} ) \frac1k
+\frac{-6 H^2 m^2 \zeta +3m^4 \zeta ^2}{16 H^4k^3}
\nn \\
&  -(\frac{17 m^4 \zeta ^2}{32 H^4}
  + \frac{m^6 \zeta ^3}{8 H^6}) \frac{1}{k^5}  +... \Big) ,
\el
at high-$k$ (see \eqref{2ndtemprho} \eqref{2ndtemppr}).
So the 2nd-order regularized temporal spectral stress tensor
at high $k$ is
\bl
\rho_{k~reg}^{T(2)}&=\frac{k^3}{2\pi^2a^4}\Big(-\frac{5 m^2 \zeta }{16 H^2 k^5}
+\frac{5 m^2 \zeta  \left(28 H^2-m^2 \zeta \right)}{64 H^4 k^7}+... \Big),
\label{regtrh2}
\\
p_{k~reg}^{T(2)}&=\frac{k^3}{2\pi^2a^4}\frac13 \Big(\frac{25 m^2 \zeta }
{16 H^2 k^5}+\frac{5 m^2 \zeta
\left(-196 H^2+39 m^2 \zeta \right)}{64 H^4 k^7}+...\Big) ,
\label{regtp2}
\el
which contain  an overall factor $\zeta$.
Fig.\ref{regrhoptee} (b) shows that $\rho_{k~reg}^{T(2)}$
is convergent and still negative,
consistent with the negative  unregularized  $\rho_{k}^{T}$
in Fig.\ref{unregular} (c).
So, the 2nd order regularization
removes all the UV divergences in the temporal stress tensor,
and keeps the sign of the temporal  spectral energy density unchanged.

If  the 0th-order regularization were adopted,
using the subtraction terms \eqref{722} \eqref{0pTad},
 $\rho_{k~reg}^{T(0)}$  would be still be UV divergent
as shown in Fig.\ref{regrhoptee} (a).
So, the 0th-order regularization is insufficient
for the temporal.
Or, if the 4th-order regularization were adopted,
using the subtraction terms \eqref{4rhotemad} \eqref{8222},
 $\rho_{k~reg}^{T(4)}$  would change its sign,
as  shown in Fig.\ref{regrhoptee} (c).
So, the 4th-order regularization subtracts too much for the temporal.

{\bf 4) the 2nd-order regularization of the GF stress tensor : }

\begin{figure}[htb]
\centering
\subcaptionbox{}
    {%
        \includegraphics[width = .41\linewidth]{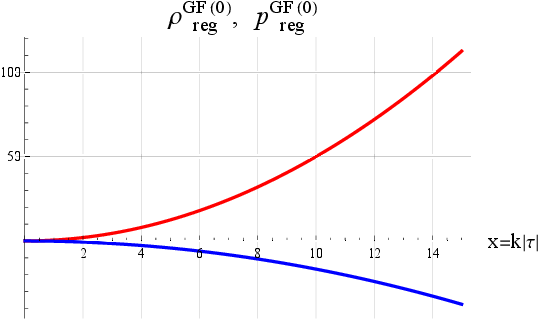}
}
\subcaptionbox{}
    {%
        \includegraphics[width = .41\linewidth]{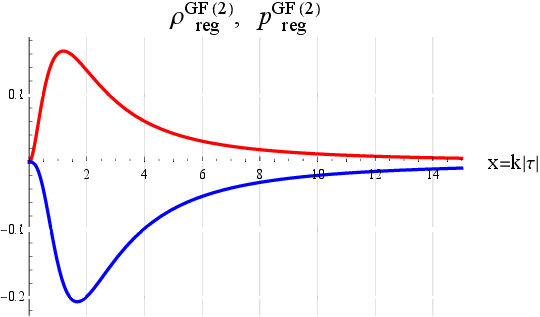}
    }
\subcaptionbox{}
    {%
        \includegraphics[width = .41\linewidth]{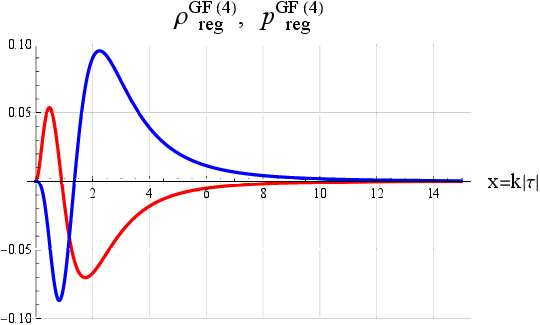}
    }
\caption{ The regularized GF spectral energy density (red) and spectral pressure (blue).
(a) 0th order.
(b) 2nd order.
(c) 4th order.
}
\label{GFrhopfey}
\end{figure}

The unregularized GF stress tensor \eqref{gfrh} \eqref{sptrpressGFVEV}
are formally similar to that of a minimally-coupling
massive scalar field \cite{ZhangYeWang2020},
so we can try the 2nd-order regularization.
The  high-$k$ expansions are
\bl
\rho^{GF}_{k}  = &
 \frac{k^3}{2\pi^2a^4} \Big( k+\frac{1}{4} (2+\frac{m^2 \zeta }{H^2})\frac1k
-\frac{m^2 \zeta  \left(-20 H^4+8 H^2 m^2 \zeta
+m^4 \zeta ^2\right)}{32 H^6 k^5} \nn \\
& +\frac{5 m^2 \zeta  \left(-112 H^6
+20 H^4 m^2 \zeta +16 H^2 m^4 \zeta ^2+m^6 \zeta ^3\right)}{128 H^8 k^7}
 \nn
\\
&      +\frac{21 m^2 \zeta  \left(2 H^2-m^2 \zeta \right)
\left(4 H^2+m^2 \zeta \right) \left(10 H^2+m^2 \zeta \right)
\left(18 H^2+m^2 \zeta \right)}{512 H^{10} k^9}
    +...   \Big) ,  \label{157}
\\
p_{k}^{GF}  = & \frac{k^3}{2\pi^2a^4} \frac13
  \Big(  k+\frac{1}{4} (-2+\frac{m^2 \zeta }{H^2})\frac1k
-\frac{m^2 \zeta  \left(100 H^4-52 H^2 m^2 \zeta
+m^4 \zeta ^2\right)}{32 H^6 k^5} \nn \\
& +\frac{5 m^2 \zeta  \left(-98 H^2+m^2 \zeta \right)
\left(-2 H^2+m^2 \zeta \right) \left(4 H^2+m^2 \zeta \right)}{128 H^8 k^7}
\nn
\\
&    -\frac{21 m^2 \zeta  \left(-162 H^2+m^2 \zeta \right)
\left(-2 H^2+m^2 \zeta \right) \left(4 H^2+m^2 \zeta \right)
\left(10 H^2+m^2 \zeta \right)}{512 H^{10} k^9} +... \Big) , \label{158}
\el
containing $k^4$ and  $k^2$  divergences, but no $\ln k$ divergence.
We adopt the 2nd-order regularization,
\be
\rho^{GF(2)}_{k\, reg} =\rho^{GF}_{k} -\rho^{GF}_{k\, A 2} ,
~~~~
p^{GF(2)}_{k\, reg} = p^{GF}_{k} - p^{GF}_{k\, A 2} ,
\ee
where the 2nd-order subtraction terms
are given by \eqref{GF2rhoad} \eqref{866},
with  the  high-$k$ expansions
\bl
\rho^{GF}_{k\, A 2}
= &
\frac{k^3}{2\pi^2a^4}   \Big(  k+\frac{1}{4}
(2+\frac{m^2 \zeta }{H^2} ) \frac{1}{k^2}
-\frac{m^4 \zeta ^2 \left(8 H^2+m^2 \zeta \right)}{32 H^6 k^5}
\nn \\
& +\frac{5 m^6 \zeta ^3 \left(16 H^2+m^2 \zeta \right)}{128 H^8 k^7}
-\frac{21 m^8 \zeta ^4 \left(30 H^2+m^2 \zeta \right)}{512 H^{10} k^9}
  +...  \Big)  ,  \label{161}
\\
p^{GF}_{k\, A 2}
= & \frac{k^3}{2\pi^2a^4}\frac13
\Big(  k+\frac{1}{4}
(-2+\frac{m^2 \zeta }{H^2})\frac{1}{k}
+\frac{m^4 \zeta ^2
\left(52 H^2-m^2 \zeta \right)}{32 H^6 k^5}
\nn \\
& +\frac{5 m^6 \zeta ^3
\left(-96 H^2+m^2 \zeta \right)}{128 H^8 k^7}
+\frac{21 m^8 \zeta ^4 \left(150 H^2-m^2 \zeta \right)}{512 H^{10} k^9}
  +...  \Big) .
 \label{162}
\el
The regularized GF spectral energy density and pressure are
\bl
&\rho_{k~reg}^{GF(2)}=\frac{k^3}{2\pi^2a^4}
 \Big(  \frac{5 m^2 \zeta }{8 H^2 k^5}
+\frac{5 m^2 \zeta  \left(-28 H^2+5 m^2 \zeta \right)}{32 H^4 k^7}
  +... \Big ),
  \label{regGFrh2}
\\
&p_{k~reg}^{GF(2)}=\frac{k^3}{2\pi^2a^4}
\Big(   -\frac{25 m^2 \zeta }{24 H^2 k^5}
+\frac{5 m^2 \zeta  \left(196 H^2-51 m^2 \zeta \right)}{96 H^4 k^7}
  +... \Big ) ,
\label{regGFp2}
\el
which also have an overall factor $\zeta$.
Fig.\ref{GFrhopfey} (b)
shows that  $\rho_{k~reg}^{GF(2)}$ is convergent and positive.
Thus, the 2nd order regularization
removes all the UV divergences in the GF stress tensor,
and keeps the sign of the GF spectral energy density unchanged.

If  the 0th-order regularization were used
for the GF stress tensor,
using the subtraction terms  \eqref{822} \eqref{855},
 $\rho_{k~reg}^{GF(0)}$  would  still be UV divergent,
as seen in Fig.\ref{GFrhopfey} (a).
If the 4th-order regularization were used,
using the subtraction terms \eqref{886} \eqref{877},
 $\rho_{k~reg}^{GF(4)}$ would become negative at $k |\tau| \gtrsim 1$,
 as seen in Fig.\ref{GFrhopfey} (c).
So, the 4th-order regularization  is improper for the GF stress tensor.

Although
$\rho^{T}_{k\, A 2}$ and $p^{T}_{k\, A 2}$
do not respect the covariant conservation,
neither do $\rho^{GF}_{k\, A 2}$ and $p^{GF}_{k\, A 2}$,
nevertheless, their  sum  does
\bl
(\rho^{T }_{k\, A2})'
   +3\frac{a'}{a}(\rho^{T }_{k\, A 2} +  p^{T }_{k\, A 2} )
     + (\rho^{GF }_{k\, A2})'
  +3\frac{a'}{a}(\rho^{GF }_{k\, A 2} +  p^{GF}_{k\, A 2} )
      =  0 .
\el
Combining with \eqref{temGF},
we find that the sum of regularized temporal and GF stress tensors
respects the covariant conservation
\bl \label{conservTGF}
(\rho^{T(2) }_{k\, reg})'
   +3\frac{a'}{a}(\rho^{T(2) }_{k\, reg} +  p^{T(2) }_{k\, reg} )
     + (\rho^{GF(2) }_{k\, reg})'
  +3\frac{a'}{a}(\rho^{GF(2) }_{k\, reg} +  p^{GF(2)}_{k\, reg} )
      =  0 .
\el

\begin{figure}[htb]
\centering
\subcaptionbox{}
    {%
        \includegraphics[width = .41\linewidth]{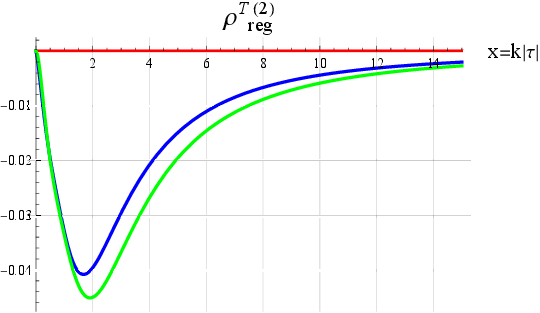}
}
\subcaptionbox{}
    {%
        \includegraphics[width = .41\linewidth]{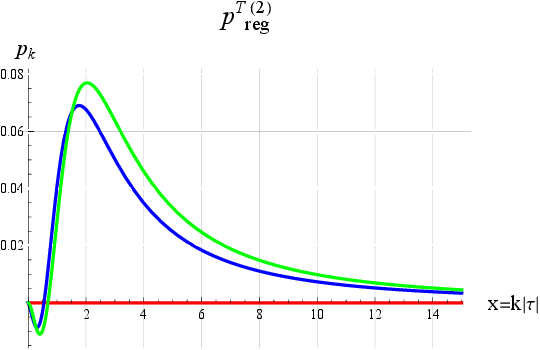}
}
\caption{The regularized temporal  spectral stress tensor depends on $\zeta$.
(a) $\rho_{k~reg}^{T(2)}$.
(b) $p_{k~reg}^{T(2)}$.
The parameters:  $\frac{m^2}{H^2}=0.1$,
$\zeta=0$ (red), $\zeta=15$ (blue), $\zeta=20$ (green),
and also for Figures 8, 9, 10.
}
\label{regrhopte2nddifzeta}
\end{figure}

\begin{figure}[htb]
\centering
\subcaptionbox{}
    {%
        \includegraphics[width = .41\linewidth]{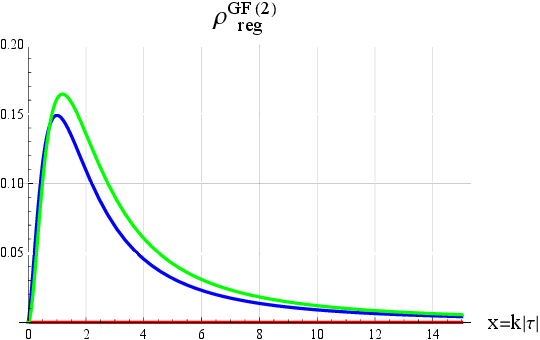}
}
\subcaptionbox{}
    {%
        \includegraphics[width = .41\linewidth]{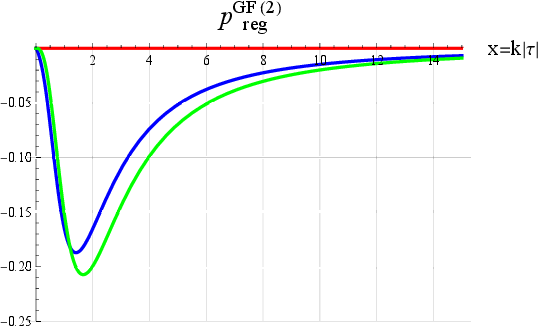}
}
\caption{
The regularized GF  spectral stress tensor depends  on $\zeta$.
(a) $\rho_{k~reg}^{GF(2)}$.
(b) $p_{k~reg}^{GF(2)}$.
}
\label{regrhopgf2nddifzeta}
\end{figure}

The $\zeta$-dependence of
the 2nd-order regularized temporal, and GF  spectral stress tensors
are shown in Fig.\ref{regrhopte2nddifzeta}
and Fig.\ref{regrhopgf2nddifzeta}, respectively.
In particular, in the Landau gauge ($\zeta=0$)
both the  temporal and GF regularized stress tensors are zero,
\bl
 \rho^{T(2) }_{k\, reg}  & = p^{T(2)}_{k\, reg}  =0,
\label{2regTemp}
\\
 \rho^{GF(2) }_{k\, reg} & = p^{GF(2)}_{k\, reg} =0 ,
\label{2regGF}
\el
as is obvious from
\eqref{regtrh2} \eqref{regtp2} \eqref{regGFrh2} \eqref{regGFp2},
so that the regularized Stueckelberg stress tensor
reduces to the regularized Proca stress tensor,
containing only the transverse and longitudinal parts. (See Sect 5.)

{\bf 5) the total regularized   stress tensor :}

\begin{figure}[htb]
\centering
\subcaptionbox{}
    {%
        \includegraphics[width = .41\linewidth]{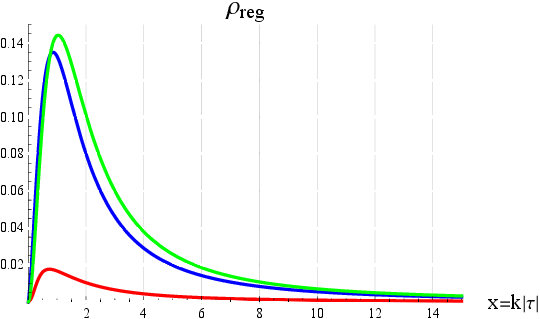}
}
\subcaptionbox{}
    {%
        \includegraphics[width = .41\linewidth]{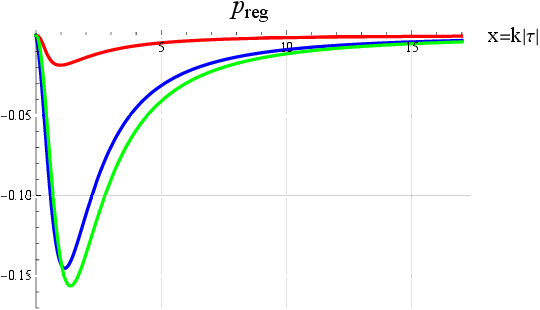}
}
\caption{ The total regularized  spectral stress tensor depends on $\zeta$.
(a) $\rho_{k~reg} >0$.
(b) $p_{k~reg} < 0$.
}
\label{regrhopTotaldiffzeta}
\end{figure}

Summing up the four parts,
the total regularized spectral stress tensor at high $k$ is
\bl
\rho_{k~reg} &  =  \rho_{k~reg}^{TR(0)}
              + \rho_{k~reg}^{L(2)}
              + \rho_{k~reg}^{T(2)}
              + \rho_{k~reg}^{GF(2)}
\nn
\\
&  = \frac{k^3}{2\pi^2a^4}
\Big(\frac{ \left(2 m^4+5 m^2H^2 (3+\zeta )\right)}{16 H^4 k^5}
\nn
\\
&   -\frac{5 m^2 \left(4 m^4+28 H^4 (3+\zeta )
       +H^2 m^2 \left(65-9 \zeta ^2\right)\right)}{64 H^6 k^7}
    +...\Big) ,
 \label{trho}
  \\
p_{k~reg} & =  p_{k~reg}^{TR(0)}
              + p_{k~reg}^{L(2)}
              + p_{k~reg}^{T(2)}
              + p_{k~reg}^{GF(2)}
\nn \\
& =  \frac{k^3}{2\pi^2a^4}
\frac13\Big(-\frac{ \left(10 m^4+25m^2 H^2 (3+\zeta )\right)}
{16 H^4 k^5}
\nn
\\
&  +\frac{35 m^2 \left(4 m^4+28 H^4 (3+\zeta )
          +H^2 m^2 \left(65-9 \zeta ^2\right)\right)}{64 H^6 k^7}
        +... \Big ) ,
             \label{stresspre}
\el
and  the trace of total spectral stress tensor  at high $k$ is
\bl
\langle T^{\beta}_{~\beta}\rangle_{k~reg}
 = & -\rho_{k~reg} +3p_{k~reg}
\nn
\\
 = & \frac{k^3}{2\pi^2a^4}
    \Big(-\frac{3 m^2 \left(2 m^2+5 H^2 (3+\zeta )\right)}{8 H^4 k^5}
  \nn
  \\
&  +\frac{5 m^2 \left(4 m^4+28 H^4 (3+\zeta )
         +H^2 m^2 \left(65-9 \zeta ^2\right)\right)}{8 H^6 k^7}
    +...  \Big) .
  \label{ttrace}
\el
The total regularized  stress tensor
respects the covariant conservation
\bl
\rho' _{k\, reg}  +3 \frac{a'}{a} (\rho_{k\, reg}+p_{k\, reg})  =0.
\el
Fig.\ref{regrhopTotaldiffzeta} shows that
the total  regularized  spectral energy densities is positive,
and the total  regularized  spectral pressure is negative for all $k$.

Setting $m=0$ in \eqref{trho} \eqref{stresspre} \eqref{ttrace}
gives the massless limit of the regularized Stueckelberg stress tensor,
\bl \label{Ststressm0}
\rho_{k\, reg}=0,
~~~
 p_{k\, reg} =0,
 ~~~
\langle T^{ \beta}\, _\beta \rangle _{k\, reg} = 0 .
\el
This result agrees with the vanishing regularized vacuum stress tensor
of Maxwell field with the GF term \cite{ZhangYe2022}.
Thus, the conflict mentioned in the Introduction disappears.
The result  \eqref{Ststressm0} can be derived
in an alternative way (see \eqref{maxreg} in Sect 5).

By numerical integration of the regularized spectral stress tensor over $k$,
we obtain the regularized vacuum stress tensor.
\bl
\rho_{reg}
& = \int^\infty_0  \rho_{k~reg} \frac{d k}{k}
  = \int^\infty_0  (\rho_{k~reg}^{TR(0)}
              + \rho_{k~reg}^{L(2)}
              + \rho_{k~reg}^{T(2)}
              + \rho_{k~reg}^{GF(2)}) \frac{d k}{k} ,
\\
p_{reg} & = \int^\infty_0  p_{k~reg} \frac{d k}{k}
  =   \int^\infty_0  (p_{k~reg}^{TR(0)}
              + p_{k~reg}^{L(2)}
              + p_{k~reg}^{T(2)}
              + p_{k~reg}^{GF(2)}) \frac{d k}{k} .
\el
For instance, for the model  $\frac{m^2}{H^2} =0.1$,
the 0th-order regularized transverse stress tensor  is
\bl
\rho_{reg}^{TR(0)} &=\int_{0}^{\infty}\rho_{k~reg}^{TR(0)}\frac{k}{dk}
   =- p_{reg}^{TR}
   = 0.0014  \times \frac{H^4}{2\pi^2}  ,
\label{eq14}
\el
the 2nd-order regularized longitudinal stress tensor is
\bl
\rho_{reg}^{L(2)}& =\int_0^\infty\rho_{k~reg}^{L(2)}\frac{dk}{k}
= - p_{reg}^{L(2)}= 0.0407  \times \frac{H^4}{2\pi^2} ,
\label{eq24}
\el
the 2nd-order regularized temporal stress tensor (for $\zeta=20$)  is
\bl
\rho_{reg}^{T(2)}& =\int_0^\infty\rho_{k~reg}^{T(2)}\frac{dk}{k}
= - p_{reg}^{T(2)}
=  -0.1041   \times \frac{H^4}{2\pi^2} ,
\label{eq34}
\el
the 2nd-order regularized GF stress tensor  (for $\zeta=20$)  is
\bl
\rho_{reg}^{GF(2)}&  =\int_0^\infty\rho_{k~reg}^{GF(2)}\frac{dk}{k}
= - p_{reg}^{GF(2)}=  0.3750 \times \frac{H^4}{2\pi^2} ,
\label{eq44}
\el
the total regularized Stueckelberg   vacuum energy density is
 constant and  positive
\bl
\rho_{reg} & = -p_{reg}
    = \rho_{reg}^{TR(0)} +\rho_{reg}^{L(2)}
             +\rho_{reg}^{T(2)} +\rho_{reg}^{GF(2)}
  = 0.3129 \times \frac{H^4}{2\pi^2} >0 .
\label{rhoreg2tot}
\el

More interesting is the regularized Proca stress tensor,
which is simply equal to the regularized Stueckelberg stress tensor
in the Landau gauge $\zeta=0$
(only the transverse and longitudinal contributions),
\bl
\rho_{reg} & =-p_{reg}
   =\int^\infty_0  (\rho_{k~reg}^{TR(0)}
         + \rho_{k~reg}^{L(2)}) \frac{d k}{k}
    = 0.0421  \times \frac{H^4}{2\pi^2} >0
    ~~~  \text{(for $\frac{m^2}{H^2}=0.1$),}
  \label{rhoLandaugm01}
    \\
\rho_{reg} & =-p_{reg}
       = 0.1395  \times \frac{H^4}{2\pi^2} >0
      ~~~  \text{(for $\frac{m^2}{H^2}=1$),}
      \label{rhoLandaugm1}
\\
\rho_{reg} & = -p_{reg}
  = 0.1922  \times \frac{H^4}{2 \pi^2 } >0
  ~~~  \text{(for $\frac{m^2}{H^2}=2$)} .
  \label{rhoLandaugm2}
\el
As is seen,  a larger mass $m$ yields a higher amplitude
of the regularized stress tensor.

Amazingly, for both the Stueckelberg and the Proca,
the regularized pressure $p_{reg}$
becomes minus of the regularized energy density.
By this important property,
the  regularized vacuum stress tensor is maximally-symmetric,
\bl
\langle T_{\mu\nu}  \rangle_{reg}
   =  \frac14 g_{\mu\nu} \langle T^{\beta} \,_{\beta }\rangle_{reg}
   = - \rho_{reg} g_{\mu\nu} ,
\el
and can be naturally regarded as the cosmological constant
that drives the de Sitter inflation.
Thus, by properly removing the UV divergences,
we have derived the cosmological constant
as the vacuum expectation of the stress tensor
of the quantum fields \cite{Weinberg1989}.
Note that the regularized vacuum spectral stress tensor
is non-uniformly distributed in the $k$-modes,
will induce  the metric perturbations,
and will form the seeds of the primordial cosmic fluctuations
\cite{WangZhang2017}.

 After regularization the next thing is renormalization
to absorb divergent terms into bare parameters.
It is  known that for a scalar field
the 0th and 2nd order divergences can be absorbed into
the cosmological constant and the gravitational constant \cite{Bunch1980}.
We expect a similar treatment may be performed on the vector field,
and this will be for future study.

\section{\bf Reduction of the Stueckelberg stress tensor}

The Stueckelberg field  can reduce to other interesting vector fields
in a simple manner.

{\bf  The Proca  stress tensor: }

For both the unregularized and the regularized,
the Stueckelberg stress tensor  reduces straightforwardly
to the Proca  stress tensor.
Note that the Proca field has no GF term.
The transverse field $B_i$ in  \eqref{f1f2}
and the transverse vacuum spectral stress tensor \eqref{spTRPRho} \eqref{spTRPRess}
are still valid for the Proca field.
By the Lorenz condition $\pi_A^0  =0$,
the expressions \eqref{ApiA} \eqref{A_0} reduce to  the following
\bl \label{A_0p}
A   = \frac{1}{a^2 m^2} ( -\partial_0 \pi_A ),
 ~~~~~~~~
A_0  = \frac{1}{a^2 m^2}    k^2  \pi_A   ,
\el
and consequently
the LT stress tensor \eqref{longittemporalrho2} \eqref{longittemporal2}
reduces to
 \bl
\rho^{L}_k   & =  \frac{k^3}{4 \pi^2a^4}
\Big[  |\Big( \partial_0 +\frac{D}{2} \Big) Y^{(L)}_k|^2
   +    \omega^2 |Y^{(L)}_k(\tau)|^2   \Big] ,
   \label{LTsptrhoProc}
   \\
p^{L}_k  & = \frac{k^3}{4 \pi^2a^4} \frac13
 \Big[ - |\Big( \partial_0 +\frac{D}{2} \Big) Y^{(L)}_k|^2
      +  3 k^2 |Y^{(L)}_k |^2   + m^2 a^2 | Y^{(L)}_k |^2
     \Big] ,
    \label{LTsptpressProc}
\el
the same as the longitudinal \eqref{rholongitudinal} and \eqref{Lonpr},
and the temporal part is absent.
So   the total unregularized Proca vacuum spectral stress tensor is
\bl
\rho_k = \rho^{TR}_k + \rho^{L}_k ,
~~~ p_k = p^{TR}_k + p^{L}_k ,
\label{Proca2parts}
\el
containing only the transverse and longitudinal parts,
and the Proca vacuum spectral trace is
\bl
\langle 0| T^\mu\, _\mu |0\rangle_k
 & = -\rho_k + 3 p_k
\nn
\\
& = \frac{k^3}{2\pi^2 a^4}
  \Big[- 2  m^2 a^2 |f_k^{(1)}(\tau)|^2 \Big]
\nn \\
& + \frac{k^3}{4 \pi^2a^4}
 \Big[ -2  | ( \partial_0 +\frac{D}{2}  ) Y^{(L)}_k|^2
     + 2 k^2 |Y^{(L)}_k|^2   \Big] .
\label{unregProc}
\el
The regularized Proca stress tensor is given by
\bl
\rho_{k\, reg} & = \rho^{TR(0)}_{k\, reg} + \rho^{L(2)}_{k\, reg} ,
\\
p_{k\, reg} & = p^{TR(0)}_{k\, reg} + p^{L(2)}_{k\, reg} ,
\el
(see \eqref{trsadrh} for the transverse,
and   \eqref{Longreg2} for the longitudinal),
which  is equal to
the regularized Stueckelberg stress tensor in the Landau gauge  $\zeta=0$.
The massless limit of the regularized Proca stress tensor is zero
\bl
\rho_{k\, reg}=0,
~~~
 p_{k\, reg} =0,
 ~~~
\langle T^{ \beta}\, _\beta \rangle _{k\, reg} = 0 ,
\label{Procstressm0}
\el
which is analogous to \eqref{Ststressm0} for
the regularized Stueckelberg stress tensor in the massless limit.

{\bf The Maxwell field with the GF term: }

The Stueckelberg  stress tensor in the massless limit
reduces to the stress tensor of Maxwell field with the GF term.
Setting  $m=0$,
the massive solutions  \eqref{f1f2} \eqref{Asol}  \eqref{A0sol}
reduce to the massless solutions (see Appendix B).
The unregularized transverse
vacuum spectral stress tensor \eqref{spTRPRho} \eqref{spTRPRess}
reduces to
\bl
\rho^{TR}_k  & = 3 p^{TR}_k
  = \frac{k^4}{2 \pi^2 a^4} ,  \label{spTRPRhomaslss}
\el
which is UV divergent and  equal to
twice of the stress tensor of a conformally-coupling massless scalar field
\cite{ZhangYeWang2020,ZhangWangYe2020}.
The unregularized longitudinal vacuum  spectral  stress tensor
\eqref{rholongitudinal} \eqref{Lonpr} at $m=0$ reduces to
\bl
\rho^{L}_k
& =  \frac{k^4}{4 \pi^2a^4}
\Big( 1 +    \frac{1}{ 2 (k\tau)^2} \Big) ,  \label{rhomasll}
\\
p^{L}_k
&     = \frac{k^4}{4 \pi^2a^4} \frac13
 \Big( 1  -    \frac{1}{2 (k\tau)^2} \Big) , \label{prmasll}
\el
which  is UV divergent and  equal to
those  of a minimally-coupling massless scalar field
 \cite{ZhangYeWang2020,ZhangWangYe2020}.
The unregularized  temporal vacuum  spectral stress tensor
 \eqref{rhotemporal} \eqref{Temppr}  at $m=0$  reduces to
\bl
\rho^{T}_k
& = \frac{k^4}{4 \pi^2a^4}
    \Big( -1 - \frac{1}{2 (k\tau)^2} \Big)  , \label{temrhomasll}
\\
p^{T}_k
 & = \frac{k^3}{4 \pi^2a^4} \frac13
 \Big( - 1 + \frac{1}{2 (k\tau)^2} \Big) ,  \label{tempmasll}
\el
which  is  independent of $\zeta$,
because $\zeta$ disappears with the zero mass.
It is seen that
the longitudinal and temporal parts cancel  out
\bl \label{prlong}
 \rho^{LT}_k=  \rho^{L}_k + \rho^{T}_k =0 ,
       ~~~~~~~~~~   p^{LT}_k = p^{L}_k + p^{T}_k = 0 .
\el
(As demonstrated in Ref.\cite{ZhangYe2022}, in the GB  states,
the longitudinal and temporal
stress tensors contributed by the photons above the vacuum
also  cancel out.)
The unregularized GF vacuum stress tensor \eqref{gfrh} \eqref{sptrpressGFVEV}
 at $m=0$  reduces to
\bl
\rho_k^{GF}
& =  \frac{k^4}{2\pi^2a^4}
  \Big( 1 + \frac{1}{2}  \frac{1}{(k\tau)^2}   \Big)
      , \label{gfmslss}
         \\
p_k^{GF}
&   =\frac{ k^4 }{6\pi^2 a^4}
  \Big(1-    \frac{1}{2 (k\tau)^2} \Big)
     , \label{gfprmlss}
\el
independent of $\zeta$,
and is the same as twice that of a minimally-coupling
massless scalar field.
 \eqref{spTRPRhomaslss} \eqref{gfmslss} \eqref{gfprmlss}
gives  the total unregularized vacuum stress tensor of
the Maxwell field  with the gauge fixing term \cite{ZhangYe2022}.
(In the GB  states,  the GF stress tensor
due to the photons above the vacuum is zero \cite{ZhangYe2022}.)

The transverse \eqref{spTRPRhomaslss}
is regularized to zero by the 0th-order  subtraction terms  \eqref{ms0th},
and the GF \eqref{gfmslss} \eqref{gfprmlss}
is regularized to zero by the 2nd-order
 subtraction terms \eqref{GF2msrho} \eqref{GF2ms},
and consequently the total regularized Maxwell vacuum stress tensor is zero,
and there is no trace anomaly \cite{AdlerLiebermanNg1977,ZhangYe2022},
\bl
\rho_{k\, reg}=0= p_{k\, reg} ,
~~~ \langle T^{ \beta}\, _\beta \rangle _{k\, reg} = 0 .
\label{maxreg}
\el
This result agrees with \eqref{Ststressm0}
that followed from taking the massless limit
of the regularized Stueckelberg stress tensor.
Thus,  both routes lead to the same result consistently.
If the 4th-order regularization were adopted
upon the Stueckelberg field
\cite{DowkerCritchley1977,BrownCassidy1977,
Endo1984,ChimentoCossarini1990,ChuKoyama2017},
its the massless limit will not be equal to \eqref{maxreg},
in conflict with the vanishing Maxwell vacuum stress tensor,
as mentioned in the Introduction.

{ \bf The limit to the Minkowski  spacetime: }

The Stueckelberg stress tensor in the flat spacetime is also revealing.
Conceptually, the stress tensor is not needed in the flat spacetime,
but it helps to reveal
the structure of UV divergences of the stress tensor in curved spacetimes.
Setting $a=1$, $D=D'=0$ in
 eqs.\eqref{Bieq1} \eqref{0aA} \eqref{Aa0m}
leads  to the equations of Stueckelberg field in flat spacetime.
The positive frequency transverse solutions are
$B_i= f^{(\sigma)}_k (\tau) \equiv \frac{1}{\sqrt{2\omega}} e^{-i \omega \tau}$,
$\sigma =1,2$, with  $\omega=(k^2+m^2)^{1/2}$,
and the longitudinal and temporal solutions  are given by
\bl
A  & = \frac{1}{m} \big( - \frac1k y^{(L)'}_k  + y^{(0)}_k \big) ,   \label{ApiAmks}
\\
A_0 & =  \frac{1}{m} \big(  y^{(0)'}_k  +k  y^{(L)}_k \big)  ,    \label{A_0mks}
\el
with $y^{(L)}_k(\tau)\equiv  \frac{1}{\sqrt{2\omega}} e^{-i \omega \tau}$,
$y^{(0)}_k(\tau) \equiv \frac{1}{\sqrt{2\omega_0 }} e^{-i \omega_0 \tau}$,
and  $ \omega_0 =(k^2+ \zeta m^2)^{1/2}$.
The  vacuum spectral stress tensor is  given by the following
\bl
\rho^{TR}_k  & =  \frac{k^3}{2\pi^2}
  \Big[ |f_k^{(1)\, '}(\tau)|^2 + k^2 |f_k^{(1)}(\tau)|^2
  + m^2  |f_k^{(1)}(\tau)|^2 \Big]
  = \frac{k^3}{2\pi^2} \,  \omega  ,
   \label{spTRPRhomk}
 \\
p^{TR}_k & = \frac{k^3}{2\pi^2} \frac13
  \bigg[  |f_k^{(1)\, '}(\tau)|^2
      +  k^2 |f_k^{(1)}(\tau)|^2
      -  m^2 |f_k^{(1)}(\tau)|^2  \bigg]
 = \frac{k^3}{2\pi^2} \frac13   \frac{k^2}{\omega}   ,
         \label{spTRPRessmk}
\el
\bl
\rho^{L}_k  & = \frac{k^3}{4 \pi^2}
\Big[  |y^{(L)\, '}_k|^2
         + k^2  |y^{(L)}_k|^2
         + m^2  |y^{(L)}_k|^2   \Big]
= \frac{k^3}{4 \pi^2}  \, \omega         ,
  \label{rholongitudinalmk}
 \\
p^{L}_k  & = \frac{k^3}{4 \pi^2 } \frac13
 \Big[    - | y^{(L)\, '}_k|^2
      +  3 k^2 |y^{(L)}_k|^2  +  m^2 |y^{(L)}_k |^2 \Big]
= \frac{k^3}{12 \pi^2 } \frac{ k^2}{\omega} ,
           \label{Lonprmk}
\el
\bl
\rho^{T}_k  & = \frac{k^3}{4 \pi^2}
\Big[ -|y^{(0)\, '}_k |^2  - k^2 |y^{(0)}_k|^2  \Big]
= \frac{k^3}{4 \pi^2}
 \Big( - \frac{1}{2} \omega_0 -\frac{k^2}{2\omega_0} \Big) ,
      \label{rhotemporalmk}
     \\
p^{T}_k  & = \frac{k^3}{4 \pi^2 } \frac13
 \Big[  -3|y^{(0)\, '}_k|^2  + k^2|y^{(0)}_k |^2 \Big]
 = \frac{k^3}{4 \pi^2 }
  \Big(    -\frac12 \omega_0 + \frac{k^2}{6 \omega_0}  \Big) ,
    \label{Tempprmk}
\el
\bl
\rho_k^{GF}  & = \frac{k^3}{2\pi^2 }\Big(
  |y^{(0\, ')}_k|^2 + k^2|y^{(0)}_k|^2
        +\frac12\zeta m^2  |y^{(0)}_k|^2   \Big)
 =  \frac{k^3}{2\pi^2 }\Big(  \frac34 \omega_0
        + \frac{k^2}{4\omega_0} \Big)  ,
\\
 p_k^{GF}  &  = \frac{k^3}{2\pi^2 }\Big(
  |y^{(0\, ')}_k|^2 -\frac13  k^2|y^{(0)}_k|^2
       - \frac12\zeta m^2  |y^{(0)}_k|^2  \Big)
=  \frac{k^3}{2\pi^2 } \Big(  \frac14 \omega_0
        + \frac{k^2}{12 \omega_0} \Big)    .
   \label{gfrhmk}
\el
Notice that  these spectra  are UV divergent,  and correspond to
the 0th-order subtraction terms for the stress tensor
(see  \eqref{0thtrrho} \eqref{pTMV0th}
\eqref{0thrhot} \eqref{0thprt}
\eqref{722} \eqref{0pTad}
\eqref{822} \eqref{855} in Appendix A).
In the Minkowski spacetime,
 these vacuum terms are commonly removed by the normal ordering.
This analysis tells that,  in the adiabatic regularization,
the 0th-order subtraction terms
remove the vacuum UV divergences in the Minkowski spacetime,
and the subtraction terms above the 0th-order
remove the UV divergences that  are associated with   curved spacetimes.

\section{ The negative energy density and the trace anomaly
         from the 4th-order regularization }

\begin{figure}[htb]
\includegraphics[width=0.6\linewidth]{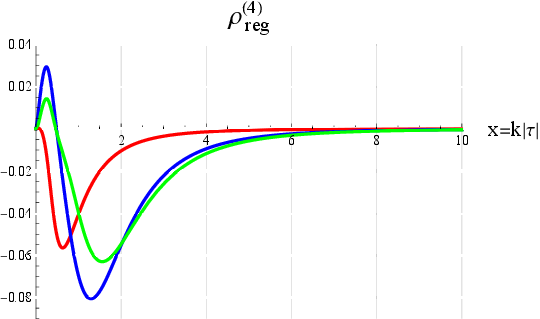}
	\centering
\caption{
  The 4th-order regularized spectral energy density
   $\rho^{(4)}_{k~reg}$  is negative in a broad  range $k\gtrsim 1$.
So the 4th-order regularization
is improper for the Stueckelberg  and Proca stress tensors
in de Sitter space.
}
\label{regrho4}
\end{figure}

In literature,
the 4th-order adiabatic regularization was performed
irrespectively  on each part of the Stueckelberg
stress tensor \cite{ChimentoCossarini1990,ChuKoyama2017,Endo1984}
  by default.
The 4th-order regularized Stueckelberg spectral stress tensor was  given by
\bl
\rho^{(4)}_{k~reg} & =  \rho_{k~reg}^{TR(4)}
              + \rho_{k~reg}^{L(4)}
              + \rho_{k~reg}^{T(4)}
              + \rho_{k~reg}^{GF(4)} , \label{4thsubrho}
 \\
p^{(4)}_{k~reg} & =  p_{k~reg}^{TR(4)}
              + p_{k~reg}^{L(4)}
              + p_{k~reg}^{T(4)}
              + p_{k~reg}^{GF(4)} ,  \label{4thsubpress}
\el
and the $k$-integrations give the 4th-order regularized stress tensor
\bl
\rho^{(4)}_{reg}
& = \int^\infty_0  \rho^{(4)}_{k~reg} \frac{d k}{k}
  = \int^\infty_0  (\rho_{k~reg}^{TR(4)}
              + \rho_{k~reg}^{L(4)}
              + \rho_{k~reg}^{T(4)}
              + \rho_{k~reg}^{GF(4)}) \frac{d k}{k} ,
\\
p^{(4)}_{reg} & = \int^\infty_0  p^{(4)}_{k~reg} \frac{d k}{k}
  =   \int^\infty_0  (p_{k~reg}^{TR(4)}
              + p_{k~reg}^{L(4)}
              + p_{k~reg}^{T(4)}
              + p_{k~reg}^{GF(4)}) \frac{d k}{k} .
\el
As it stands, the 4th-order regularization  has several problems.
First, the 4th-order regularization
does not obey the minimal subtraction rule \cite{ParkerFulling1974},
would subtract off more terms than necessary
and generally lead to a negative spectral energy density
shown in Fig.\ref{regrho4},
and a negative energy density as well.

For instance, for the model $\frac{m^2}{H^2} =0.1$ and $\zeta=20$,
our properly regularized total energy density \eqref{rhoreg2tot} is positive,
while  the 4th-order regularized total energy density would be negative
\bl
\rho_{reg}^{TR(4)}&=\int_{0}^{\infty}\rho_{k~reg}^{TR(4)}\frac{dk}{k}
  = 0.0014  \times \frac{H^4}{2\pi^2} ,\label{eqa4}
\\
\rho_{reg}^{L(4)}&=\int_{0}^{\infty}\rho_{k~reg}^{L(4)}\frac{dk}{k}
   = -0.0822  \times \frac{H^4}{2\pi^2} ,\label{eqb4}
\\
\rho_{reg}^{T(4)}&=\int_{0}^{\infty}\rho_{k~reg}^{T(4)}\frac{dk}{k}
   =  0.0229   \times \frac{H^4}{2\pi^2} ,\label{eqc4}
\\
\rho_{reg}^{GF(4)}&=\int_{0}^{\infty}\rho_{k~reg}^{GF(4)} \frac{dk}{k}
  \sim  10^{-6} \times \frac{H^4}{2\pi^2} ,
 \label{eqd4}
\\
\rho_{reg}^{(4)} & =  \rho_{reg}^{TR(4)} +\rho_{reg}^{L(4)}
+\rho_{reg}^{T(4)}  +\rho_{reg}^{GF(4)}
   =-0.0579 \times \frac{H^4}{2\pi^2} <0 .
\label{rhoreg4}
\el

For the Proca stress tensor,  our properly regularized total energy density
is positive (see \eqref{rhoLandaugm01} \eqref{rhoLandaugm1}  \eqref{rhoLandaugm2}),
while the  4th-order regularization would give a negative energy density,
respectively
\bl
\rho^{(4)}_{reg}
&  =  -p^{(4)}_{reg}
=\int^\infty_0  (\rho_{k~reg}^{TR(4)}
         + \rho_{k~reg}^{L(4)}) \frac{d k}{k}
         =  -0.0808   \times \frac{H^4}{2\pi^2} <0
  ~~~  \text{(for $\frac{m^2}{H^2}=0.1$),}
  \label{4rhoregul01}
\\
\rho^{(4)}_{reg}
&  =  -p^{(4)}_{reg}
   = -0.0209  \times \frac{H^4}{2\pi^2} < 0
        ~~~  \text{(for $\frac{m^2}{H^2}=1$),}
  \label{4rhoregul}
\\
\rho^{(4)}_{reg} & =  -p^{(4)}_{reg}
  = -0.0098 \times \frac{H^4}{2 \pi^2} <0
  ~~~  \text{(for $\frac{m^2}{H^2}=2$).}
     \label{4rhoregul2}
\el
This is also equal to  the 4th-order regularized Stueckelberg stress tensor
in the Landau gauge $\zeta=0$,
since the temporal and GF stress tensors with $\zeta=0$ are effectively massless,
so that the 4th-order regularization
yields $\rho_{reg}^{T(4)} = p_{reg}^{T(4)}=0$
and $\rho_{reg}^{GF(4)}= p_{reg}^{GF(4)} =0$,
similar to the 2nd-order \eqref{2regTemp} \eqref{2regGF}.

The issue of the negative energy density
of the 4th-order regularized Stueckelberg  and Proca  stress tensors
has not been noticed,  nor addressed in the literature
that adopted the 4th-order regularization.
Our analysis shows that the  unphysical, negative energy density
was caused by the oversubtraction of the 4th-order regularization
and is in conflict with the de Sitter inflation.

Next, the 4th-order regularization simultaneously causes
the so-called trace anomaly in the massless limit in literature.
Under our proper regularization,
there is no  trace anomaly for the Stueckelberg and Proca stress tensors
(see \eqref{Ststressm0} and \eqref{Procstressm0}).
In the following we show that,
in the improper 4th-order regularization,
the trace anomaly is carried by the 4th-order subtraction terms
above the 2nd-order ones.

Consider the de Sitter space,
in which the divergent $1/\omega^3$ and  $1/\omega_0^3$ terms,
 \eqref{rhoom3},  \eqref{pom3},  \eqref{rhoTom},  \eqref{preTemom},
  \eqref{rhovGF},  \eqref{pvGF},
are all vanishing in  the 4th-order subtraction terms.
We evaluate the trace of the 4th-order subtraction terms
as the following
\bl
  \lim_{m=0} \int_0^\infty
\Big( -(\rho_{k\, A4}-\rho_{k\, A2})
          +3 (p_{k\, A4}-p_{k\, A2}) \Big) \frac{dk}{k} ,
\label{trsub4}
\el
where the 2nd-order subtraction terms  are taken away
since they do not contribute to the trace anomaly in the massless limit.
The procedure of \eqref{trsub4} is
 $k$-integrating first and then taking the massless limit.
If  the massless limit is taken before the $k$-integration,
 \eqref{trsub4}  will give a  zero trace.
For the Stueckelberg stress tensor,
integrating first and then taking the massless limit,
\eqref{trsub4} is found to be
\bl
 \frac{H^4}{120\pi^2}
   -\frac{59 H^4}{240 \pi^2}
   + \frac{61H^4}{240\pi^2}
    -\frac{3H^4}{4\pi^2}
    = -\frac{11 H^4}{15 \pi^2},
\label{Stuecktrace}
\el
independent of $m$ and $\zeta$,
where four terms are contributed respectively by
the transverse, longitudinal,  temporal, and GF parts.
(See \eqref{desttiertrs} \eqref{longesi} \eqref{desitsu}
\eqref{desitsugf}
in Appendix A for details.)
The trace \eqref{Stuecktrace} is associated with
the negative energy density \eqref{rhoreg4},
and both  origin from the 4th-order regularization.

For the Proca  stress tensor
with only the transverse and longitudinal contributions,
the trace of the 4th-order subtraction terms \eqref{trsub4} yields
\bl
\frac{H^4}{120\pi^2}    -\frac{59 H^4}{240 \pi^2}
 = -\frac{19 H^4}{80 \pi^2} ,
\label{Proctrace}
\el
which is the same as (5.7) (5.8) (5.9) of Ref.\cite{Salas2023}
that contained a typo $H^2$.
The trace \eqref{Proctrace} is associated with
the negative energy density \eqref{4rhoregul01} \eqref{4rhoregul} \eqref{4rhoregul2}.

If a complex ghost field was additionally introduced
\cite{ChimentoCossarini1990,ChuKoyama2017,Endo1984},
under the 4th-order adiabatic regularization,
the trace of the 4th-order subtraction terms
for the system of the Stueckelberg and the ghost fields
in  generic fRW spacetimes would be given by  \eqref{rtana}
as the following  (see Appendix A for details)
\bl
 \frac{1}{2880\pi^2}\Big(-62(R_{\mu\nu}R^{\mu\nu}
-\frac{1}{3}R^2)+( 18 +15\log \zeta) \square R\Big) .
\label{Stgentrace}
\el
independent of $m$, but depending on $\zeta$,
and in the de Sitter space, it becomes
\bl
 -\frac{11 H^4}{15 \pi^2}
    + \frac{119 H^4}{120\pi^2}
    = \frac{31 H^4}{120\pi^2}   ,
\label{StueckGhosttrace}
\el
independent of $m$ and $\zeta$,
where the term $\frac{119 H^4}{120\pi^2}$  is
the ghost contribution (see \eqref{ghosttrdestt} \eqref{desittrtana}).
By definition,  the trace anomalies
are minus of \eqref{Stuecktrace}, \eqref{Proctrace},
\eqref{Stgentrace},   \eqref{StueckGhosttrace},  respectively.

A conformally-coupling massive scalar field  in the massless limit
has a zero trace ($T^\mu\, _\mu= 0$) of the unregularized stress tensor
and thus becomes conformally invariant.
Under the 4th-order regularization
the trace would become nonzero
and the conformal invariance would be broken,
and this has been referred to as
the so-called conformal trace anomaly in literature.
This kind of search for trace anomaly was indiscriminatively generalized to
other non-conformally-invariant fields.
The Stueckelberg and Proca fields are not conformally invariant
in the massless limit.
This is evidenced by the nonzero trace $T^\mu\, _\mu \ne 0$
in the massless limit (see \eqref{unregstrSt} and \eqref{unregProc})
of the unregularized stress tensors,
and also evidenced by the fact that the equation and the stress tensor
of the longitudinal component are structurally different from
those of the comformally-coupling scalar field.
In this regards,  the phrase ``the conformal trace anomaly" is improper
 for the Stueckelberg and Proca fields,
and  it is meaningless to search for a nonzero regularized trace
of the stress tensor  of these fields in the massless limit.

Here we emphasize again that the trace anomalies
for the scalar and vector fields
are merely artifacts arising from the improper 4th-order regularization,
and do not exist in the proper regularization.

\section{Conclusions and Discussions}

We have studied  the Stueckelberg field in de Sitter space.
The equations of components of Stueckelberg  field are  different.
The transverse equation \eqref{Bieq1} is analogous to
that of a conformally-coupling  massive scalar field \cite{ZhangYeWang2020}.
The longitudinal and temporal equations \eqref{0aA} \eqref{Aa0m} are mixed up,
and we have been able to separate into two fourth order differential equations
\eqref{0aA3} \eqref{st4eqA0}
and obtained the exact  solutions \eqref{Asol} \eqref{A0sol}.
Upon the covariant canonical quantization,
we give the unregularized Stueckelberg spectral stress tensor
containing UV divergences.

In literature the Stueckelberg vacuum stress tensor
was conventionally  regularized to the 4th-order, by default.
We find that this would give an unphysical negative energy density,
a fact that has not been noticed nor addressed in literature before.
Moreover, in the massless limit the 4th-order regularized stress tensor
would lead to a nonzero trace,
which is in conflict with the zero trace of
the regularized  Maxwell stress tensor.
To avoid these difficulties,
we have decomposed the Stueckelberg stress tensor
into the transverse, longitudinal, temporal, and GF parts,
each containing its respective UV divergences,
in analogy to the Maxwell field \cite{ZhangYe2022,YeZhang2024}.
According to the minimal subtraction rule \cite{ParkerFulling1974},
we have regularized each part
to its pertinent adiabatic order,
so that the regularized total spectral energy density is positive.
Specifically,
for the transverse stress tensor,
the 0th-order regularization is sufficient to remove all the UV divergences,
while for the longitudinal, temporal and GF stress tensors,
the 2nd-order regularization removes all the UV divergences.
Other schemes of regularization of orders different from ours
would either be insufficient, or would subtract more than necessary,
leading to a negative total energy density.
We like to point out that,
although the 4th-order regularization is improper
for the Stueckelberg stress tensor in de Sitter space,
it was necessarily used for the minimally-coupling massless scalar field
and for the Maxwell stress tensor with the GF term
in the MD stage  \cite{YeZhang2024}.
All these results tell that the pertinent order of regularization
depends upon the type of fields
(the components of a vector  field and the coupling of a scalar field),
as well as upon the background spacetimes \cite{ZhangYe2022,YeZhang2024}.

Under  the Lorenz condition, for both the unregularized and the regularized,
the Stueckelberg stress tensor
reduces to the Proca stress tensor
that consists of only the transverse and longitudinal parts.
Besides, in the massless limit,
the regularized Stueckelberg (Proca as well)
vacuum stress tensor becomes zero,
and this agrees  with the regularized Maxwell vacuum stress tensor
with the GF term \cite{ZhangYe2022,YeZhang2024}.
Thus, under our proper regularization,
two alternative routes (starting with the massive or the massless)
lead to  the same result for the massless field,
and the conflict disappears.

In literature the so-called trace anomaly came out of
the following scheme:
starting with a massive  field,
applying the 4th-order regularization on the stress tensor,
performing $k$-integration,
and then taking the massless limit of the trace
of the regularized stress tensor.
In contrast, our proper regularization is lower than the 4th-order,
so that neither the trace anomaly nor the negative energy density occur.
As we check explicitly,
the trace anomaly was actually carried in
by the 4th-order subtraction terms of the Stueckelberg and Proca stress tensors.
Hence,  the trace anomaly and the associated negative energy density
are merely the artifacts caused by the improper 4th-order regularization.
This conclusion holds for
the vector fields \cite{ZhangYe2022,YeZhang2024},
 as well as  for the scalar fields
\cite{ZhangYeWang2020,ZhangWangYe2020,ZhangYe2022PointSpl}.

We have also demonstrated that, in fRW spacetimes,
the 4th-order subtraction terms
contain new divergences in the longitudinal
\cite{ChimentoCossarini1990,Salas2023},
as well as in the temporal and GF parts.
These new divergent terms happen to be vanishing in de Sitter space
and in the RD  stage \cite{ZhangYe2022,YeZhang2024},
but will be nonvanishing in generic fRW  spacetimes.
For instance, for the MD stage,
these 4th-order divergent terms
are necessary in removing the IR divergences of
the unregularized Maxwell stress tensor with the GF term \cite{YeZhang2024}.

In literature, a ghost field was often  introduced
to cancel the GF stress tensor of a vector field.
However, as we notice,
the GF stress tensor of the Stueckelberg field with a general $\zeta$
will not be canceled out by that of a massive ghost field
since their masses differ.
Moreover, the ghost particles in the excited states above the vacuum
may cause additional unwanted effects.
As for the massless case,
the regularized  GF vacuum stress tensor
is already zero \cite{ZhangYe2022,YeZhang2024},
so that a massless ghost field is not needed.

Under our proper regularization,
the regularized Stueckelberg vacuum spectral stress tensor is
UV and IR convergent and covariantly conserved,
and its spectral energy density is positive
in the whole range of $k$.
After $k$-integration,
the total regularized vacuum stress tensor is
constant and maximally-symmetric, with a positive energy density,
and can be regarded  as the cosmological constant
that drives the de Sitter inflation,
in contrast to the 4th-order regularized total energy density
which is negative, and inconsistent with de Sitter inflation.
From cosmological point of view,
the cosmological constant generally can be
contributed by several quantum fields,
such as the massive scalar fields \cite{ZhangYeWang2020,ZhangYe2022PointSpl}
and the massive vector fields.
The regularized vacuum stress tensors of these fields
are constant and maximally-symmetric,
and jointly drive the inflationary expansion.
The $k$-modes of these fields
in the vacuum state will induce   the metric perturbations,
and form  the seeds of the primordial cosmic fluctuations.

\

{\textbf{Acknowledgements}}

Y. Zhang is supported by NSFC Grant
No. 11675165,
No.  11961131007,
No. 12261131497,
and in part by National Key RD Program of China (2021YFC2203100).

\appendix
\numberwithin{equation}{section}

\section{ The adiabatic subtraction terms of
the Stueckelberg  stress tensor }

The WKB  approximate  solutions \cite{Chakraborty1973,ParkerFulling1974,
FullingParkerHu1974,Bunch1980,AndersonParker1987,BirrellDavies1982}
of each component of the Stueckelberg field
are written as the following
\be \label{Yvn}
Y_k^{(a)}(\tau)
  = \frac{1}{ \sqrt{ 2 W_a(\tau)}}
   \exp \Big[  -i \int^{\tau} W_a(\tau')d\tau' \Big],
\ee
where the effective frequency is
\be \label{YequaWk}
  W_a(\tau)     = \Big[  \omega_a ^2 +\alpha_a
-\frac12 \l( \frac{ W_a  '' }{ W_a}
- \frac32 \big( \frac{W_a  '}{W_a} \big)^2 \r)  \Big]^{1/2}
\ee
with
\bl
\omega^2_{T} &=  k^2  + m^2 a^2,
        ~~~~~\alpha_{T}=0 , \\
\omega^2_{L} &=  k^2   + m^2 a^2 ,
          ~~~~~ \alpha_{L}= \frac{a''}{a} - 2 \frac{a'\, ^2}{a^2},
      \\
\omega^2_0 &= k^2   + \zeta m^2  a^2 ,
          ~~~~ \alpha_{0}= -\frac{a''}{a}   ,
\el
 respectively for the transverse, longitudinal, and temporal modes.
$W_a(\tau)$ is obtained by iteratively solving eq.(\ref{YequaWk})
to a desired adiabatic  order.
The WKB solutions describes the exact solution at high $k$  approximately,
and  the description is becoming more accurately with the increasing order.
Take  the 0th-order  effective  frequency and its reciprocal  to be
\bl
W^{(0)}_a =\omega_a ,
 ~~~~~~ \frac{1}{W^{(0)}_a} =\frac{1}{\omega_a} \, .
\el
Then, we obtain the 2nd-order
\bl
W^{(2)}_a & = \omega_a
 + \frac{\alpha_a}{2 \omega_a }
  -\frac{m_a^2 a'^2}{4 \omega_a ^3}
    -\frac{m_a^2 a a''}{4 \omega_a ^3}
    +\frac{5 m_a^4 a^2 a'^2}{8 \omega_a ^5},
\el
\bl
\frac{1}{W^{(2)}_a} & = \frac{1}{\omega_a } -\frac{\alpha_a}{2 \omega_a ^3}
    +\frac{m_a^2 a'^2}{4 \omega_a ^5}
    +\frac{m_a^2 a a''}{4 \omega_a ^5}
    -\frac{5 m_a^4 a^2 a'^2}{8 \omega_a ^7}     ,
\el
and the 4th-order
\bl
W^{(4)}_a & = \Big[  \omega_a + \frac{\alpha_a}{2 \omega_a }
  -\frac{m_a^2 a'^2}{4 \omega_a ^3}
  +\frac{5 m_a^4 a^2 a'^2}{8 \omega_a ^5}
  -\frac{m_a^2 a a''}{4 \omega_a ^3}   \Big]
  \nn \\
&  -\frac{\alpha_a '' }{8 \omega_a ^3}
-\frac{\alpha_a^2}{8 \omega_a ^3}
+\frac{3 m_a^2 \alpha_a a'^2}{8 \omega_a ^5}
-\frac{25 m_a^4 a^2 \alpha_a a'^2}{16 \omega_a ^7}
-\frac{19 m_a^4 a'^4}{32 \omega_a ^7}
+\frac{221 m_a^6 a^2 a'^4}{32 \omega_a ^9}\nn
\\
&-\frac{1105 m_a^8 a^4 a'^4}{128 \omega_a ^{11}}
+\frac{5 m_a^2 a a' \alpha_a '}{8 \omega_a ^5}
+\frac{3 m_a^2 a \alpha_a  a''}{8 \omega_a ^5}
-\frac{61 m_a^4 a a'^2 a''}{16 \omega_a ^7}
+\frac{221 m_a^6 a^3 a'^2 a''}{32 \omega_a ^9}
\nn
\\
&+\frac{3 m_a^2 a''^2}{16 \omega_a ^5}
-\frac{19 m_a^4 a^2 a''^2}{32 \omega_a ^7}
+\frac{m_a^2 a' a^{(3)}}{4 \omega_a ^5}
-\frac{7 m_a^4 a^2 a' a'''}{8 \omega_a ^7}
+\frac{m_a^2 a a''''}{16 \omega_a ^5},
\el
\bl
\frac{1}{W^{(4)}_a} & =\Big[  \frac{1}{\omega_a }
    -\frac{\alpha_a}{2 \omega_a ^3}
    +\frac{m_a^2 a'^2}{4 \omega_a ^5}
    -\frac{5 m_a^4 a^2 a'^2}{8 \omega_a ^7}
    +\frac{m_a^2 a a''}{4 \omega_a ^5}      \Big]
       \nn \\
&   -\frac{5 m_a^2\alpha_a a'^2}{8 \omega_a ^7}
   + \frac{3 \alpha_a ^2}{8 \omega_a ^5}
    +\frac{35 m_a^4 a^2 \alpha_a a'^2}{16 \omega_a ^9}
    +\frac{21 m_a^4 a'^4}{32 \omega_a ^9}
    -\frac{231 m_a^6 a^2 a'^4}{32 \omega_a ^{11}}\nn
\\
&+\frac{1155 m_a^8 a^4 a'^4}{128 \omega_a ^{13}}
   -\frac{5m_a^2 a a' \alpha_a '}{8 \omega_a ^7}
   -\frac{5 m_a^2 a \alpha_a  a''}{8 \omega_a ^7}
    +\frac{63 m^4 a a'^2 a''}{16 \omega_a ^9}
    -\frac{231 m^6 a^3 a'^2 a''}{32 \omega_a ^{11}}
   \nn
\\
& -\frac{3 m^2 a''^2}{16 \omega_a ^7}
   +\frac{21 m_a^4 a^2 a''^2}{32 \omega_a ^9}
   +\frac{\alpha_a ''}{8 \omega_a ^5}
   -\frac{m_a^2 a' a'''}{4 \omega_a ^7}
   +\frac{7 m_a^4 a^2 a' a'''}{8 \omega_a ^9}
   -\frac{m_a^2 a a''''}{16 \omega_a ^7}.
\el

{\bf 1) The adiabatic subtraction terms
of the transverse stress tensor }

Substituting the transverse WKB solution
into the transverse spectral stress tensor \eqref{spTRPRho} \eqref{spTRPRess}
to replace the mode $f^{(1)}_k$ gives
\bl
\rho_k^{TR}
&=\frac{k^3}{2\pi^2a^4} \Big(\frac{ W_{T}}{2}
 +\frac{\omega^2}{2W_{T}}+\frac18\frac{W_{T}'^2}{W_{T}^3}
         \Big),\label{rhorhoTMV}
\\
p^{TR}_k
&=\frac13\frac{k^3}{2\pi^2a^4} \Big(  \frac{ W_{T}}{2}
 + \frac{(k^2  -m^2a^2) }{2W_T} +\frac{W_{T}'^2}{8 W_{T}^3}
      \Big) .   \label{ppTMV}
\el
Plugging  $W_T$ of $n$th-order into the above
will give the $n$-th order subtraction terms of transverse stress tensor.
The 0th-order adiabatic subtraction terms
of the transverse spectral stress tensor are
\bl
\rho_{k\, A0}^{TR}
&=\frac{k^3}{2\pi^2a^4}  \omega , \label{0thtrrho}
\\
p_{k\, A0}^{TR}
&=\frac{k^3}{2\pi^2a^4}\frac{1}{3} \Big( \omega-\frac{m^2a^2}{\omega}\Big) ,
   \label{pTMV0th}
\el
which are sufficient to remove all the UV divergences
in the transverse stress tensor.
The 2nd-order subtraction terms of the transverse spectral stress tensor are
\bl
\rho_{k\, A2}^{TR}
&=\frac{k^3}{2\pi^2a^4}
\Big(\omega +  \frac{m^4 a^2 a'^2}{8 \omega^5} \Big)   ,
\label{2thtrrho}
\\
p_{k\, A2}^{TR}
&=\frac{k^3}{2\pi^2a^4}
  \frac13\Big(\omega   -\frac{m^2 a^2}{\omega}
-\frac{m^4 a^3 a''}{4 \omega ^5}
-\frac{m^4 a^2 a'^2}{8 \omega^5}
+\frac{5 m^6 a^4 a'^2}{8 \omega^7} \Big) .
\label{pTMV2nd}
\el
The 4th-order subtraction terms  are
\bl
\rho_{k\, A4}^{TR}
&=\frac{k^3}{2\pi^2a^4}
   \Big(
\omega
+\frac{ m^4 a^2 a'^2}{8 \omega^5}
+\frac{m^4 a'^4}{32 \omega ^7}
-\frac{m^4 a a'^2 a''}{8 \omega ^7}
+\frac{m^4 a^2 a''^2}{32 \omega ^7}
-\frac{m^4 a^2 a' a'''}{16 \omega^7}
   \nn  \\
& ~~~
+\frac{7 m^6 a^2 a'^4}{16 \omega ^9}
+\frac{7 m^6 a^3 a'^2 a''}{16 \omega ^9}
-\frac{105 m^8 a^4 a'^4}{128 \omega ^{11}}
 \Big) ,
\label{tr4thadrho}
\\
p_{k\, A 4}^{TR}
&=\frac{k^3}{2\pi^2a^4} \frac13\Big(\omega-\frac{m^2 a^2}{\omega }
-\frac{m^4 a^3 a''}{4 \omega ^5}
-\frac{m^4 a^2 a'^2}{8 \omega ^5}
+\frac{5 m^6 a^4 a'^2}{8 \omega ^7}
\nn
\\
& ~~~ +\frac{m^4 a^3 a^{(4)}}{16 \omega ^7}
+\frac{m^4 a'^4}{32 \omega ^7}
+\frac{7 m^4 a^2 a''^2}{32 \omega ^7}
+\frac{3 m^4 a^2 a^{(3)} a'}{16 \omega ^7}
-\frac{m^4 a a'^2 a''}{8 \omega ^7}
\nn
\\
& ~~~ -\frac{21 m^6 a^4 a''^2}{32 \omega ^9}
-\frac{7 m^6 a^2 a'^4}{32 \omega ^9}
-\frac{7 m^6 a^3 a'^2 a''}{2 \omega ^9}
 -\frac{7 m^6 a^4 a^{(3)} a'}{8 \omega ^9}
\nn
 \\
& ~~~ +\frac{231 m^8 a^5 a'^2 a''}{32 \omega ^{11}}
+\frac{819 m^8 a^4 a'^4}{128 \omega ^{11}}
-\frac{1155 m^{10} a^6 a'^4}{128 \omega ^{13}}
\Big)  , \label{pTMV4th}
\el
Notice that
extra terms, $1/\omega^{5}$, $1/\omega^{7}$, ..., $1/\omega^{13}$
in $\rho_{k\, A2}^{TR}$,  $p_{k\, A2}^{TR}$,
$\rho_{k\, A4}^{TR}$,  $\rho_{k\, A4}^{TR}$ are convergent,
 and need not to subtract,
according to the minimum subtraction rule.
In the massless limit $m=0$, the transverse subtraction terms are
\bl
\rho_{k\, A0}^{TR} &  =3  p_{k\, A0}^{TR}  =\frac{k^4}{2\pi^2a^4},
\label{ms0th}
\\
\rho_{k\, A2}^{TR} &  =3  p_{k\, A2}^{TR}  =\frac{k^4}{2\pi^2a^4},
\\
\rho_{k\, A4}^{TR} &  =3  p_{k\, A4}^{TR}  =\frac{k^4}{2\pi^2a^4},
\el
 all the higher order terms being equal to the 0th-order one.
This property is similar to that of
the conformally-coupling scalar field in a general fRW spacetime
 \cite{ZhangWangYe2020}.

In the improper 4th-order adiabatic regularization,
the  trace anomaly
is contributed by the 4th-order subtraction terms above the 2nd-order,
$\rho_{k\, trace}^{TR} \equiv \rho_{k\, A4}^{TR}-\rho_{k\, A2}^{TR}$
 and $p_{k\, trace}^{TR} \equiv p_{k\, A4}^{TR}-p_{k\, A2}^{TR}$.
 (For instance, $\frac{m^4 a^2 a'^2}{8 \omega^5}$ in \eqref{2thtrrho}
 and $-\frac{m^4 a^3 a''}{4 \omega ^5} -\frac{m^4 a^2 a'^2}{8 \omega^5}
+\frac{5 m^6 a^4 a'^2}{8 \omega^7}$ in \eqref{pTMV2nd}
do not contribute to the trace anomaly at $m=0$.)
First analytically performing $k$-integration
and then taking the massless limit,
one would  come up with the trace of the transverse subtraction terms
\bl
& \lim_{m\rightarrow 0}
  \int_0^\infty
  \Big(- \rho_{k\, trace }^{TR} +3 p_{k\, trace }^{TR} \Big) \frac{dk }{k}
\label{inttran4rh}
\\
&  = \frac{a^{(4)}}{240 \pi ^2 a^5}
-\frac{a^{(3)} a'}{60 \pi ^2 a^6}
-\frac{a''^2}{80 \pi ^2 a^6}
+\frac{a'^2 a''}{30 \pi ^2 a^7}
-\frac{a'^4}{120 \pi ^2 a^8}   ,
\label{iontran4rh}
\\
&
=\frac{1}{2880\pi^2}\Big(-2(R_{\mu\nu}R^{\mu\nu}-\frac{1}{3}R^2)+2 \square R\Big),
\label{cortravet1}
\\
&   = \frac{H^4}{120\pi^2}  ~~~  (\text{for the de Sitter space}),
\label{desttiertrs}
\el
where
\bl
R_{\mu\nu}R^{\mu\nu} &=\frac{12 a''^2}{a^6}+\frac{12 a'^4}{a^8}
-\frac{12 a'^2 a''}{a^7},\label{wchu1}
\\
R^2 &=\frac{36 a''^2}{a^6},\label{wchu2}
\\
\square R &= -\Big( \frac{6 a^{(4)}}{a^5}-\frac{18 a''^2}{a^6}
   -\frac{24 a^{(3)} a'}{a^6}+\frac{36 a'^2 a''}{a^7} \Big) .
   \label{wchu3}
\el
We remark that if the massless limit is taken first,
$ \lim_{m\rightarrow 0}\rho_{k\, trace}^{TR} =0$,
$ \lim_{m\rightarrow 0} p_{k\, trace}^{TR} =0$,
 the $k$-integration \eqref{inttran4rh} will give a zero trace.

{\bf 2) The adiabatic subtraction terms of the longitudinal stress tensor}

In analogy to  the transverse part,
 substituting   the longitudinal WKB solution
into the longitudinal spectral stress tensor
\eqref{rholongitudinal} \eqref{Lonpr} gives
\bl
\rho_{k}^{L}
&=\frac{k^3}{4 \pi^2a^4}
\Big(   \frac{\omega^2}{2W_L}
+\frac{W_L}{2}+\frac{a'^2}{2 a^2 W_L}
-\frac{a' W_L'}{2 a W_L^2}+\frac{W_L'^2}{8 W_L^3}\Big) ,
\label{rhorhoLTMV}
\\
p^{L}_k
&=\frac{k^3}{4 \pi^2a^4}
\Big(\frac13\frac{m^2a^2}{2W_L}
+\frac{k^2}{2 W_L}
-\frac13(-\frac{a' W_L'}{2 a W_L^2}
+\frac{a'^2}{2 a^2 W_L}+\frac{W_L'^2}{8 W_L^3}
+\frac{W_L}{2})\Big) . \label{ppLTMV}
\el
The 0th-order subtraction terms of the longitudinal spectral stress tensor is
\bl
\rho^{L}_{k\, A 0}
&=\frac{k^3}{2\pi^2a^4}\Big(\frac{\omega }{2}\Big) ,
\label{0thrhot}
\\
p^{L}_{k\, A 0}
&=\frac{k^3}{2\pi^2a^4} \Big(\frac{\omega }{6}-\frac{m^2 a^2}{6 \omega}\Big),
\label{0thprt}
\el
The 2nd-order subtraction terms  are
\bl
\rho^{L}_{k\, A 2}
=& \frac{k^3}{2\pi^2a^4}\Big(\frac{\omega}{2}
+\frac{a'^2}{4 a^2 \omega }-\frac{m^2 a'^2}{4 \omega ^3}
+\frac{m^4 a^2 a'^2}{16 \omega ^5}\Big),
\label{2ndrhoL}
\\
p^{L}_{k\, A 2}
=& \frac{k^3}{2\pi^2a^4} \Big(\frac{\omega }{6}
-\frac{m^2 a^2}{6 \omega}+(\frac{a'^2}{4 a^2 }
-\frac{a''}{6 a })\frac{1}{ \omega}+\frac{m^2 a a''}{6 \omega ^3}
\nn \\
& -(\frac{m^4 a^3 a''}{24}+\frac{13 m^4 a^2 a'^2}{48})\frac{1}{\omega^5}
+\frac{5 m^6 a^4 a'^2}{48 \omega ^7}\Big) ,
\label{2ndadpL}
\el
which are sufficient to remove UV divergences in the longitudinal stress tensor,
and are actually used in our regularization in Sect 4.
The 4th-order subtraction terms are
\bl
\rho^{L}_{k\, A 4}
= & \frac{k^3}{2\pi^2a^4}\Big(\frac{\omega}{2}
+\frac{a'^2}{4 a^2 \omega }-\frac{m^2 a'^2}{4 \omega ^3}
+\frac{m^4 a^2 a'^2}{16 \omega ^5}
\nn
\\
&+(\frac{3 a'^2 a'' m^2}{8 a }+\frac{a' a^{(3)} m^2}{8 }
-\frac{a''^2 m^2}{16 }-\frac{5 a'^4 m^2}{16 a^2 })\frac{1}{\omega ^5}
\nn \\
& +(\frac{a^2 a''^2 m^4}{64 }-\frac{11 a a'^2 a'' m^4}{16 }
-\frac{a^2 a' a^{(3)} m^4}{32 }-\frac{49 a'^4 m^4}{64 })\frac{1}{\omega ^7}
\nn
\\
&+\frac{21 a^2 a'^4 m^6}{16 \omega ^9}
+\frac{7 a^3 a'^2 a'' m^6}{32 \omega ^9}
-\frac{105 a^4 a'^4 m^8}{256 \omega ^{11}}
\nn
\\
& + (\frac{a''^2}{16 a^2 } +\frac{a'^2 a''}{4 a^3}
  -\frac{a' a^{(3)}}{8 a^2} ) \frac{1}{\omega ^3}
  \Big)  ,
\label{4thadrhoL}
\\
p^{L}_{k\, A 4}
=& \frac{k^3}{2\pi^2a^4} \Big(\frac{\omega }{6}
-\frac{m^2 a^2}{6 \omega}+(\frac{a'^2}{4 a^2}
-\frac{a''}{6 a })\frac{1}{\omega }
+\frac{m^2 a a''}{6 \omega ^3}
\nn
\\
&   -( \frac{m^4 a^3 a''}{24}+\frac{13 m^4 a^2 a'^2}{48 } )\frac{1}{\omega ^5}
  +\frac{5 m^6 a^4 a'^2}{48 \omega ^7}
\nn \\
&   +(\frac{11 a'^2 a'' m^2}{12 a }
-\frac{5 a''^2 m^2}{24 }-\frac{5 a' a^{(3)} m^2}{24 }
-\frac{a a^{(4)} m^2}{24 }
-\frac{5 a'^4 m^2}{16 a^2 })\frac{1}{\omega ^5}
\nn
\\
& +(\frac{67 a^2 a''^2 m^4}{192 }
+\frac{79 a a'^2 a'' m^4}{48 }
+\frac{43 a^2 a' a^{(3)} m^4}{96 }
+\frac{a^3 a^{(4)} m^4}{96 }-\frac{149 a'^4 m^4}{192 })\frac{1}{\omega ^7}
\nn
\\
& -(\frac{7 a^3 a'^2 a'' m^6}{2}+\frac{7 a^4 a' a^{(3)} m^6}{48 }
+\frac{7 a^4 a''^2 m^6}{64 }
+\frac{427 a^2 a'^4 m^6}{192 })\frac{1}{\omega ^{9}}\nn
\\
& +(\frac{1113 a^4 a'^4 m^8}{256 }
+\frac{77 a^5 a'^2 a'' m^8}{64 })\frac{1}{\omega ^{11}}
-\frac{385 a^6 a'^4 m^{10}}{256 \omega ^{13}}
\nn
\\
&  + (\frac{a'^2 a''}{3 a^3 }
+\frac{a^{(4)}}{24 a }
-\frac{5 a' a^{(3)}}{24 a^2 }
-\frac{5 a''^2}{48 a^2})\frac{1}{\omega ^3}
\Big).
\label{1088}
\el
The last lines in \eqref{4thadrhoL} and \eqref{1088}
 are respectively the following
\be
\frac{k^3}{2\pi^2a^4}  (\frac{a''^2}{16 a^2 } -\frac{a' a^{(3)}}{8 a^2}
 +\frac{a'^2 a''}{4 a^3}
   ) \frac{1}{\omega ^3} ,
  \label{rhoom3}
\ee
\be
\frac{k^3}{2\pi^2a^4}  (\frac{a'^2 a''}{3 a^3 }
+\frac{a^{(4)}}{24 a }
-\frac{5 a' a^{(3)}}{24 a^2 }
-\frac{5 a''^2}{48 a^2})\frac{1}{\omega ^3} ,
\label{pom3}
\ee
which give rise to new UV divergences ($\ln k$ as $k\rightarrow \infty$)
 and IR divergences ($\ln k$ as $k\rightarrow 0$) at $m=0$
 in the integrations $\int \rho^{L} _{k \,  A4} \frac{dk }{k}$
and $\int p^{L} _{k \,  A4} \frac{dk }{k}$,
as noticed in Ref.\cite{ChimentoCossarini1990}.
These terms happen to be vanishing in de Sitter space \cite{Salas2023}
and in the RD stage \cite{YeZhang2024},
but are generally nonvanishing in fRW spacetimes.
For instance, in the MD stage  \cite{YeZhang2024},
 at $m=0$  the Stueckelberg field will reduce to the Maxwell field,
and correspondingly the IR divergences of \eqref{rhoom3} and \eqref{pom3}
just remove the IR divergences of
the unregularized Maxwell stress tensor   \cite{YeZhang2024}.

In the massless limit $m=0$,
\eqref{0thrhot}$\sim$\eqref{1088}   become
\bl
& \rho_{k\, A0} ^{L}   =3 p_{k\, A0} ^{L} = \frac{k^4}{ 4\pi^2a^4},
  \label{m0longsub}
\\
& \rho_{k\, A2} ^{L} = \rho_{k\, A4} ^{L} = ,,,=
    \frac{k^3}{2\pi^2a^4}\frac{k}{2}\Big(1+\frac{1}{2k^2\tau^2}\Big) ,
     \label{1200rho}
    \\
& p_{k\, A2} ^{L} = p_{k\, A4} ^{L}  =,,, =
  \frac{k^3}{2\pi^2a^4}\frac13\frac{k}{2}\Big(1-\frac{1}{2k^2\tau^2}\Big) .
     \label{1200}
\el
All higher order subtraction terms
are equal to the 2nd-order ones,
similar to those of a massless   minimally-coupling scalar field
 in a general fRW spacetime \cite{ZhangWangYe2020}.

Analogously to the transverse part,
in the improper 4th-order   regularization,
one would come up with the trace of the longitudinal subtraction terms
in the massless limit
\bl
& \lim_{m\rightarrow 0}\int_{0}^{\infty}
 \Big( -\rho^{L}_{k\, trace }+3p^{L}_{k\, trace} \Big)  \frac{dk}{k}
\label{intlongit4rh}
 \\
& =  -\frac{3 a^{(4)}}{160 \pi ^2 a^5}-\frac{3 a''^2}{80 \pi ^2 a^6}
-\frac{61 a'^4}{240 \pi ^2 a^8}-\frac{a^{(3)} a'}{20 \pi ^2 a^6}
+\frac{109 a'^2 a''}{240 \pi ^2 a^7}
\nn
\\
&  + \Big(\frac{a^{(4)}}{8 a^5 } -\frac{3 a''^2}{8 a^6}
   -\frac{a^{(3)} a'}{2 a^6}+\frac{3 a'^2 a''}{4 a^7} \Big)
    \int\frac{k^3}{2\pi^2}\frac{1}{\omega^3}\frac{d k}{k} ,
\label{longitrace}
\\
&   = -\frac{59 H^4}{240 \pi^2} ~~~ \text{(for the de  Sitter  space)}
\label{longesi}
\el
where
$\rho_{k\, trace}^{L} \equiv \rho_{k\, A4}^{L}-\rho_{k\, A2}^{L}$
 and $p_{k\, trace}^{L} \equiv p_{k\, A4}^{L}-p_{k\, A2}^{L}$.
In \eqref{intlongit4rh},
integration is performed before taking the massless limit.

{\bf 3) The adiabatic subtraction terms of the temporal stress tensor}

Substituting the temporal WKB solution into
  \eqref{rhotemporal} \eqref{Temppr} gives
\bl
\rho_{k}^{T}
&=\frac{k^3}{4\pi^2a^4}
\Big( -\frac{W_0}{2}-\frac{a'^2}{2 a^2 W_0}-\frac{a' W_0'}{2 a W_0^2}
      -\frac{W_0'^2}{8 W_0^3}-\frac{k^2}{2W_0} \Big) ,
        \label{rhorhoLTMV}
\\
p^{T}_k
&=\frac{k^3}{4\pi^2a^4}
 \Big( - (\frac{a'W_0'}{2 a W_0^2} +\frac{a'^2}{2 a^2 W_0}
    +\frac{W_0'^2}{8 W_0^3}+\frac{W_0}{2})
    +\frac13\frac{k^2}{2W_0} \Big) ,
         \label{ppLTMV}
\el
 depending on the GF parameter $\zeta$.
The 0th-order subtraction terms of the temporal spectral stress tensor are
\bl
\rho^{T}_{k\, A 0}
&=\frac{k^3}{2\pi^2a^4}\Big(-\frac{\omega _0}{2}
+\frac{\zeta  m^2 a^2}{4 \omega _0}\Big) ,
\label{722}
\\
p^{T}_{k\, A 0}
&=\frac{k^3}{2\pi^2a^4} \Big(-\frac{\omega _0}{6}
-\frac{\zeta  m^2 a^2}{12 \omega _0}\Big) .
\label{0pTad}
\el
The 2nd-order subtraction terms are
\bl
\rho^{T}_{k\, A 2}
&=\frac{k^3}{2\pi^2a^4}\Big( -\frac{\omega_0}{2}+(\frac{a^2\zeta  m^2}{4}
-\frac{a'^2}{4 a^2})\frac{1}{ \omega _0}
+(\frac{a \zeta m^2 a''}{8}-\frac{\zeta m^2 a'^2}{4})\frac{1}{ \omega _0{}^3}
\nn \\
& ~~~  +\frac{a^3 \zeta^2m^4 a''}{16 \omega _0{}^5}
-\frac{5 a^4 \zeta^3m^6 a'^2}{32 \omega _0{}^7}\Big) ,
\label{2ndtemprho}
\\
p^{T}_{k\, A 2}
&=\frac{k^3}{2\pi^2a^4} \Big(-\frac{\omega _0}{6}
-\frac{\zeta  m^2 a^2}{12 \omega _0}
-(\frac{a'^2}{4 a^2}-\frac{a''}{6 a })\frac{1}{\omega _0}
+(\frac{\zeta  m^2 a a''}{24 }-\frac{\zeta  m^2 a'^2}{6 })\frac{1}{\omega _0^3}
\nn
\\
&~~~ -(\frac{\zeta ^2 m^4 a^3 a''}{48 }
+\frac{7 \zeta ^2 m^4 a^2 a'^2}{24 })\frac{1}{\omega _0{}^5}
+\frac{5 \zeta ^3 m^6 a^4 a'^2}{96 \omega _0{}^7} \Big) ,
\label{2ndtemppr}
\el
which are sufficient to remove all UV divergences
of the  temporal stress tensor,
and are used in our regularization in Sect 4.
The 4th-order subtraction terms are
\bl
\rho^{T}_{k\, A 4}
= & \frac{k^3}{2\pi^2a^4}\Big(-\frac{\omega_0}{2}
+( \frac{a^2\zeta  m^2}{4 }-\frac{a'^2}{4 a^2 })\frac{1}{\omega _0}
  + (\frac{a \zeta m^2 a''}{8 }-\frac{\zeta m^2 a'^2}{4})\frac{1}{\omega _0{}^3}
\nn  \\
&     +\frac{a^3 \zeta^2m^4 a''}{16 \omega _0{}^5}
      -\frac{5 a^4 \zeta^3m^6 a'^2}{32 \omega _0{}^7}
\nn
\\
&  +(\frac{\zeta  a''^2 m^2}{16}
+\frac{3 \zeta  a' a^{(3)} m^2}{16 }
-\frac{\zeta  a a^{(4)} m^2}{32 }
-\frac{7 \zeta  a'^2 a'' m^2}{16 a}
-\frac{\zeta  a'^4 m^2}{16 a^2})\frac{1}{\omega _0{}^5}
\nn
\\
&  +(\frac{3 \zeta ^2 a^2 a''^2 m^4}{32}
+\frac{\zeta ^2 a^2 a' a^{(3)} m^4}{8 }
-\frac{9 \zeta ^2 a a'^2 a'' m^4}{16 }
-\frac{31 \zeta ^2 a'^4 m^4}{64 }
-\frac{\zeta ^2 a^3 a^{(4)} m^4}{64 })\frac{1}{ \omega _0{}^7}\nn
\\
& +(\frac{133 \zeta ^3 a^2 a'^4 m^6}{128 }
+\frac{21 \zeta ^3 a^4 a''^2 m^6}{128 }
+\frac{7 \zeta ^3 a^3 a'^2 a'' m^6}{32 }
+\frac{7 \zeta ^3 a^4 a' a^{(3)} m^6}{32 })\frac{1}{\omega _0{}^9}
\nn
\\
&  -(\frac{231 \zeta ^4 a^5 a'^2 a'' m^8}{128 }
+\frac{357 \zeta ^4 a^4 a'^4 m^8}{256})\frac{1}{\omega _0{}^{11}}
+\frac{1155 \zeta ^5 a^6 a'^4 m^{10}}{512 \omega _0{}^{13}}
\nn \\
& - (\frac{a''^2}{16  a^2 } +\frac{a'^2 a''}{4  a^3}
  -\frac{a' a^{(3)}}{8  a^2} ) \frac{1}{\omega_0^3}
 \Big)  ,
\label{4rhotemad}
\\
p^{T}_{k\, A 4}
= &\frac{k^3}{2\pi^2a^4} \Big(-\frac{\omega _0}{6}
-\frac{\zeta  m^2 a^2}{12 \omega_0}
+(\frac{a''}{6 a }-\frac{a'^2}{4 a^2})\frac{1}{\omega _0}
+(\frac{\zeta  m^2 a a''}{24}
-\frac{\zeta  m^2 a'^2}{6 })\frac{1}{\omega _0{}^3}
\nn \\
& -(\frac{\zeta ^2 m^4 a^3 a''}{48 }
+\frac{7 \zeta ^2 m^4 a^2 a'^2}{24 })\frac{1}{\omega _0{}^5}
+\frac{5 \zeta ^3 m^6 a^4 a'^2}{96 \omega _0{}^7}
\nn
\\
&  -(\frac{m^2 \zeta  a'^4}{16 a^2}
+\frac{17 m^2 \zeta  a'^2 a''}{48 a}
-\frac{m^2 \zeta  a''^2}{24}-\frac{11 m^2 \zeta  a' a^{(3)}}{48}
+\frac{m^2 \zeta  a a^{(4)}}{96})\frac{1}{\omega _0{}^5}\nn
\\
&  +(\frac{\zeta ^2 a^2 a''^2 m^4}{6 }
+\frac{\zeta ^2 a a'^2 a'' m^4}{48 }+\frac{7 \zeta ^2 a^2 a' a^{(3)} m^4}{24 }
+\frac{\zeta ^2 a^3 a^{(4)} m^4}{192 }
-\frac{17 \zeta ^2 a'^4 m^4}{64 })\frac{1}{\omega _0{}^7}\nn
\\
&  -(\frac{133 \zeta ^3 a^3 a'^2 a'' m^6}{48 }
+\frac{7 \zeta ^3 a^4 a' a^{(3)} m^6}{96 }
+\frac{203 \zeta ^3 a^2 a'^4 m^6}{128 }
+\frac{7 \zeta ^3 a^4 a''^2 m^6}{128 })\frac{1}{\omega _0{}^9}\nn
\\
& +\frac{1029 \zeta ^4 a^4 a'^4 m^8}{256 \omega _0{}^{11}}
+\frac{77 \zeta ^4 a^5 a'^2 a'' m^8}{128 \omega _0{}^{11}}
-\frac{385 \zeta ^5 a^6 a'^4 m^{10}}{512\omega_0{}^{13}}
\nn \\
& + (\frac{5 a''^2}{48 a^2 }
+\frac{5 a' a^{(3)}}{24 a^2}
-\frac{a^{(4)}}{24 a}
-\frac{a'^2 a''}{3 a^3})\frac{1}{\omega _0{}^3}
\Big) .
\label{8222}
\el
The last lines of \eqref{4rhotemad} and \eqref{8222}
are respectively  the following
\bl
-  \frac{k^3}{2\pi^2a^4}
(\frac{a''^2}{16  a^2 }  -\frac{a' a^{(3)}}{8  a^2}
      +\frac{a'^2 a''}{4  a^3}  ) \frac{1}{\omega_0^3} ,
  \label{rhoTom}
\\
-  \frac{k^3}{2\pi^2a^4}  (\frac{a'^2 a''}{3 a^3 }
+\frac{a^{(4)}}{24 a }
-\frac{5 a' a^{(3)}}{24 a^2 }
-\frac{5 a''^2}{48 a^2})\frac{1}{\omega ^3} ,
  \label{preTemom}
\el
which are vanishing for the de Sitter space,
like \eqref{rhoom3} \eqref{pom3},
and the remarks below \eqref{pom3} also apply here.
It is noticed that,
in the Landau gauge $\zeta=0$
the terms involving $\zeta$
in $\rho^{T}_{k\, A 4}$ and  $p^{T}_{k\, A 4}$ are vanishing
in the de Sitter space,
so that  $\rho^{T}_{k\, A 4} = \rho^{T}_{k\, A 2} $ and
 $p^{T}_{k\, A 4}= p^{T}_{k\, A 2}$,
consequently the 4th-order regularization
is effectively the same as the 2nd-order regularization
for the temporal stress tensor.

In the massless limit $m=0$,
 \eqref{722}$\sim$\eqref{8222} become
\bl
\rho^{T}_{k\, A 0} & =3  p^{T}_{k\, A 0} =\frac{k^3}{2\pi^2a^4}(-\frac{k}{2}),
\\
\rho^{T}_{k\, A 2}   & = \rho^{T}_{k\, A 4}  =,,,
        = \frac{k^3}{2\pi^2a^4}(-\frac{k}{2})
\Big(1+\frac{1}{2k^2\tau^2}\Big),
       \label{rhotem}
\\
p^{T}_{k\, A 2}   & = p^{T}_{k\, A 4}  =,,,
=\frac{k^3}{2\pi^2a^4}\frac13(-\frac{k}{2})\Big(1-\frac{1}{2k^2\tau^2}\Big),
\label{ptem}
\el
which have  opposite signs to
the longitudinal \eqref{m0longsub}---\eqref{1200}.

Analogously to the transverse and longitudinal parts,
in the improper 4th-order   regularization,
one would come up with the trace of the temporal subtraction terms
(for $\zeta\ne 0$ only)
\bl
& \lim_{m\rightarrow 0}\int_{0}^{\infty}
\Big(-\rho^{T}_{k\, trace} +3 p^{T}_{k\, trace} \Big)\frac{dk}{k}
\label{tempodesu}
\\
& =\frac{a^{(4)}}{480 \pi ^2 a^5}+\frac{a''^2}{40 \pi ^2 a^6}
-\frac{a'^4}{240 \pi ^2 a^8}+\frac{7 a^{(3)} a'}{60 \pi ^2 a^6}
-\frac{71 a'^2 a''}{240 \pi ^2 a^7}
\nn
\\
&   -\Big(\frac{a^{(4)}}{8 a^5 }-\frac{3 a''^2}{8 a^6 }-\frac{a^{(3)} a'}{2 a^6 }+
\frac{3 a'^2 a''}{4 a^7 }\Big)
   \int\frac{k^3}{2\pi^2}\frac{1}{\omega_0^3}\frac{d k}{k},
\label{desu}
\\
&  =  \frac{61H^4}{240\pi^2} ~~~ \text{(for the de  Sitter  space)} .
\label{desitsu}
\el
In \eqref{tempodesu},
integration is performed before taking the massless limit.

{\bf 4) The adiabatic subtraction terms of the GF stress tensor}

Substituting  the temporal WKB solution
into \eqref{gfrh} \eqref{sptrpressGFVEV} gives
\bl
\rho_k^{GF}
&=  \frac{k^3}{2\pi^2 a^4}  \Big(  (\frac{W_0}{2}
+\frac{a'^2}{2 a^2 W_0}+\frac{a' W'_0}{2 a W_0^2}+\frac{W^{'2}_0}{8 W_0^3})
      + \frac{(\omega_0^2-\zeta  m^2a^2)}{2W_0}
      +   \frac{\zeta     m^2 a^2}{4W_0}   \Big),\label{rhorhogf}
\\
p^{GF}
&=\frac{k^3}{2\pi^2a^4}\Big(-\frac13\frac{\omega_0^2-\zeta m^2a^2}{2W_0}
+(\frac{a'W_0'}{2 a W_0^2}+\frac{a'^2}{2 a^2 W_0}
+\frac{W_0'^2}{8 W_0^3}+\frac{W_0}{2})
-\frac12\frac{\zeta m^2a^2}{2W_0}\Big) ,
\label{ppgf}
\el
 depending on  $\zeta$.
The 0th order subtraction terms for
the GF spectral  stress tensor  are
\bl
\rho^{GF}_{k\, A0}
&=\frac{k^3}{2\pi^2a^4}\Big(\omega _0-\frac{\zeta  m^2 a^2}{4 \omega _0}\Big),
\label{822}
\\
p^{GF}_{k\, A0}
&=\frac{k^3}{2\pi^2a^4}\Big(\frac{\omega _0}{3}
-\frac{\zeta  m^2 a^2}{12 \omega _0}\Big) . \label{855}
\el
The 2nd-order  adiabatic subtraction terms for
the GF stress tensor are
\bl
\rho^{GF}_{k\, A2}
 = & \frac{k^3}{2\pi^2a^4}
\Big(   \omega _0  -\frac{\zeta  m^2 a^2}{4 \omega _0}
+\frac{a'^2}{2 a^2 \omega _0}
-\frac{\zeta  m^2 a a''}{8 \omega _0{}^3}
+\frac{\zeta  m^2 a'^2}{2 \omega _0{}^3}
\nn
\\
&-\frac{\zeta ^2 m^4 a^3 a''}{16 \omega _0{}^5}
+\frac{\zeta ^2 m^4 a^2 a'^2}{16 \omega _0{}^5}
+\frac{5 \zeta ^3 m^6 a^4 a'^2}{32 \omega _0{}^7}  \Big),
       \label{GF2rhoad}
\\
p^{GF}_{k\, A 2}
=& \frac{k^3}{2\pi^2a^4}\Big( \frac{\omega _0}{3}
-\frac{\zeta  m^2 a^2}{12 \omega _0}
-\frac{a''}{3 a \omega _0}
+\frac{a'^2}{2 a^2 \omega _0}
-\frac{5 \zeta  m^2 a a''}{24 \omega _0{}^3}
+\frac{\zeta  m^2 a'^2}{3 \omega _0{}^3}
\nn \\
& -\frac{\zeta ^2 m^4 a^3 a''}{48 \omega _0{}^5}
+\frac{25 \zeta ^2 m^4 a^2 a'^2}{48 \omega _0{}^5}
+\frac{5 \zeta ^3 m^6 a^4 a'^2}{96 \omega _0{}^7}
\Big) ,
\label{866}
\el
which are sufficient to remove all UV divergences
of the GF stress tensor.
The 4th order subtraction terms for  the GF stress tensor are
{\allowdisplaybreaks
\bl
\rho^{GF}_{k\, A4}
&=\frac{k^3}{2\pi^2a^4}\Big( \omega _0-\frac{\zeta  m^2 a^2}{4 \omega _0}
+\frac{a'^2}{2 a^2 \omega _0}
-(\frac{\zeta  m^2 a a''}{8}
-\frac{\zeta  m^2 a'^2}{2 })\frac{1}{\omega _0{}^3}
\nn \\
&~~~~~~~ -(\frac{\zeta ^2 m^4 a^3 a''}{16 }
-\frac{\zeta ^2 m^4 a^2 a'^2}{16})\frac{1}{\omega _0{}^5}
+\frac{5 \zeta ^3 m^6 a^4 a'^2}{32 \omega _0{}^7}
\nn
\\
&~~~~~~~+(\frac{\zeta  m^2 a a^{(4)}
}{32 }+\frac{\zeta  m^2 a'^4}{8 a^2 }
-\frac{5 \zeta  m^2 a^{(3)} a'}{16 }
+\frac{13 \zeta  m^2 a'^2 a''}{16 a})\frac{1}{\omega _0{}^5}\nn
\\
&~~~~~~~+(\frac{\zeta ^2 m^4 a^3 a^{(4)}}{64}
-\frac{5 \zeta ^2 m^4 a^2 a''^2}{64 }
+\frac{31 \zeta ^2 m^4 a'^4}{32 }
-\frac{5 \zeta ^2 m^4 a^2 a^{(3)} a'}{32 }
+\frac{9 \zeta ^2 m^4 a a'^2 a''}{8 })\frac{1}{\omega _0{}^7}\nn
\\
&~~~~~~~-(\frac{21 \zeta ^3 m^6 a^4 a''^2}{128 }
+\frac{245 \zeta ^3 m^6 a^2 a'^4}{128 }
+\frac{7 \zeta ^3 m^6 a^4 a^{(3)} a'}{32})\frac{1}{\omega _0{}^9}\nn
\\
&~~~~~~~ +(\frac{63 \zeta ^4 m^8 a^4 a'^4}{64 }
+\frac{231 \zeta ^4 m^8 a^5 a'^2 a''}{128 })
    \frac{1}{\omega _0{}^{11}}-\frac{1155 \zeta ^5
    m^{10} a^6 a'^4}{512 \omega _0{}^{13}}
\nn
\\
& ~~~~~~~  + (\frac{ a'^2 a''}{2  a^3 }
+ \frac{ a''^2}{8  a^2 }
 -\frac{ a^{(3)} a'}{4  a^2 }
  ) \frac{1}{\omega_0{}^3}
 \Big )  ,
   \label{886}
\\
p^{GF}_{k\, A4}
&=\frac{k^3}{2\pi^2a^4}\Big( \frac{\omega _0}{3}
-\frac{\zeta  m^2 a^2}{12 \omega _0}
-(\frac{a''}{3 a }-\frac{a'^2}{2 a^2})\frac{1}{\omega _0}
-(\frac{5 \zeta  m^2 a a''}{24}-\frac{\zeta  m^2 a'^2}{3 })\frac{1}{ \omega _0{}^3}
\nn
\\
&~~~~~~~ -(\frac{\zeta ^2 m^4 a^3 a''}{48 }
-\frac{25 \zeta ^2 m^4 a^2 a'^2}{48 })\frac{1}{\omega _0{}^5}
+\frac{5 \zeta ^3 m^6 a^4 a'^2}{96 \omega _0{}^7}
\nn
\\
&~~~~~~~ +(\frac{5 \zeta  m^2 a a^{(4)}}{96 }
-\frac{5 \zeta  m^2 a''^2}{24 }+\frac{\zeta  m^2 a'^4}{8 a^2 }
-\frac{25 \zeta  m^2 a^{(3)} a'}{48 }
+\frac{37 \zeta  m^2 a'^2 a''}{48 a })\frac{1}{\omega _0{}^5}\nn
\\
&~~~~~~~ +(\frac{\zeta ^2 m^4 a^3 a^{(4)}}{192 }
-\frac{85 \zeta ^2 m^4 a^2 a''^2}{192 }
+\frac{17 \zeta ^2 m^4 a'^4}{32 }
-\frac{65 \zeta ^2 m^4 a^2 a^{(3)} a'}{96}
-\frac{\zeta ^2 m^4 a a'^2 a''}{24 })\frac{1}{\omega _0{}^7}\nn
\\
&~~~~~~~ -(\frac{7 \zeta ^3 m^6 a^4 a''^2}{128}
-\frac{385 \zeta ^3 m^6 a^2 a'^4}{128 }
+\frac{7 \zeta ^3 m^6 a^4 a^{(3)} a'}{96 }
-\frac{245 \zeta ^3 m^6 a^3 a'^2 a''}{48 })\frac{1}{\omega _0{}^9}\nn
\\
&~~~~~~~ -(\frac{399 \zeta ^4 m^8 a^4 a'^4}{64 }
-\frac{77 \zeta ^4 m^8 a^5 a'^2 a''}{128})\frac{1}{\omega _0{}^{11}}
-\frac{385 \zeta ^5 m^{10} a^6 a'^4}{512 \omega _0{}^{13}}
\nn \\
&  ~~~~~ +
(\frac{a^{(4)}}{12 a }-\frac{5 a''^2}{24 a^2 }
-\frac{5 a^{(3)} a'}{12 a^2 }+\frac{2 a'^2 a''}{3 a^3 })\frac{1}{\omega _0{}^3}
\Big ).
\label{877}
\el
}
The last lines in  $\rho^{GF}_{k\, A4}$ and  $p^{GF}_{k\, A4}$
are  respectively  the following
\bl
 \frac{k^3}{2\pi^2a^4}  ( \frac{ a''^2}{8  a^2 }
 -\frac{ a^{(3)} a'}{4  a^2 }
 + \frac{ a'^2 a''}{2  a^3 }) \frac{1}{\omega _0{}^3} ,
  \label{rhovGF}
\\
\frac{k^3}{2\pi^2a^4}
(\frac{2 a'^2 a''}{3 a^3 }
+\frac{a^{(4)}}{12 a }
-\frac{5 a^{(3)} a'}{12 a^2 }
-\frac{5 a''^2}{24 a^2 }
 )\frac{1}{\omega _0{}^3} ,
\label{pvGF}
\el
which are vanishing in de Sitter space,
like \eqref{rhoom3} \eqref{pom3},
and the remarks below \eqref{pom3} also apply here.
In the Landau gauge $\zeta=0$
the terms involving $\zeta$
in $\rho^{GF}_{k\, A 4}$ and  $p^{GF}_{k\, A 4}$ are vanishing
in the de Sitter space,
so that  $\rho^{GF}_{k\, A 4} = \rho^{GF}_{k\, A 2} $ and
 $p^{GF}_{k\, A 4}= p^{GF}_{k\, A 2}$,
and the 4th-order regularization
is effectively the same as the 2nd-order regularization
for the GF stress tensor.

In the massless limit $m=0$,  \eqref{822}$\sim$\eqref{877} become
\bl
& \rho_{k\, A0}^{GF} =\frac{k^4}{2\pi^2a^4} =3 p_{k\, A0}^{GF} ,
\\
& \rho_{k\, A2}^{GF}=\rho_{k\, A4}^{GF}=,,,=
\frac{k^4}{2\pi^2a^4} \Big(1+\frac{1}{2k^2\tau^2}\Big),
\label{GF2msrho}
\\
& p_{k\, A2}^{GF}=p_{k\, A4}^{GF}=,,,=
\frac{k^4}{2\pi^2a^4}\frac13 \Big(1-\frac{1}{2k^2\tau^2}\Big) .
\label{GF2ms}
\el

Analogously,
in the improper 4th-order adiabatic regularization,
one would come up with the trace of the GF subtraction terms
(for $\zeta\ne 0$ only)
\bl
& \lim_{m\rightarrow 0}\int_{0}^{\infty}
\Big(-\rho^{GF}_{k\, A 4} +3 p^{GF}_{k\, A 4}  \Big)\frac{dk}{k}
\label{GFtr}
\\
& = \frac{a^{(4)}}{48\pi^2 a^5}-\frac{3 a''^2}{16 \pi^2 a^6}
-\frac{a^{(3)} a'}{3 \pi^2  a^6}+\frac{3 a'^2 a''}{4 \pi^2 a^7}
\nn
\\
& ~~~ +  \Big(\frac{a^{(4)}}{4 a^5}-\frac{3 a''^2}{4 a^6 }
-\frac{a^{(3)} a'}{a^6 }+\frac{3 a'^2 a''}{2 a^7 }\Big)
\int\frac{k^3}{2\pi^2}\frac{1}{\omega_0^3}\frac{d k}{k}
\label{desugf}
\\
& =  -\frac{3H^4}{4\pi^2}  ~~~ \text{(for the de  Sitter  space)} .
\label{desitsugf}
\el
In  \eqref{GFtr},
integration is performed before taking the massless limit.

To compare with the literature
\cite{Endo1984, ChimentoCossarini1990,ChuKoyama2017,Salas2023},
we also list the 4th-order subtraction terms of
the complex ghost field $\chi$ in the following.
The Lagrangian of the ghost field is
\bl\label{ghlgrx}
{\cal L}^{gh} =  \sqrt{-g} \Big( i g^{\mu\nu} \bar \chi_{;\mu}  \chi_{;\nu}
      +i m^2_\chi \bar\chi \chi  \Big)
\el
the ghost field equation is
\bl
 \eta^{\mu\nu}\partial_{\mu}\partial_{\nu}\chi
   -D \partial_{0}\chi
   - a^2  m_{\chi}^2  \chi=0 ,
   \label{chieqx}
\el
the ghost stress tensor is
\bl\label{ghosttmunux}
T^{gh}_{\mu\nu} = -i \big( \bar\chi_{,\mu} \chi_{,\nu}
                       +  \bar\chi_{,\nu} \chi_{,\mu} \big)
      +i g_{\mu\nu} \big( g^{\sigma\rho}\bar\chi_{,\sigma} \chi_{,\rho}
      + m^2_\chi \, \bar\chi \chi  \big) .
\el
In the de Sitter space,
the  $k$ mode solution of the ghost field is
\be\label{ghosYx}
\chi_k (\tau) \equiv   \frac{1}{a}Y^{(\chi)}_k (\tau) ,
~~ Y^{(\chi)}_k(\tau) = \sqrt{\frac{\pi}{2}} \sqrt{\frac{x}{2k}}
     e^{i \frac{\pi}{2}(\nu_\chi+\frac12)}  H^{(1)}_{\nu_\chi}(x),
\ee
with $\nu_\chi = \sqrt{\frac94 -\frac{m^2_\chi}{ H^2}}$,
the ghost vacuum spectral stress tensor is
\bl \label{ghrx}
 \rho^{gh}_k
& =  \frac{1}{2\pi^2 a^4}
  \Big( -| \Big(\partial_0 -\frac{D}{2} \Big) Y^{(\chi)}_k |^2
    - k^2 | Y^{(\chi)}_k |^2 - a^2 m^2_\chi | Y^{(\chi)}_k  |^2  \Big) ,
\\
  p^{gh}_k
& =  \frac{1}{2\pi^2  a^4}
  \Big( -  | \Big(\partial_0 -\frac{D}{2} \Big) Y^{(\chi)}_k |^2
    + \frac13 k^2 | Y^{(\chi)}_k |^2  + a^2 m^2_\chi | Y^{(\chi)}_k |^2
    \Big) ,
\el
which is $-2$ times that of a minimally-coupling scalar field \cite{ZhangYeWang2020}.
By the  WKB method described in \eqref{Yvn} \eqref{YequaWk},
with $\omega_{\chi} = (k^2  + m_\chi^2 a^2)^{1/2}$ and $\alpha_{\chi}= -\frac{a''}{a}$,
we derive the 0th-order subtraction terms of the ghost spectral stress tensor
\bl
\rho^{gh}_{k\, A 0}
&=\frac{k^3}{2\pi^2a^4}\Big( -\omega_{\chi} \Big) ,
\\
p^{gh}_{k\, A 0}
&=\frac{k^3}{2\pi^2a^4}\Big( -\frac{\omega _{\chi }}{3}
           +\frac{a^2 m_{\chi }^2}{3 \omega _{\chi }} \Big) ,
\el
the 2nd-order subtraction terms
\bl
\rho^{gh}_{k\, A 2}
&=\frac{k^3}{2\pi^2a^4}\Big( -\omega_{\chi}
-\frac{a^2 m_{\chi }^4 a'^2}{8 \omega _{\chi }{}^5}
-\frac{m_{\chi }^2 a'^2}{2 \omega _{\chi }{}^3}
-\frac{a'^2}{2 a^2 \omega _{\chi }} ,
\\
p^{gh}_{k\, A 2}
&=\frac{k^3}{2\pi^2a^4}\Big(
-\frac{5 a^4 m_{\chi }^6 a'^2}{24 \omega _{\chi }{}^7}
-\frac{11 a^2 m_{\chi }^4 a'^2}{24 \omega _{\chi }{}^5}
-\frac{m_{\chi }^2 a'^2}{3 \omega _{\chi }{}^3}
+\frac{a^3 m_{\chi }^4 a''}{12 \omega _{\chi }{}^5}
+\frac{a m_{\chi }^2 a''}{3 \omega _{\chi }{}^3}
\nn
\\
& ~~~ -\frac{a'^2}{2 a^2 \omega _{\chi }}
+\frac{a''}{3 a \omega _{\chi }}
\Big) ,
\el
the 4th-order subtraction (above the 2nd-order) terms
\bl
\rho^{gh}_{k\, trace } & \equiv \rho^{gh}_{k\, A 4} - \rho^{gh}_{k\, A 2}
\nn \\
&
=\frac{k^3}{2\pi^2a^4} \Big(
  -\frac{m_ \chi^4 a^2 a''^2}{32 \omega_\chi ^7}
  -\frac{m_ \chi^2 a''^2}{8 \omega_\chi ^5}
  +\frac{105 m_ \chi^8 a^4 a'^4}{128 \omega_\chi ^{11}}
  +\frac{7 m_ \chi^6 a^2 a'^4}{4 \omega_\chi ^9}-\frac{31 m_ \chi^4 a'^4}{32 \omega_\chi ^7}\nn
  \\
  &-\frac{m_ \chi^2 a'^4}{8 a^2 \omega_\chi ^5}+\frac{m_ \chi^4 a^2 a^{(3)} a'}{16 \omega_\chi ^7}
  +\frac{m_ \chi^2 a^{(3)} a'}{4 \omega_\chi ^5}-\frac{7 m_ \chi^6 a^3 a'^2 a''}{16 \omega_\chi ^9}
  -\frac{9 m_ \chi^4 a a'^2 a''}{8 \omega_\chi ^7}-\frac{3 m_ \chi^2 a'^2 a''}{4 a \omega_\chi ^5}
  \nn
  \\
&   -\frac{a''^2}{8 a^2 \omega_\chi ^3}
  +\frac{a^{(3)} a'}{4 a^2 \omega_\chi ^3}
  -\frac{a'^2 a''}{2 a^3 \omega_\chi ^3}
  \Big),\label{rghc}
\el
and
\bl
p^{gh}_{k\, trace } & \equiv p^{gh}_{k\, A 4} - p^{gh}_{k\, A 2}
\nn
\\
&=\frac{k^3}{2\pi^2a^4} \Big(
  -\frac{m_ \chi^4 a^3 a^{(4)}}{48 \omega_ \chi ^7}
  -\frac{m_ \chi^2 a a^{(4)}}{12 \omega_ \chi ^5}
  +\frac{7 m_ \chi^6 a^4 a''^2}{32 \omega_ \chi ^9}
  +\frac{53 m_ \chi^4 a^2 a''^2}{96 \omega_ \chi ^7}
  +\frac{m_ \chi^2 a''^2}{3 \omega_ \chi ^5}
  \nn
  \\
  &  +\frac{385 m_ \chi^{10} a^6 a'^4}{128 \omega_ \chi ^{13}}
  +\frac{567 m_ \chi^8 a^4 a'^4}{128 \omega_ \chi ^{11}}
  -\frac{91 m_ \chi^6 a^2 a'^4}{32 \omega_ \chi ^9}
  -\frac{17 m_ \chi^4 a'^4}{32 \omega_ \chi ^7}
  -\frac{m_ \chi^2 a'^4}{8 a^2 \omega_ \chi ^5}
  +\frac{7 m_ \chi^6 a^4 a^{(3)} a'}{24 \omega_ \chi ^9}
  \nn
  \\
  &+\frac{37 m_ \chi^4 a^2 a^{(3)} a'}{48 \omega_ \chi ^7}
  +\frac{7 m_ \chi^2 a^{(3)} a'}{12 \omega_ \chi ^5}
  -\frac{77 m_ \chi^8 a^5 a'^2 a''}{32 \omega_ \chi ^{11}}
  -\frac{14 m_ \chi^6 a^3 a'^2 a''}{3 \omega_ \chi ^9}
  +\frac{m_ \chi^4 a a'^2 a''}{24 \omega_ \chi ^7}
  -\frac{5 m_ \chi^2 a'^2 a''}{6 a \omega_ \chi ^5}
  \nn
  \\
& + \frac{5 a''^2}{24 a^2 \omega_ \chi ^3}
  +\frac{5 a^{(3)} a'}{12 a^2 \omega_ \chi ^3}
  -\frac{a^{(4)}}{12 a \omega_ \chi ^3}
  -\frac{2 a'^2 a''}{3 a^3 \omega_ \chi ^3}
  \Big).\label{pghc}
\el
Analogously to the  Stueckelberg field,
in the improper 4th-order   regularization,
one would come up with the trace of the subtraction terms of the ghost field
\bl
& \lim_{m_{\chi} \rightarrow 0}\int_{0}^{\infty}
\Big(-\rho^{gh}_{k\, trace } +3 p^{gh}_{k\, trace }  \Big)\frac{dk}{k}
\nn
\\
& =-\frac{11 a^{(4)}}{240 \pi ^2 a^5}+\frac{13 a''^2}{40 \pi ^2 a^6}
+\frac{a'^4}{120 \pi ^2 a^8}+\frac{13 a^{(3)} a'}{30 \pi ^2 a^6}
-\frac{109 a'^2 a''}{120 \pi ^2 a^7}
\nn
\\
&  ~~~ -  \Big(\frac{a^{(4)}}{4 a^5}-\frac{3 a''^2}{4 a^6 }
  -\frac{a^{(3)} a'}{a^6 }
  +\frac{3 a'^2 a''}{2 a^7 }\Big)
  \int\frac{k^3}{2\pi^2}\frac{1}{\omega_\chi^3}\frac{d k}{k}
\label{ghosttrace}
\\
&  = \frac{119H^4 }{120\pi^2}   ~~~  \text{(for the de  Sitter space)}.
\label{ghosttrdestt}
\el
Summing up the results  \eqref{iontran4rh}  \eqref{longitrace}
\eqref{desu}  \eqref{desugf}  \eqref{ghosttrace},
one would come up with the trace of the 4th-order subtraction terms
for the Stueckelberg + the ghost fields
in the 4th-order adiabatic regularization
\bl
& \lim_{m\rightarrow 0}\int_{0}^{\infty}
\Big(  (-\rho^{TR}_{k\, trace } +3 p^{TR}_{k\, trace } )
+ (-\rho^{L}_{k\, trace } +3 p^{L}_{k\, trace } )
+( -\rho^{T}_{k\, trace } +3 p^{T}_{k\, trace } )
 \nn
 \\
& +( -\rho^{GF}_{k\, trace } +3 p^{GF}_{k\, trace } ) \Big)\frac{dk}{k}
  +\lim_{m_{\chi} \rightarrow 0}\int_{0}^{\infty}
  ( -\rho^{gh}_{k\, trace } +3 p^{gh}_{k\, trace } )  \Big)
\frac{dk}{k}
\nn
\\
& =  - \frac{1}{48}\square R
\Big(
\lim_{m  \rightarrow 0}
\int_{0}^{\infty} \frac{k^3}{2\pi^2}
(\frac{1}{\omega ^3}
-\frac{1}{\omega_0 ^3}
+\frac{2}{\omega_0 ^3} )  \frac{dk}{k}
- \lim_{m_{\chi} \rightarrow 0} \int_{0}^{\infty}
    \frac{k^3}{2\pi^2} \frac{2}{\omega_{\chi} ^3} \frac{dk}{k} \Big)
\nn
\\
&  + \frac{1}{2880\pi^2}\Big( -62(R_{\mu\nu}R^{\mu\nu}-\frac{1}{3}R^2)
   + 18 \square R
     \Big).
\label{tracanmalytotal}
\el
Choosing $m_{\chi}=m$,
so that the divergences in the $k$-integration terms cancel each other,
\bl
\lim_{m  \rightarrow 0}
\int_{0}^{\infty} \frac{k^3}{2\pi^2}
(\frac{1}{\omega ^3}
-\frac{1}{\omega_0 ^3}
+\frac{2}{\omega_0 ^3} )  \frac{dk}{k}
- \lim_{m_{\chi} \rightarrow 0} \int_{0}^{\infty}
    \frac{k^3}{2\pi^2} \frac{2}{\omega_{\chi} ^3} \frac{dk}{k}
 =-\frac{1}{4 \pi^2 }\log \zeta ,
\el
and \eqref{tracanmalytotal} becomes
\bl
& = \frac{1}{2880\pi^2}\Big(-62(R_{\mu\nu}R^{\mu\nu} -\frac{1}{3}R^2)
+(18 + 15\log \zeta) \square R\Big)
\label{rtana}
\\
&=    \frac{31H^4}{120\pi^2}  ~~~  \text{(for the de Sitter space)} .
\label{desittrtana}
\el

\section{ The massless limit of the modes $A$ and $A_0$}

We demonstrate that the solutions \eqref{Asol} \eqref{A0sol}
of the Stueckelberg field
reduce to the solutions of the Maxwell field with the GF term.
At small $m$, by the definitions \eqref{Y3sol} \eqref{Y0solmo},
$\lim_{m\rightarrow0} \nu \simeq \frac12 - \frac{m^2}{H^2}$,
$\lim_{m\rightarrow0}\nu_0 \simeq \frac32 -\frac13 \zeta\frac{m^2}{ H^2}$,
and the Hankel  functions give \cite{NISTHandbook2010},
\bl
\lim_{m^2\rightarrow0}H^{(1)}_{\nu}(z)
&=-i\sqrt{\frac{2}{\pi z}}\Big[
1+\frac{m^2}{H^2}Ei[2iz]e^{-2iz}
 +i\frac\pi2\frac{m^2}{H^2}  \Big] e^{iz}  ,
\\
\lim_{m^2\rightarrow0}H^{(1)}_{\nu_0}( z)
& =-i\sqrt{\frac{2}{\pi z}}\frac1z\Big[(1-iz)
-\frac13\frac{\zeta m^2}{H^2}
\big( 2 -(1+iz )e^{-2 iz}  Ei[2 iz] \big)
\nn \\
& ~~~ +\frac13\frac{\zeta m^2}{H^2}\frac{\pi}{2} (i+z)
 \Big]  e^{iz},
\el
with $ Ei[x] \equiv -\int^\infty_{-x} t^{-1} e^{-t} dt$,
and, by  \eqref{pisoldestt} \eqref{pia0},
\bl
\lim_{m^2\rightarrow0}\pi_A
& \simeq c_1\frac{m a}{k}\frac{1}{\sqrt{2k}}
\Big[ 1+\frac{m^2}{H^2}Ei[-2ik\tau]e^{2ik\tau} \Big]
e^{-ik\tau},
\label{pit}
\\
\lim_{m^2\rightarrow0}\pi_A^0 &\simeq  c_2\frac{m  a}{\sqrt{2k}}
\Big[(1-\frac{i}{k\tau})
+\frac13\frac{\zeta m^2}{H^2}\frac{i}{k\tau}
\big( 2 -(1-ik\tau )e^{2ik\tau}
 Ei[-2 ik\tau] \big) \Big]e^{-ik\tau}. \label{pi0t}
\el
Substituting \eqref{pit} \eqref{pi0t}
into  \eqref{ApiA}  \eqref{A_0} gives
\bl
\lim_{m^2\rightarrow0} A
= &  (c_1-ic_2)  \frac{ k}{m}\frac1a(1-\frac{i}{k\tau} )
\frac{i}{k}\frac{1}{\sqrt{2k}}e^{-i k \tau }
\nn \\
& -c_2 \frac{m}{H} \frac{ 2 i\zeta-3 \frac{c_1}{c_2} +e^{2 i k \tau }
(-3 i \frac{c_1}{c_2}- \zeta ) (i+k \tau )
       Ei[-2 i k \tau ]}{3 k}\frac{1}{\sqrt{2k}}e^{-i k \tau },
   \label{Amless}
\\
\lim_{m^2\rightarrow0} A_0
= & (c_1-i c_2)\frac{ k }{ m}
\frac{1}{a}\frac{1}{\sqrt{2k}}e^{-i k \tau }
\nn \\
& -c_2 \frac{m}{H} \frac{  \zeta  ( k \tau -i)-e^{2 i k \tau }
(3\frac{c_1}{c_2}- i \zeta ) k^2 \tau ^2
    Ei[-2 i k \tau ]}{3 k \tau }\frac{1}{\sqrt{2k}}e^{-i k \tau } .
 \label{A0mless}
\el
Denoting $b_1 \equiv (c_1-ic_2)\frac{k}{m}$, $b_2 \equiv c_2 \frac{m}{H}$,
 \eqref{Amless} \eqref{A0mless} are written as
\bl
\lim_{m^2\rightarrow0} A&=b_1  \frac1{a(\tau)}\frac{i}{k}
\frac{1}{\sqrt{2k}}(1-\frac{i}{k\tau} )e^{-i k \tau }
-b_2 \frac{ 2 i\zeta-3 i +e^{2 i k \tau}(3- \zeta)
  (i+k \tau) Ei[-2 i k \tau ]}{3 k}\frac{1}{\sqrt{2k}}e^{-i k \tau},
        \label{maxwellA}
 \\
\lim_{m^2\rightarrow0} A_0&=b_1 \frac1{a(\tau)}
\frac{1}{\sqrt{2k}}e^{-i k \tau }
-b_2 \frac{\zeta ( k \tau -i)
-e^{2 i k \tau}(3i- i \zeta) k^2 \tau ^2 Ei[-2 i k \tau ]}{3 k \tau }
   \frac{1}{\sqrt{2k}} e^{-i k \tau } .
   \label{maxwellA0}
\el
where $\frac{c_1}{c_2}=i +\frac{m}{k c_2} b_1 \simeq i$
for small $m$ has been used in the inhomogeneous parts.
\eqref{maxwellA} \eqref{maxwellA0} are exactly the solutions (24) (25)
of the Maxwell field with the GF term  in Ref.\cite{ZhangYe2022}.
Substituting \eqref{maxwellA} \eqref{maxwellA0}
into  \eqref{piacm} \eqref{pi0def} give  the canonical momenta
\bl \label{maxp stt}
\pi_A & =  b_2 \frac{i H}{k} a(\tau)
             \frac{1}{\sqrt{2k}} e^{-i k\tau} ,
\\
\label{maxpia0}
\pi_A^0 & =  b_2  H a(\tau)
     \frac{1}{\sqrt{2k}} (1- \frac{i}{k \tau})e^{- i k\tau} .
\el
which are just  (26) (27)  in Ref.\cite{ZhangYe2022}.

\end{document}